\documentclass[printer]{aa}
\usepackage[utf8]{inputenc}
\usepackage[T1]{fontenc}
\usepackage{natbib}
\usepackage{amsmath}
\usepackage{amssymb}
\usepackage{geometry}
\usepackage{graphicx}
\usepackage{bm}
\usepackage{indentfirst}
\usepackage{hyperref}
\usepackage[capitalise]{cleveref} % Nice fancy references
\creflabelformat{equation}{#2\textup{#1}#3}
\usepackage{caption}
\usepackage{subcaption}
\usepackage{xcolor}
\usepackage{txfonts}
\usepackage{ulem}
\bibpunct{(}{)}{;}{a}{}{,}
\usepackage{braket}
\usepackage{relsize}
\usepackage{stackengine,wasysym}
\usepackage{scalerel}
\usepackage{enumitem}

\hypersetup{
    bookmarksopen=true,
    bookmarksopenlevel=0,
    hypertexnames=false,
    colorlinks=true,% Set to false to disable coloring links
    citecolor=blue,% The color of citations
    linkcolor=blue,% The color of references to document elements (sections, figures, etc)
    breaklinks=true,
}

\newcommand{\ii}{\mathrm{j}}
\renewcommand*\d{\mathop{}\!\mathrm{d}}

\newcommand\musum[1]{{\mathlarger{\sum}}_{#1}}
\newcommand\xiosc{\bm{\xi}_\mathrm{osc}}
\newcommand\xiosci[1]{\xi_\mathrm{osc,#1}}
\newcommand\uosc{\mathbf{u}_\mathrm{osc}}
\newcommand\uosci[1]{u_\mathrm{osc,#1}}
\newcommand\ut{\mathbf{u}_\text{t}}
\newcommand\uti[1]{u_{\text{t},#1}}
\newcommand\xit{\bm{\xi}_\text{t}}
\newcommand\xiti[1]{\xi_{\text{t},#1}}

\newcommand\acc[2]{\left\{\vphantom{\dfrac{1}{A}} #1 ~ ; ~ #2\right\}}
\newcommand\lsin[1]{\sin\left(\vphantom{f_1^\tau} #1 \right)}
\newcommand\lcos[1]{\cos\left(\vphantom{f_1^\tau} #1 \right)}
\newcommand\inttau{\displaystyle\int_{-\infty}^0 \d\tau~}

\begin{document}

\title{Coupling between turbulence and solar-like oscillations: A combined Lagrangian PDF/SPH approach}
\subtitle{II -- Mode driving, damping and modal surface effect}
\titlerunning{Coupling between turbulence and solar-like oscillations}

\author{J. Philidet\inst{\ref{inst1}, \ref{inst2}} \and K. Belkacem\inst{\ref{inst2}} \and M.-J. Goupil\inst{\ref{inst2}}}

\institute{Max-Planck-Institut für Sonnensystemforschung, Justus-von-Liebig-Weg 3, 37077 Göttingen, Germany \label{inst1} \and LESIA, Observatoire de Paris, PSL Research University, CNRS, Universit\'e Pierre et Marie Curie, Universit\'e Paris Diderot, 92195 Meudon, France \label{inst2}}

\date{Received 18 December 2021 / Accepted 21 April 2022}

\abstract
{The ever-increasing quality of asteroseismic measurements offers a unique opportunity to use the observed global acoustic modes to infer the physical properties of stellar interiors. In solar-like oscillators, the finite lifetime of the modes allows their amplitudes and linewidths to be estimated, which provide invaluable information on the highly turbulent motions at the top of the convective envelope. But exploiting these observables requires a realistic theoretical framework for the description of the turbulence--oscillation coupling.
}
{The first paper of this series established a linear stochastic wave equation for solar-like $p$-modes, correctly taking the effect of turbulence thereon into account. In this second paper, we aim at deriving simultaneous expressions for the excitation rate, damping rate, and modal surface effect associated with any given $p$-mode, as an explicit function of the statistical properties of the turbulent velocity field.
}
{We reduce the stochastic wave equation to complex amplitude equations for the normal oscillating modes of the system. We then derive the equivalent Fokker-Planck equation that governs the evolution of the probability density function jointly associated with the real amplitudes and phases of all the oscillating modes of the system simultaneously. The effect of the finite-memory time of the turbulent fluctuations (comparable to the period of the modes) on the modes themselves is consistently and rigorously accounted for, by means of the simplified amplitude equation formalism. This formalism accounts for mutual linear mode coupling in full, and we then turn to the special single-mode case. This allows us to derive evolution equations for the mean energy and mean phase of each mode, from which the excitation rate, the damping rate, and the modal surface effect naturally arise.
}
{The expressions obtained here 1) are written as explicit functions of the statistical properties of turbulence, thus allowing for any prescription thereof to be tested against observations, 2) include the contribution of the turbulent dissipation more realistically, and 3) concern the excitation rate, the damping rate, and the modal surface effect of the modes simultaneously. We show that the expression for the excitation rate of the modes is identical to previous results obtained through a different modelling approach, thus supporting the validity of the formalism presented here. We also recover the fact that the damping rate and modal surface effect correspond to the real and imaginary part of the same single complex quantity. We explicitly separate the different physical contributions to these observables, in particular the turbulent pressure contribution and the joint effect of the pressure-rate-of-strain correlation and the turbulent dissipation. We show that the former dominates for high-frequency modes and the latter for low-frequency modes. To illustrate the usefulness of this formalism, we apply it to a simplified case where we can quantify the relative importance of these two contributions, and in particular the threshold between the two frequency regimes, as a function of the turbulent frequency and the degree of anisotropy of both the Reynolds-stress tensor and the dissipation of turbulent energy.
}
{The formalism developed in these first two papers, applied to the case of a simplified Lagrangian stochastic model for proof-of-concept purposes, indeed proves to be viable, relevant, and useful for addressing the issue of turbulence--oscillation coupling in the context of solar-like oscillators. It opens the door to subsequent studies physically more appropriate to the stellar case. It will also allow, once mode coupling is included (i.e. by going beyond the single-mode case), for a realistic description of mode-mode scattering and its influence on mode damping, mode frequency, and the energy distribution across the solar $p$-mode eigenspectrum.
}


\maketitle

\section{Introduction}

Stellar oscillations are a powerful tool for probing the interior of stars. In solar-type stars, these oscillations have a finite lifetime, meaning that it is possible to measure not only the frequency of the resonant modes, but also their line profile in the Fourier domain. While the frequencies give invaluable information about the equilibrium structure of the star, the line profiles carry the signature of the energetic aspects of the modes, which are tightly related to the physics of the turbulence occurring at the top of the convective envelope. As a result, solar-like oscillations offer a unique opportunity to constrain the properties of stellar turbulent convection, which remains to this day one of the major challenges in stellar physics \citep[e.g.][]{kupka17review}.

The coupling between solar-like oscillations and turbulent convection manifests itself in the seismic observables in several ways. First, it is responsible for the discrepancy between the observed $p$-mode frequencies and those theoretically computed from oscillation calculations –- commonly referred to as the `surface effect’ \citep{dziembowski88,jcd96,rosenthal99}. This discrepancy is the most important hurdle on the road to a full exploitation of the asteroseismic diagnosis potential of solar-like $p$-mode frequencies, and it prevents us from accurately inferring the internal equilibrium structure of these stars or their global parameters. It is therefore of utmost importance to correct the surface effect, a task to which numerous studies have been devoted, using either theoretical prescriptions \citep[e.g.][]{gabriel75,balmforth92b,houdek96thesis,rosenthal99,grigahcene05} or empirical formulations \citep[e.g.][]{kjeldsen08,jcd12,ballG14,sonoi15}.

Secondly, the turbulent convective motions in the superadiabatic region are responsible for both the stochastic excitation and the linear damping of solar-like oscillations. As a result, the signature of turbulent convection is also carried by the observed amplitudes and linewidths of the modes, and the latter can be used to constrain the former. This task calls for the development of theoretical prescriptions for both the stochastic driving \citep[e.g.][]{goldreich77a,goldreich77b,balmforth92a,balmforth92c,samadi01a,chaplin05,samadi05,samadi06,belkacem06,belkacem08,belkacem10a} and the linear damping \citep[e.g.][]{goldreich91,balmforth92a,grigahcene05,dupret06,belkacem12} of solar-like $p$-modes.

But because the coupling between turbulent convection and solar-like oscillations occurs predominantly in the superadiabatic region, where heat transfers, convective motions, and oscillations all share the same typical spatial scales and timescales, establishing a theoretical prescription for this coupling constitutes a challenging task. As a result, simplifying assumptions are very often necessary to obtain tractable relations. Many previous theoretical efforts are based on time-dependent mixing-length theories \citep[e.g.][]{gabriel75,houdek96thesis,grigahcene05,sonoi17,houdek17,houdek19}, which amounts to reducing the entire turbulent flow to a single typical spatial scale and timescale, in direct contradiction with the picture of the turbulent cascade in high Reynolds number flows. This constitutes a major flaw of these formalisms, as the turbulent pressure and the turbulent dissipation both derive directly from this turbulent cascade and are deemed to play a major role in turbulence--oscillation coupling. Furthermore, while the mixing-length hypothesis is relevant in the bulk of the convective envelope, it ceases to be valid closer to the surface of the star, as highlighted by 3D hydrodynamic simulations \citep[see][for a review]{nordlund09review}. Another problem with these approaches is the necessity to separate the equations of hydrodynamics into distinct sets of equations for the turbulent convective motions and the oscillations, respectively, either in the form of a separation between typical wavenumbers \citep[e.g.][]{grigahcene05} or by performing horizontal averaging in a simulation \citep[e.g.][]{nordlundS01}. While other approaches have been investigated as an alternative, such as Reynolds-stress models \citep[e.g.][]{xiong00} or the direct extraction of modes from 3D large-eddy simulations \citep[e.g.][]{belkacem19,zhou20}, no ideal solution seems to stand out. Therefore, it remains imperative to seek a more adapted theoretical framework in which to describe this coupling.

In this series of papers, we investigate a novel modelling approach designed to address the issue of turbulence--oscillation coupling in the context of solar-like oscillators, based on Lagrangian stochastic models of turbulence. In the first paper of this series \citep{paper1}, we used a Lagrangian stochastic model to obtain a linear stochastic wave equation that is representative of solar-like $p$-modes and simultaneously contains the effect of the turbulence on the modes. This linear stochastic wave equation constitutes our baseline theoretical framework for quantifying turbulence--oscillation coupling. In this second paper, we exploit it to directly relate the asteroseismic observables -- namely, the amplitudes and linewidths of the modes, as well as the surface effect incurred by their eigenfrequencies -- to the statistical properties of the underlying turbulence. The present formalism differs from previous approaches in several ways: 1) the turbulence is included in its most general form, which allows for any prescription of turbulent convection to be tested against observations; 2) the treatment of turbulent dissipation is more realistic than in mixing-length theory, which allows its impact on turbulence--oscillation coupling to be quantified; 3) all seismic observables (surface effect, mode amplitudes, and linewidths) are quantified simultaneously, thus strengthening the constraints we can put on the properties of turbulent convection.

This paper is structured as follows. In \cref{sec:Prologue} we recall the linear stochastic wave equation we obtained in the first paper of this series \citep{paper1}, and we give a succinct physical interpretation of the various terms that it exhibits. In \cref{sec:Methods} we detail how this stochastic wave equation can be reduced to coupled equations for the complex amplitude of all the individual normal modes of the system, through the application of the simplified amplitude equation formalism, and we explicitly derive this amplitude equation in the case of solar-like $p$-modes. In doing so, we keep the simplifying assumption that the entropy fluctuations of the gas can be neglected. In \cref{sec:SingleMode} we consider the single-mode case, which, for main-sequence solar-like oscillators, is likely to constitute the dominant contribution, even in the presence of mode coupling. This yields simultaneous analytical expressions for the excitation rate, the damping rate, and the modal surface effect associated with any given mode, as an explicit function of the statistical properties of the underlying turbulent velocity field in their most general form, and allows us to separate the contributions to these quantities. To illustrate the usefulness of the present formalism, we apply it to a simplified case in \cref{sec:SimpleApplication}, where the relative importance of the physical contributions to both mode damping and the modal surface effect can be quantified. Finally, we summarise our findings and draw our conclusions in \cref{sec:Conclusion}.

\section{The stochastic wave equation\label{sec:Prologue}}

In this section we summarise the results obtained in \citet{paper1}, hereafter referred to as Paper I. Paper I established a stochastic linear wave equation for solar-like acoustic modes, which consistently encompass the effect of the turbulence, and therefore naturally contains the information on the coupling between turbulence and oscillations. As such, this linear wave equation is stochastic by nature (i.e. it contains an intrinsically random part), and the stochastic part represents the effect of the random turbulent fluctuations on the behaviour of solar-like $p$-modes. To derive the linear stochastic wave equation, we adopt the three following steps.

First, we consider a Lagrangian stochastic model of turbulence, whereby the flow is represented by a large set of individual fluid particles. As a first step, these fluid particles are only characterised by their position and velocity. This Lagrangian stochastic model is then modified to yield stochastic differential equations for Eulerian variables, which are more suited to the description of oscillations. This operation leads us to two stochastic differential equations governing the evolution of the fluid displacement $\bm{\xi}(\mathbf{x},t)$ and velocity $\mathbf{u}(\mathbf{x},t)$ (see Eqs. 25 and 26 of Paper I).

Second, these two equations rely on the estimate of the instantaneous ensemble average of several fluid quantities -- namely the density, pressure, velocity and Reynolds-stress tensor. These instantaneous ensemble averages are estimated directly from the set of fluid particles, in a similar fashion to the smoothed particle hydrodynamics (SPH) formalism. In particular, this step requires the introduction of a kernel function $K(\mathbf{r})$, which serves as a spatial weighting function designed to filter the neighbouring fluid particles for the estimation of the local instantaneous ensemble averages. This step leads us to explicit expressions for these averages, as a function of $\bm{\xi}(\mathbf{x},t)$ and $\mathbf{u}(\mathbf{x},t)$ only (see Eqs. 32 to 35 in Paper I).

Third, we then treat the turbulent part of the displacement variable $\bm{\xi}_\text{t}$ (resp. $\mathbf{u}_\text{t}$ for the velocity variable) as an external input, while the residual $\bm{\xi}_\text{osc} \equiv \bm{\xi} - \bm{\xi}_\text{t}$ (resp. $\mathbf{u}_\text{osc} \equiv \mathbf{u} - \mathbf{u}_\text{t}$) represent the oscillatory component of these variables, including their coupling with turbulence. We note that the equations themselves are not split into turbulence equations and oscillation equations, so that we retain one system of equations for both at the same time, instead of two separate sets. The full system of equations -- namely the fluid equations obtained in step 1) and the mean field expressions obtained in step 2) -- is then linearised in terms of the two wave variables $\bm{\xi}_\text{osc}$ and $\mathbf{u}_\text{osc}$. This is done by adopting a certain number of hypotheses and approximations, all of which are explicitly itemised in Paper I (see hypotheses H1 to H5 in Section 3.1 therein, as well as H6 to H8 in Section 4 therein). Of particular importance is the adiabatic approximation, which we adopted as a first step, and whereby we considered any fluid parcel to conserve its entropy during its evolution. This freed us from having to include an entropy or energy equation in the system, and we instead adopted a polytropic relation between gas pressure and density. In the following, this is what we mean by `adiabatic'. By contrast, we do not consider the oscillations to be adiabatic in the mechanical sense of the term, meaning that they can still exchange energy with the background, thus allowing the contributions of both the driving and the damping of the modes to be included (see discussion in \cref{sec:SingleMode}).

Ultimately, this allows us to establish the following linear stochastic wave equation for solar-like acoustic modes in the presence of turbulent convection (see Eqs. 42 and 43 in Paper I):
\begin{align}
    & \dfrac{\partial \xiosc}{\partial t} - \uosc - (\xiosc \cdot \bm{\nabla})\ut - (\bm{\xi_t} \cdot \bm{\nabla})\uosc = (\bm{\xi_t} \cdot \bm{\nabla})\ut~, \label{eq:DisplacementWaveEquation} \\
    & \dfrac{\partial \uosc}{\partial t} - \mathbf{L}_1^d - \mathbf{L}_1^s = \mathbf{L}_0~.
    \label{eq:VelocityWaveEquation}
\end{align}
The vectors $\mathbf{L}_0$, $\mathbf{L}_1^d$, and $\mathbf{L}_1^s$ are given by (see Eqs. 44 to 46 in Paper I)
\begin{multline}
    L_{1,i}^d = \dfrac{1}{\rho_0} \left( \dfrac{1}{\rho_0}\partial_i p_0 - \partial_i c_0^2 \right) \mathfrak{I}\left( \xiosci{j} \partial_j K^{\mathbf{x}} \right) + \dfrac{c_0^2}{\rho_0} \mathfrak{I} \left( \xiosci{j} \partial_j \partial_i K^{\mathbf{x}} \right) \\
    + G_{ij,0}\left( \uosci{j} - \dfrac{1}{\rho_0} \mathfrak{I}\left( \uosci{j} K^{\mathbf{x}} \right) \right)~,
    \label{eq:DeterministicLinearOperator}
\end{multline}
\begin{multline}
    L_{1,i}^s = -\uosci{j} \partial_j \uti{i} - \uti{j} \partial_j \uosci{i} - G_{ij,0} \dfrac{1}{\rho_0}\mathfrak{I}\left( \xiosci{k} \partial_k \left(\uti{j} K^{\mathbf{x}} \right) \right) \\
    + \left(\dfrac{\partial G_{ij}}{\partial \widetilde{u_k'' u_l''}} \widetilde{u_k'' u_l''}_{(1)} + \dfrac{\partial G_{ij}}{\partial (\partial_k\widetilde{u_l})} \partial_k \widetilde{u_l}_{(1)} + \dfrac{1}{2} \dfrac{\partial G_{ij}}{\partial \epsilon} \omega_t \widetilde{u_i''u_i''}_{(1)} \right) \uti{j} \\
    + \dfrac{1}{4}\sqrt{\dfrac{2 C_0 \omega_t}{\widetilde{u_i''u_i''}_0}} \widetilde{u_i''u_i''}_{(1)} \eta_i
    \label{eq:StochasticLinearOperator}
\end{multline}
and
\begin{equation}
    L_{0,i} = -\dfrac{1}{\rho_0} \partial_j \left( \vphantom{\dfrac{1}{2}} \rho_0 \uti{i}\uti{j} - \rho_0 \overline{\uti{i} \uti{j}} \right)~,
    \label{eq:StochasticDriving}
\end{equation}
where we have defined
\begin{equation}
    \mathfrak{I}(f) \equiv \displaystyle\int \d^3\mathbf{y} ~ \rho_0(\mathbf{y}) f(\mathbf{y})~.
    \label{eq:OperatorIntegral}
\end{equation}
The indices `0' and `1' refer to equilibrium values and fluctuations around the equilibrium values, respectively. The notation $\partial_i$ denotes the $i$-th component of the gradient operator, and we have used Einstein summation convention. The dimensionless quantity $C_0$ denotes the Kolmogorov constant, for which a commonly accepted experimental value is $C_0 = 2.1$ \citep[e.g.][]{haworth86}. We have introduced the centred kernel function $K^{\mathbf{x}}(\mathbf{y}) \equiv K(\mathbf{y} - \mathbf{x})$, and the perturbations of the Reynolds-stress tensor and mean shear tensor are given by (see Eqs. 47 and 48 in Paper I)
\begin{multline}
    \rho_0 \widetilde{u_i'' u_j''}_{(1)} = - \widetilde{u_i''u_j''}_0 \mathfrak{I}\left( \xiosci{k} \partial_k K^{\mathbf{x}} \right) \\
     + \mathfrak{I}\left( \xiosci{k} \partial_k \left(\uti{i} \uti{j} K^{\mathbf{x}}\right) + \uti{i} \uosci{j} K^{\mathbf{x}} + \uti{j} \uosci{i} K^{\mathbf{x}} \right)
    \label{eq:PerturbationReynoldsStressTensor}
\end{multline}
and
\begin{multline}
    \rho_0 (\partial_i \widetilde{u_j})_{(1)} = - \mathfrak{I}\left( \uosci{j} \partial_i K^{\mathbf{x}} + \xiosci{k} \partial_k \left(\uti{j} \partial_i K^{\mathbf{x}}\right) \right) \\
    - \dfrac{1}{\rho_0} \partial_i \rho_0 ~ \mathfrak{I}\left( \uosci{j} K^{\mathbf{x}} + \xiosci{k} \partial_k \left(\uti{j} K^{\mathbf{x}}\right) \right)~.
    \label{eq:PerturbationMeanShearTensor}
\end{multline}
The quantities $\rho_0$, $p_0$, and $\gamma$ are the equilibrium density, equilibrium gas pressure, and polytropic exponent, respectively, and $c_0^2 \equiv p_0\gamma / \rho_0$ is the equilibrium sound speed squared. Finally, by construction, $\eta_i(\mathbf{x},t)$ is defined as a multivariate Gaussian process whose values at two distinct locations are completely uncorrelated, and which verifies
\begin{align}
    & \overline{\bm{\eta}} = \mathbf{0}~, \\
    & \overline{\eta_i(\mathbf{x},t) \eta_j(\mathbf{x},t')} = \delta(t-t')\delta_{ij}~,
\end{align}
where $\delta_{ij}$ is the Kronecker symbol, $\delta$ is the Dirac distribution, and $\overline{\vphantom{a^a} ~.~ }$ denotes the ensemble average.

\Cref{eq:DisplacementWaveEquation,eq:VelocityWaveEquation} put together govern the time evolution of the wave variables $\xiosc$ and $\uosc$ as an explicit function of either the wave variables themselves or the turbulent fields $\xit$ and $\ut$. More specifically, \cref{eq:VelocityWaveEquation} is split three ways. $\mathbf{L}_1^d$ contains all the terms that are linear in the wave variables, but independent of the turbulent fields (they are therefore deterministic). $\mathbf{L}_1^s$ contains all the terms that are linear in the wave variables, and also depend on the turbulent fields (they are therefore stochastic). Finally, $\mathbf{L}_0$ contains all the terms that are independent of the wave variables, and encompass all the inhomogeneous forcing terms.

Formally, the oscillations can be described in a Hilbert space of infinite dimension, where the ket $\ket{\bm{z}}$, in Dirac notation, is defined as\footnote{We note that the oscillatory displacement, $\xiosc$, is multiplied by the angular frequency, $\omega$, of the mode under consideration such that all components of $\ket{\bm{z}}$ have the same dimension, i.e. that of a velocity. This frequency must not be confused with the turbulent frequency $\omega_t$ introduced in the wave equation (see Paper I for more details).}
\begin{equation}
    \ket{\bm{z}} \equiv \Ket{\left\{ \omega\xiosc(\mathbf{x}) ~;~ \uosc(\mathbf{x}) \right\}_{\mathbf{x} \in \mathcal{V}}}~,
    \label{eq:DefinitionKet}
\end{equation}
where $\mathcal{V}$ is the volume of the star, and $\omega$ refers to the angular frequency of the mode under consideration. Then, \cref{eq:DisplacementWaveEquation,eq:VelocityWaveEquation} can be rewritten in the following way:
\begin{equation}
    \dfrac{\d\Ket{\bm{z}}}{\d t} = \mathcal{L}^d\ket{\bm{z}} + \mathcal{L}^s(t)\ket{\bm{z}} + \ket{\Theta(t)}~,
    \label{eq:GeneralWaveEquation}
\end{equation}
where the deterministic, time-independent linear operator $\mathcal{L}^d$ is given by
\begin{equation}
    \Ket{\mathcal{L}^d | \bm{z} } = \Ket{\left\{ \omega\uosc ~;~ \mathbf{L}_1^d \right\}_{\mathbf{x} \in \mathcal{V}}}~,
    \label{eq:KetDeterministicLinearOperator} 
\end{equation}
the stochastic, time-dependent linear operator $\mathcal{L}^s(t)$ is given by
\begin{equation}
    \Ket{\mathcal{L}^s(t) | \bm{z} } = \Ket{\left\{ \omega(\xiosc \cdot \bm{\nabla})\ut + \omega(\xit \cdot \bm{\nabla})\uosc ~;~ \mathbf{L}_1^s \right\}_{\mathbf{x} \in \mathcal{V}}}~,
    \label{eq:KetStochasticLinearOperator}
\end{equation}
the stochastic vector $\Ket{\Theta(t)}$ is given by
\begin{equation}
    \Ket{\Theta(t)} = \Ket{\left\{ \omega (\xit \cdot \bm{\nabla})\ut ~;~ \mathbf{L}_0 \right\}_{\mathbf{x} \in \mathcal{V}}}~,
    \label{eq:KetStochasticDriving}
\end{equation}
and we recall that the vectors $\mathbf{L}_1^d$, $\mathbf{L}_1^s$, and $\mathbf{L}_0$ are given respectively by \cref{eq:DeterministicLinearOperator,eq:StochasticLinearOperator,eq:StochasticDriving}. We note that, since the space dependence of the wave variables is already contained in the infinite dimension of the Hilbert space where the oscillations are described, $\ket{\bm{z}}$ only depends on time, such that $\d ~/~ \d t$ is a total time derivative.

The impact of turbulent convection on the frequency and the damping rate of the acoustic modes is contained within the linear stochastic operator $\mathcal{L}^s(t)$, whereas the driving of the acoustic modes is contained within the additive noise vector $\ket{\Theta(t)}$. Because of their stochastic nature, a large portion of these two quantities is actually incoherent or statistically uncorrelated with the acoustic modes themselves, and therefore incapable of efficiently impacting the oscillations. It is therefore necessary to filter in the part of these stochastic perturbations that does lead to a significant impact on the modes. A naive way to do this would be to simply take the ensemble average of \cref{eq:GeneralWaveEquation} (or, equivalently, of \cref{eq:DisplacementWaveEquation,eq:VelocityWaveEquation}). However, this would not work, because even the zero-mean stochastic part of the turbulent perturbations, being auto-correlated over timescales comparable to the period of the modes, can impact the behaviour of the oscillations. It is therefore necessary to incorporate the effective impact of the finite memory time of the stochastic perturbations on the oscillations. This can be done in the framework of the simplified amplitude equation formalism \citep{stratonovich65,buchlerG84,buchlerGK93}.

\section{The simplified amplitude equations\label{sec:Methods}}

The goal of this section is to use the simplified amplitude equation formalism to properly account for the effect of convective noise on the excitation and damping rates of the normal modes of oscillation. To do so, we proceeded in a two-step approach.

First, the linear stochastic wave equation is transformed into a finite set of stochastic differential equations governing the evolution of the complex amplitudes of the normal modes of oscillation. This is done by treating the stochastic part of the wave equation as perturbations to an otherwise deterministic wave equation. This leads to a set of coupled amplitude equations \citep{buchlerG84}. We present this first step in \cref{sec:AmplitudeEquationFormalism}.

The amplitude equations contain a stochastic part that, as mentioned above, has a finite memory time. In a second step, these are substituted for simplified amplitude equations, which are also stochastic differential equations governing the evolution of the complex amplitude of the modes, but whose stochastic part, unlike the original amplitude equations, has zero memory time, and therefore has no long-term impact on the oscillations. This step involves two procedures. First the original stochastic amplitude equations are substituted for an equation for the probability density function (PDF) of the complex amplitude of the modes, in the form of a Fokker-Planck equation. We present this Fokker-Planck equation in \cref{sec:FokkerPlanckAmplitude}. Then another set of stochastic differential equations is constructed, under the constraints that 1) it must reduce to the exact same Fokker-Planck equation, while 2) only involving Markov processes (i.e. stochastic processes with zero memory time). We derive these simplified amplitude equations in \cref{sec:SimplifiedAmplitudeEquationFormalism}.

\subsection{The full amplitude equations\label{sec:AmplitudeEquationFormalism}}

The stochastic amplitude equations are obtained in two steps \citep{stratonovich65}: first deterministic amplitude equations are derived as if there were no stochastic perturbations; then the stochastic part is treated in a perturbative framework, and stochastic corrections are added to the deterministic amplitude equations after the fact. The non-perturbed wave equation is obtained by setting both $\mathcal{L}^s(t)$ and $\Ket{\Theta(t)}$ to zero in \cref{eq:GeneralWaveEquation}, which yields
\begin{equation}
    \dfrac{\d\Ket{\bm{z}}}{\d t} = \mathcal{L}^d\ket{\bm{z}}~.
    \label{eq:NonPerturbedGeneralWaveEquation}
\end{equation}
The linear operator $\mathcal{L}^d$ can be diagonalised, and any vector $\ket{\bm{z}}$ can be decomposed on a basis of eigenvectors for $\mathcal{L}^d$. Because this linear operator is real, the associated eigenvalues are either real or come in pairs of complex conjugates. Each such pair of eigenvectors with complex-conjugated eigenvalues correspond to a single mode of oscillation of the system; we denote them as $\ket{\Psi^\mu}$ and $\ket{\Psi^{\dagger \mu}}$ (where the exponent notation $\mu$ is used to index the different modes) and their associated eigenvalues as $\kappa_\mu \pm \ii\omega_\mu$, where $\ii$ denotes the imaginary unit.

Without loss of generality, we consider that the eigenvectors are normalised to unity,
\begin{equation}
    \Braket{\Psi^\mu | \Psi^\mu} = \Braket{\Psi^{\dagger\mu} | \Psi^{\dagger\mu}} = 1~,
    \label{eq:NormalisationCondition}
\end{equation}
where $\braket{.|.}$ refers to the scalar product in the $\ket{\bm{z}}$ Hilbert space. This scalar product needs to be specified. A necessary condition is that distinct eigenvectors $\ket{\Psi^\mu}$ of the non-perturbed wave equation (\cref{eq:NonPerturbedGeneralWaveEquation}) -- that is, eigenvectors of the linear operator $\mathcal{L}^d$ -- must be orthogonal to one another in the sense of this scalar product. For adiabatic, non-radial oscillations, this is verified by the following scalar product \citep{unno89}:
\begin{equation}
    \braket{\psi|\phi} \equiv \displaystyle\int \d^3\mathbf{x} ~ \rho_0(\mathbf{x}) \left(\vphantom{f_1^\tau} \bm{\psi}_{\xi}(\mathbf{x}) \cdot \bm{\phi}_{\xi}^\star(\mathbf{x}) + \bm{\psi}_{u}(\mathbf{x}) \cdot \bm{\phi}_{u}^\star(\mathbf{x}) \right)~,
    \label{eq:ScalarProduct}
\end{equation}
where $\bm{\psi}_{\xi}(\mathbf{x})$ (resp. $\bm{\psi}_u(\mathbf{x})$) is the part of $\ket{\psi}$ associated with the oscillatory displacement (resp. velocity) at location $\mathbf{x}$. For any eigenvector $\ket{\Psi}$ of the problem, these two quantities are simply the normalised displacement and velocity eigenfunctions. Because of the normalisation condition (\cref{eq:NormalisationCondition}), they are related to the non-normalised eigenfunctions $\xiosc$ and $\uosc$ through (see \cref{sec:appB-Normalisation} for more details)
\begin{equation}
    \bm{\Psi}^\mu_u = \ii \bm{\Psi}^\mu_\xi = \dfrac{\uosc^\mu}{\sqrt{2\mathcal{I}\omega_\mu^2}} = \ii \dfrac{\xiosc^\mu}{\sqrt{2\mathcal{I}_\mu}}~,
\end{equation}
where $\mathcal{I}_\mu$ is the inertia of the mode, defined by
\begin{equation}
    \mathcal{I}_\mu \equiv \displaystyle\int \d^3\mathbf{x} ~ \rho_0(\mathbf{x}) \left|\xiosc^\mu(\mathbf{x})\right|^2~.
\end{equation}

The general solution of \cref{eq:NonPerturbedGeneralWaveEquation} is \citep{buchlerGK93}
\begin{equation}
    \Ket{\bm{z}(t)} = \musum{\mu} \dfrac{1}{2} a_\mu(t) \exp^{\ii \omega_\mu t} \Ket{\Psi^\mu} + \text{c.c}~,
    \label{eq:NonPerturbedSolution}
\end{equation}
where `c.c' denotes the complex conjugate. In general, $\ket{\bm{z}}$ would be written as an arbitrary linear combination of $\ket{\Psi^\mu}$ and $\ket{\Psi^{\dagger\mu}}$; however, if $\ket{\bm{z}}$ is initially real, \cref{eq:NonPerturbedGeneralWaveEquation} shows that it will remain real at all later times (because $\mathcal{L}^d$ itself is real), so the $\ket{\Psi^{\dagger\mu}}$ component of $\ket{\bm{z}}$ is necessarily the complex conjugate of its $\ket{\Psi^\mu}$ component. As for $a_\mu(t)$, it denotes the slowly varying complex amplitude of the mode $\mu$. In the general case, plugging the solution given by \cref{eq:NonPerturbedSolution} into \cref{eq:NonPerturbedGeneralWaveEquation}, one finds the following, very simple amplitude equation for $a_\mu(t)$:
\begin{equation}
    \dfrac{\d a_\mu}{\d t} = \kappa_\mu a_\mu~.
    \label{eq:AmplitudeEquationUnperturbed}
\end{equation}
As expected, while the imaginary part $\omega_\mu$ of the eigenvalue associated with the mode represents its oscillatory angular frequency, its real part $\kappa_\mu$ represents the rate at which its amplitude varies -- either a growing rate or a damping rate, depending on the intrinsic stability of the mode in the absence of turbulence. However, in the present adiabatic case, since the operator $\mathcal{L}^d$ is Hermitian, we actually have $\kappa_\mu = 0$. The quantity $\kappa_\mu$ therefore contains the contribution of all non-adiabatic effects to the damping of the modes. As it contributes to defining the variation timescale of the mode amplitudes, we keep this term; however, we do not consider it more explicitly, as our present goal is to estimate the additional contribution coming from the interaction with turbulent convection.

Turning now to the full, stochastically perturbed wave equation (\cref{eq:GeneralWaveEquation}), the addition of the two terms $\mathcal{L}^s(t)\ket{\bm{z}}$ and $\ket{\Theta(t)}$ leads to the following modifications to the amplitude equations \citep{buchlerGK93}:
\begin{multline}
    \dfrac{\d a_\mu}{\d t} = \kappa_\mu a_\mu + c_{1\mu} ~ \exp^{- \ii \omega_\mu t} + \musum{\nu} c_{2\mu\nu} ~ a_\nu ~ \exp^{-\ii(\omega_\mu - \omega_\nu)t} \\
    + \musum{\nu} c_{3\mu\nu} ~ a_\nu^\star \exp^{- \ii (\omega_\mu + \omega_\nu) t}~,
\end{multline}
where
\begin{equation}
    c_{1\mu} \equiv 2 \Braket{\Psi^\mu | \Theta}~, ~~ c_{2\mu\nu} \equiv \Braket{\Psi^\mu | \mathcal{L}^s | \Psi^\nu} ~~ \text{and} ~~ c_{3\mu\nu} \equiv \braket{\Psi^\mu | \mathcal{L}^s | \Psi^{\dagger\nu}}~.
\end{equation}
The form of the stochastic linear operator $\mathcal{L}_s(t)$ and stochastic vector $\Ket{\Theta(t)}$ being completely specified by \cref{eq:KetStochasticLinearOperator,eq:KetStochasticDriving}, respectively, and the scalar product being defined by \cref{eq:ScalarProduct}, the stochastic processes $c_i(t)$ ($i = 1, 2, 3$) constitute `knowns' of the system. These are given by
\begin{align}
    & c_{1\mu} = 2 ~ \mathfrak{I}\left(\vphantom{f_1^\tau} \omega_\mu \Psi_{\xi,i}^\mu \xiti{j} \partial_j \uti{i} + \Psi_{u,i}^\mu L_{0,i} \right) \label{eq:ExpressionC1} \\
    & c_{3\mu\nu} = \mathfrak{I}\left( \vphantom{f_1^\tau} \Psi_{\xi,i}^\mu \Psi_{\xi,j}^\nu \partial_j \uti{i} + \omega_\mu \Psi_{\xi,i}^\mu \xiti{j} \partial_j \Psi_{u,i}^\nu + \Psi_{u,i}^\mu L_{1,i}^{s,\nu} \right)~,
    \label{eq:ExpressionC3}
\end{align}
where $L_{1,i}^{s,\nu}$ is obtained by replacing $\xiosc$ and $\uosc$ by the eigenfunctions $\bm{\Psi}_\xi^\nu$ and $\bm{\Psi}_u^\nu$ in \cref{eq:StochasticLinearOperator}.

It will be more practical, in the following, to separate the complex amplitude of the modes into their amplitude $A_\mu(t)$ and phase $\Phi_\mu(t)$, both being real functions of time, such that
\begin{equation}
    a_\mu(t) = A_\mu(t)\exp^{\ii\Phi_\mu(t)}~.
\end{equation}
In place of a single complex stochastic equation for the evolution of $a_\mu(t)$, we obtain two real stochastic equations for the evolution of $A_\mu(t)$ and $\Phi_\mu(t)$, which take the following form:
\begin{align}
    & \dfrac{\d A_\mu}{\d t} \equiv G_\mu(A_\nu, \Phi_\nu, t) \nonumber \\
    & = \kappa_\mu A_\mu + \mathrm{Re}\left[ c_{1\mu} \exp^{-\ii(\omega_\mu t + \Phi_\mu)} + \musum{\nu} A_\nu c_{2\mu\nu} \exp^{-\ii(\omega_\mu - \omega_\nu)t - \ii(\Phi_\mu - \Phi_\nu)} \right. \nonumber \\
    & \hspace{0.5cm} \left. + \musum{\nu} A_\nu c_{3\mu\nu} \exp^{-\ii(\omega_\mu + \omega_\nu) t - \ii(\Phi_\mu + \Phi_\nu)} \right]~,
    \label{eq:RealAmplitudeEquation}
\end{align}
and
\begin{align}
    & \dfrac{\d \Phi_\mu}{\d t} \equiv H_\mu(A_\nu, \Phi_\nu, t) \nonumber \\
    & = \dfrac{1}{A_\mu} \mathrm{Im}\left[ c_{1\mu} \exp^{-\ii(\omega_\mu t + \Phi_\mu)} + \musum{\nu} A_\nu c_{2\mu\nu} \exp^{-\ii(\omega_\mu - \omega_\nu)t - \ii(\Phi_\mu - \Phi_\nu)} \right. \nonumber \\
    & \hspace{0.5cm} \left. + \musum{\nu} A_\nu c_{3\mu\nu} \exp^{-\ii(\omega_\mu + \omega_\nu) t - \ii(\Phi_\mu + \Phi_\nu)} \right]~,
    \label{eq:RealPhaseEquation}
\end{align}
where $\mathrm{Re}$ and $\mathrm{Im}$ denote the real and imaginary parts, respectively. We warn the reader that the function $G_\mu(A_\nu, \Phi_\nu, t)$ defined by the right-hand side of \cref{eq:RealAmplitudeEquation} must not be confused with the drift tensor $G_{ij}$ appearing in \cref{eq:DeterministicLinearOperator,eq:StochasticLinearOperator}.

\subsection{The Fokker-Planck equation for mode amplitude\label{sec:FokkerPlanckAmplitude}}

A fundamental hypothesis we need to make concerning $\mathcal{L}^s(t)$ and $\ket{\Theta(t)}$ is that their correlation timescale is very small compared to the timescale over which the amplitude and the phase of the mode typically vary. In the context of solar-like oscillations, the correlation timescale of these stochastic perturbations correspond to the turnover time of the turbulent eddies, which, close to the surface of the star, is similar to the period of the $p$-modes (i.e. $\sim 5$ minutes). On the other hand, the amplitude and phase of the modes vary over typical timescales that correspond to their life time, which is indeed much longer than their period ($\tau \sim 3$ hours for the shortest-lived solar modes). As such, this hypothesis is largely verified in solar-like oscillators. Consequently, all the turbulent perturbations in the wave equation (\cref{eq:GeneralWaveEquation}) can be approximated by Markov processes (in the sense that their memory time, while finite, is much smaller than the evolution timescale of the amplitude of the modes), and subsequently, so can the stochastic processes $A_\mu(t)$ and $\Phi_\mu(t)$. It is then possible to replace the set of $2N$ (where $N$ is the total number of modes) stochastic differential equations on these quantities (i.e. \cref{eq:RealAmplitudeEquation,eq:RealPhaseEquation}) with an equivalent, single Fokker-Planck equation governing the evolution of their joint PDF $w(A_\mu, \Phi_\mu, t)$, but whose coefficients are carefully defined to incorporate in a rigorous manner the effect of the finite memory time of the processes $A_\mu(t)$ and $\Phi_\mu(t)$. The Fokker-Planck equation takes the general form \citep{stratonovich65}
\begin{equation}
    \dfrac{\partial w}{\partial t} = -\dfrac{\partial w\mathcal{G}_\mu}{\partial A_\mu} - \dfrac{\partial w\mathcal{H}_\mu}{\partial \Phi_\mu} + \dfrac{1}{2}\dfrac{\partial^2 w\mathcal{D}_{\mu\nu}}{\partial A_\mu\partial A_\nu} + \dfrac{1}{2}\dfrac{\partial^2 w\mathcal{E}_{\mu\nu}}{\partial A_\mu\partial \Phi_\nu} + \dfrac{1}{2}\dfrac{\partial^2 w\mathcal{F}_{\mu\nu}}{\partial \Phi_\mu \partial\Phi_\nu}~,
    \label{eq:FokkerPlanckAmplitudePhase}
\end{equation}
where $\mathcal{G}_\mu$ and $\mathcal{H}_\mu$ represent the probability fluxes in ($A_\mu, \Phi_\mu$) space, and $\mathcal{D}_{\mu\nu}$, $\mathcal{E}_{\mu\nu}$ and $\mathcal{F}_{\mu\nu}$ are the elements of the $2N \times 2N$ probability diffusion matrix. The explicit computation of the probability fluxes and diffusion coefficients is detailed in \cref{appA}, and these coefficients read
\begin{multline}
    \mathcal{G}_\mu = A_\mu \left[ \kappa_\mu + \musum{\nu\neq\mu} \mathrm{Re}\left( \alpha_{2\mu\nu}^{(a)} \right) + \alpha_{2\mu}^{(R)} + \musum{\nu} \mathrm{Re}\left( \alpha_{3\mu\nu}^{(a)} \right) \right] \\
    + \dfrac{1}{2A_\mu} \left[ \mathrm{Re}\left( \alpha_{1\mu} \right) + \musum{\nu\neq\mu} A_\nu^2 \mathrm{Re}\left( \alpha_{2\mu\nu}^{(b)} \right) + \musum{\nu} A_\nu^2 \mathrm{Re}\left( \alpha_{3\mu\nu}^{(b)} \right) \right]~,
    \label{eq:ProbabilityFluxAmplitude}
\end{multline}
\begin{equation}
    \mathcal{H}_\mu = \musum{\nu\neq\mu} \mathrm{Im}\left( \alpha_{2\mu\nu}^{(a)} \right) + \musum{\nu} \mathrm{Im}\left( \alpha_{3\mu\nu}^{(a)} \right)~,
    \label{eq:ProbabilityFluxPhase}
\end{equation}
\begin{multline}
    \mathcal{D}_{\mu\nu} = \delta_{\mu\nu}\left[ \mathrm{Re}\left( \alpha_{1\mu} \right) + \musum{\lambda\neq\mu} A_\lambda^2 \mathrm{Re}\left( \alpha_{2\mu\lambda}^{(b)} \right) \right. \\
    \left. + 2 A_\mu^2 \alpha_{2\mu}^{(R)} + \musum{\lambda} A_\lambda^2 \mathrm{Re}\left( \alpha_{3\mu\lambda}^{(b)} \right) \right] \\
    + \dfrac{1}{2}(1 - \delta_{\mu\nu}) A_\mu A_\nu \left[ \mathrm{Re}\left( \alpha_{2\mu\nu}^{(a)} \right) + \mathrm{Re}\left( \alpha_{3\mu\nu}^{(a)} \right) + \mathrm{sym.} \right]~,
    \label{eq:DiffusionCoefficientAmplitude}
\end{multline}
\begin{equation}
    \mathcal{E}_{\mu\nu} = 0~,
    \label{eq:DiffusionCoefficientOffDiagonal}
\end{equation}
\begin{multline}
    \mathcal{F}_{\mu\nu} = \delta_{\mu\nu} \left[ \dfrac{1}{A_\mu^2} \mathrm{Re}\left( \alpha_{1\mu} \right) + \dfrac{1}{A_\mu^2} \musum{\lambda\neq\mu} A_\lambda^2 \mathrm{Re}\left( \alpha_{2\mu\lambda}^{(b)} \right) \right. \\
    \left. + 2 \alpha_{2\mu}^{(I)} + \dfrac{1}{A_\mu^2} \musum{\lambda} A_\lambda^2 \mathrm{Re}\left( \alpha_{3\mu\lambda}^{(b)} \right) \right] \\
    \dfrac{1}{2}(1 - \delta_{\mu\nu}) \left[ \mathrm{Re}\left( \alpha_{3\mu\nu}^{(a)} \right) - \mathrm{Re}\left( \alpha_{2\mu\nu}^{(a)} \right) + \mathrm{sym.} \right]~,
    \label{eq:DiffusionCoefficientPhase}
\end{multline}
where `sym.' refers to a swapping between indices $\mu$ and $\nu$, and we have introduced the following autocorrelation spectra:
\begin{align}
    & \alpha_{1\mu} \equiv \inttau \left\langle c_{1\mu} c_{1\mu}^{\tau\star} \right\rangle \exp^{\ii \omega_\mu \tau}~, \\
    & \alpha_{2\mu\nu}^{(a)} \equiv \inttau \left\langle c_{2\mu\nu} c_{2\nu\mu}^\tau \right\rangle \exp^{\ii(\omega_\mu - \omega_\nu)\tau}~, \\
    & \alpha_{2\mu\nu}^{(b)} \equiv \inttau \left\langle c_{2\mu\nu} c_{2\mu\nu}^{\tau\star} \right\rangle \exp^{\ii(\omega_\mu - \omega_\nu)\tau}~, \\
    & \alpha_{2\mu}^{(R)} \equiv \inttau \left\langle \mathrm{Re}\left( c_{2\mu\mu} \right) \mathrm{Re}\left( c_{2\mu\mu}^\tau \right) \right\rangle~, \\
    & \alpha_{2\mu}^{(I)} \equiv \inttau \left\langle \mathrm{Im}\left( c_{2\mu\mu} \right) \mathrm{Im}\left( c_{2\mu\mu}^\tau \right) \right\rangle~, \\
    & \alpha_{3\mu\nu}^{(a)} \equiv \inttau \left\langle c_{3\mu\nu} c_{3\nu\mu}^{\tau\star} \right\rangle \exp^{\ii(\omega_\mu + \omega_\nu)\tau}~, \\
    & \alpha_{3\mu\nu}^{(b)} \equiv \inttau \left\langle c_{3\mu\nu} c_{3\mu\nu}^{\tau\star} \right\rangle \exp^{\ii(\omega_\mu + \omega_\nu)\tau}~.
\end{align}
It can already be seen, without having to specify the stochastic processes $c_i(t)$, that their autocorrelation spectra are independent of the two stochastic variables $A$ and $\Phi$. Furthermore, under the assumption that the turbulence characterising the convective motions at the top of the envelope of solar-like oscillators is stationary, the autocorrelation functions appearing in the definition of these coefficients are only dependent on the time increment $\tau$, and not on the absolute time $t$. As such, the $\alpha_i$ are also independent of time $t$. All things considered, they are therefore simply complex constants.

\subsection{The equivalent simplified amplitude equations\label{sec:SimplifiedAmplitudeEquationFormalism}}

As we have mentioned above, either the stochastic differential equations (\cref{eq:RealAmplitudeEquation,eq:RealPhaseEquation}) together or the Fokker-Planck equation (\cref{eq:FokkerPlanckAmplitudePhase}), both of which are equivalent to one another, can be used to model the time evolution of the real amplitude and phase of the modes. However, both are equally impractical to use, albeit for different reasons. Numerically integrating the stochastic equations proves extremely expensive, because a large range of very different timescales must be resolved. Indeed, the total integration time must far exceed the typical timescale of the slowly varying mode amplitude. But at the same time, the rapidly varying phase $\omega t$ appearing in \cref{eq:RealAmplitudeEquation,eq:RealPhaseEquation} must be accurately resolved. Finally, the whole range of memory timescales associated with the processes $c_i(t)$ must also be resolved, which is problematic, as it corresponds to the range of timescales in the turbulent cascade, and is therefore very wide in high Reynolds number flows. Consequently, the numerical integration of \cref{eq:RealAmplitudeEquation,eq:RealPhaseEquation} requires an unreasonably small time step compared to the total integration time.

By contrast, the Fokker-Planck equation (\cref{eq:FokkerPlanckAmplitudePhase}) does not have this timescale problem. Indeed, in computing its probability fluxes and diffusion coefficients, we have filtered out all rapidly oscillating features (see \cref{appA} for more details), and as such, the numerical integration of the Fokker-Planck equation does not require these very short timescales to be resolved. However, it poses other difficulties inherent to the integration of Fokker-Planck equations in general. Indeed, because the PDF $w$ is a function of the entire parameter space (i.e. not only of time, but also of the stochastic variables $A_\mu$ and $\Phi_\mu$), its numerical integration would require the discretisation of all three variables. For that reason, it is usually very impractical to directly integrate the Fokker-Planck equation in time.

Fortunately, while a given stochastic differential equation is equivalent to a single Fokker-Planck equation, the opposite is not true. Indeed, there exists an infinite number of stochastic models for the mode amplitudes $A_\mu$ and phases $\Phi_\mu$ that possess the same Fokker-Planck equation (\cref{eq:FokkerPlanckAmplitudePhase}) and therefore contain the exact same statistical information as \cref{eq:RealAmplitudeEquation,eq:RealPhaseEquation} while having a much simpler form. They are the simplified amplitude equations (\citealt{stratonovich65}; see also \citealt{buchlerGK93})
\begin{align}
    & \d A_\mu = \left(\mathcal{G}_\mu - \dfrac{1}{2}\musum{\lambda\nu} \dfrac{\partial \mathcal{D}^{1/2}_{\mu\lambda}}{\partial A_\nu} \mathcal{D}^{1/2}_{\lambda\nu} \right) ~ \d t + \musum{\nu} \mathcal{D}^{1/2}_{\mu\nu} \circ \d W_{A\nu}~, \label{eq:GeneralSimplifiedRealAmplitudeEquation} \\
    & \d \Phi_\mu = \left(\mathcal{H}_\mu - \dfrac{1}{2}\musum{\lambda\nu} \dfrac{\partial \mathcal{F}^{1/2}_{\mu\lambda}}{\partial \Phi_\nu} \mathcal{F}^{1/2}_{\lambda\nu} \right) ~ \d t + \musum{\nu} \mathcal{F}^{1/2}_{\mu\nu} \circ \d W_{\Phi\nu}~,
    \label{eq:GeneralSimplifiedPhaseEquation}
\end{align}
where $\mathcal{D}^{1/2}$ and $\mathcal{F}^{1/2}$ are the square-root of the positive definite matrices\footnote{Because the `off-diagonal' matrix $\mathcal{E}$ is zero, the square-root of the blocks $\mathcal{D}$ and $\mathcal{F}$ correspond to the blocks of the square-root of the total diffusion matrix. We note that this would not be the case if $\mathcal{E}$ had not been zero.} $\mathcal{D}$ and $\mathcal{F}$, and $\d W_{A\nu}$ and $\d W_{\Phi\nu}$ are the increment over $\d t$ of independent Wiener processes, that is, Gaussian processes with zero mean and autocorrelation $\langle \d W_{A\mu} \d W_{A\nu} \rangle = \langle \d W_{\Phi\mu} \d W_{\Phi\nu} \rangle = \delta_{\mu\nu} \d t$. The symbol $\circ$ is the differential notation of the Stratonovich stochastic integral. We note that, in order for the equivalence between the Fokker-Planck equation (\cref{eq:FokkerPlanckAmplitudePhase}) and the stochastic differential equations (\cref{eq:GeneralSimplifiedRealAmplitudeEquation,eq:GeneralSimplifiedPhaseEquation}) to be justified, a sufficient condition is that the probability fluxes $\mathcal{G}_\nu$ and $\mathcal{H}_\nu$, as well as the elements of the diffusion matrix $\mathcal{D}_{\mu\nu}$ and $\mathcal{F}_{\mu\nu}$, only depend on time $t$ and on the variables $A_\mu$ and $\Phi_\mu$ themselves. In fact, in the present case, the situation is even simpler, since these coefficients actually only depend on the real amplitudes $A_\nu$.

These equations are much more practical to handle than either the exact amplitude equations derived in \cref{sec:AmplitudeEquationFormalism} or the Fokker-Planck equation derived in \cref{sec:FokkerPlanckAmplitude}, in the sense that they allow us to circumvent the problems outlined above. Indeed, similarly to the Fokker-Planck equation, all rapidly oscillating terms have been averaged out, which means that the short timescales $\sim \omega^{-1}$ need not be resolved. Furthermore, the stochastic part of these equations now has zero memory -- because the effect of the finite width of the turbulent cascade timescale range has been properly and rigorously incorporated in the coefficients of the Fokker-Planck equation -- which also drastically reduces the range of timescales that we need to resolve. In addition, \cref{eq:GeneralSimplifiedRealAmplitudeEquation,eq:GeneralSimplifiedPhaseEquation} are much easier to integrate numerically than the corresponding Fokker-Planck equation, as only the time variable needs to be discretised. The only cost is that a large number of independent realisations must be integrated in order to reconstruct the moments of the amplitude and phase of the modes.

\section{Single-mode case\label{sec:SingleMode}}

The final result of \cref{sec:Methods} consists in the coupled simplified amplitude equations (\cref{eq:GeneralSimplifiedRealAmplitudeEquation,eq:GeneralSimplifiedPhaseEquation}). One important aspect of these equations is the fact that they are coupled: even in the limit of small amplitudes, where the modes behave linearly, the non-linear nature of the turbulent convection with which they are coupled makes it possible to transfer energy from mode to mode, and for their respective phases to evolve in a dependent manner. This was already noted by previous studies concerning mode scattering in solar-like oscillators \citep[e.g][]{goldreich94} or non-linear mode interactions \citep[e.g][]{kumar89}. This can potentially have an impact on the energy distribution across the $p$-mode spectrum of the star, and may also impact their frequency and their lifetime. That being said, as a first application of the present formalism, we temporarily discard the question of mode coupling and instead focus on the single-mode case. The coupled-mode case, to which we briefly return at the end of \cref{sec:Conclusion}, will be the subject of a later paper in this series.

\subsection{Simplified amplitude equations in the single-mode case}

In the single-mode case, we obtain
\begin{align}
    & \d A = \left(\mathcal{G} - \dfrac{1}{4} \dfrac{\partial \mathcal{D}}{\partial A} \right) ~ \d t + \sqrt{\mathcal{D}} \circ \d W_A~, \label{eq:SingleModeAmplitude} \\
    & \d\Phi = \left(\mathcal{H} - \dfrac{1}{4} \dfrac{\partial \mathcal{F}}{\partial \Phi} \right) ~ \d t + \sqrt{\mathcal{F}} \circ \d W_\Phi~, \label{eq:SingleModePhase}
\end{align}
where the probability fluxes and diffusion matrix elements become
\begin{align}
    & \mathcal{G} = A \left[ \kappa + \alpha_2^{(R)} + \dfrac{3}{2} \mathrm{Re}\left( \alpha_3 \right) \right] + \dfrac{\mathrm{Re}\left(\alpha_1\right)}{2A}~, \label{eq:SingleModeG} \\
    & \mathcal{H} = \mathrm{Im}\left( \alpha_3 \right)~, \label{eq:SingleModeH} \\
    & \mathcal{D} = A^2 \left[ 2 \alpha_2^{(R)} + \mathrm{Re}\left( \alpha_3 \right) \right] + \mathrm{Re}\left( \alpha_1 \right)~, \label{eq:SingleModeD} \\
    & \mathcal{F} = 2 \alpha_2^{(I)} + \mathrm{Re}\left( \alpha_3 \right) + \dfrac{\mathrm{Re}\left( \alpha_1 \right)}{A^2}~, \label{eq:SingleModeF}
\end{align}
and the $\alpha_i$ are defined by
\begin{align}
    & \alpha_1 \equiv \inttau \left\langle c_1 c_1^{\tau\star} \right\rangle \exp^{\ii \omega \tau}~, \label{eq:AutocorrelationSpectrumC1} \\
    & \alpha_2^{(R)} \equiv \inttau \left\langle \mathrm{Re}\left( c_2 \right) \mathrm{Re}\left( c_2^\tau \right) \right\rangle~, \\
    & \alpha_2^{(I)} \equiv \inttau \left\langle \mathrm{Im}\left( c_2 \right) \mathrm{Im}\left( c_2^\tau \right) \right\rangle~, \\
    & \alpha_3 \equiv \inttau \left\langle c_3 c_3^{\tau\star} \right\rangle \exp^{2 \ii \omega \tau}~, \label{eq:AutocorrelationSpectrumC3}
\end{align}
$\omega$ now being the angular frequency of the single mode under consideration. Plugging \cref{eq:SingleModeG,eq:SingleModeH,eq:SingleModeD,eq:SingleModeF} into \cref{eq:SingleModeAmplitude,eq:SingleModePhase}, we obtain
\begin{multline}
    \d A = \left( A\left(\kappa + \mathrm{Re}\left( \alpha_3 \right) \right) + \dfrac{\mathrm{Re}\left(\alpha_1\right)}{2A} \right) ~ \d t  \\
    + \left( A^2 \left[ 2 \alpha_2^{(R)} + \mathrm{Re}\left( \alpha_3 \right) \right] + \mathrm{Re}\left( \alpha_1 \right) \right)^{1/2} \circ \d W_A~,
    \label{eq:SingleModeAmplitudeFinal}
\end{multline}
\begin{equation}
    \d \Phi = \mathrm{Im}\left( \alpha_3 \right) ~ \d t + \left( 2 \alpha_2^{(I)} + \mathrm{Re}\left( \alpha_3 \right) + \dfrac{\mathrm{Re}\left( \alpha_1 \right)}{A^2} \right)^{1/2} \circ \d W_\Phi~.
    \label{eq:SingleModePhaseFinal}
\end{equation}
Alternatively, we can merge these two equations into a single stochastic differential equation on the complex amplitude $a = A\exp(\ii\Phi)$ of the mode, which yields\footnote{Since the diffusion part of the stochastic differential equation is interpreted in the Stratonovich sense, the usual chain rule of differentiation applies.}
\begin{multline}
    \d a = \left[ a \left( \kappa + \alpha_3 \right) + \dfrac{\mathrm{Re}\left( \alpha_1 \right)}{a^\star} \right] ~ \d t \\
    + a \left( 2 \alpha_2^{(R)} + \mathrm{Re}\left( \alpha_3 \right) + \dfrac{\mathrm{Re}\left( \alpha_1 \right)}{|a|^2} \right)^{1/2} \circ \d W_A \\
    + \ii a \left( 2 \alpha_2^{(I)} + \mathrm{Re}\left( \alpha_3 \right) + \dfrac{\mathrm{Re}\left( \alpha_1 \right)}{|a|^2} \right)^{1/2} \circ \d W_\Phi~.
    \label{eq:SingleModeComplexFinal}
\end{multline}

While \cref{eq:SingleModeAmplitudeFinal,eq:SingleModePhaseFinal} -- or, equivalently, \cref{eq:SingleModeComplexFinal} -- feature all of the above autocorrelation spectra, $\alpha_i$, the first terms on their respective right-hand sides (which, as will be discussed in more detail in \cref{subsec:Driving,subsec:DampingSurfaceEffect}, governs the mean evolution of the mode amplitudes) only depend on $\alpha_1$ and $\alpha_3$. These two autocorrelation spectra can be computed by plugging the explicit expressions of $\mathbf{L}_1^s$ (\cref{eq:StochasticLinearOperator}) and $\mathbf{L}_0$ (\cref{eq:StochasticDriving}) into the expressions for $c_1(t)$ and $c_3(t)$ (\cref{eq:ExpressionC1,eq:ExpressionC3}), and then the latter into \cref{eq:AutocorrelationSpectrumC1,eq:AutocorrelationSpectrumC3}. Since $c_1(t)$ and $c_3(t)$ depend on the turbulent velocity field, $\ut$, their autocorrelation spectrum can be described in terms of the autocorrelation spectrum of the turbulent velocity itself. We introduce
\begin{equation}
    \mathcal{C}_{\omega,\mathbf{k}}(f_1 ~;~ f_2) \equiv \displaystyle\int_{-\infty}^0 \d\tau \displaystyle\int \d^3\delta\mathbf{x} ~ \left\langle f_1 f_2^{\tau,\delta\mathbf{x}} \right\rangle \exp^{\ii(\omega\tau - \mathbf{k}\cdot\delta\mathbf{x})}~,
\end{equation}
where
\begin{equation}
    f^{\tau,\delta\mathbf{x}} \equiv f(\mathbf{X}+\delta\mathbf{x}, T+\tau)~.
\end{equation}
Then we can define the second-order turbulent velocity spectrum as
\begin{equation}
    \phi_{ij}^{(2)}(\mathbf{k},\omega) \equiv \mathcal{C}_{\omega,\mathbf{k}}( \uti{i} ~;~ \uti{j} )~,
\end{equation}
as well as the following fourth-order spectra
\begin{align}
    & \phi_{ijkl}^{(4a)}(\mathbf{k},\omega) \equiv \mathcal{C}_{\omega,\mathbf{k}}( \uti{i} ~;~ \uti{j} \uti{k} \uti{l} )~, \\
    & \phi_{ijkl}^{(4b)}(\mathbf{k},\omega) \equiv \mathcal{C}_{\omega,\mathbf{k}}( \uti{i} \uti{j} ~;~ \uti{k} \uti{l} )~, \\
    & \phi_{ijklmn}^{(4c)}(\mathbf{k},\omega) \equiv \mathcal{C}_{\omega,\mathbf{k}}( \uti{i} \partial_m \uti{j} ~;~ \uti{k} \partial_n \uti{l} )~, \\
    & \phi_{ijklm}^{(4d)}(\mathbf{k},\omega) \equiv \mathcal{C}_{\omega,\mathbf{k}}( \uti{i} \partial_m \uti{j} ~;~ \uti{k} \uti{l} )~.
\end{align}

The derivation of the autocorrelation spectra $\alpha_1$ and $\alpha_3$ is detailed in \cref{appB}, and we eventually obtain
\begin{equation}
    \alpha_1 = \dfrac{2}{\mathcal{I}} \mathfrak{I}\left( \rho_0 k_j k_l \xiosci{i} \xiosci{k}^\star \phi_{ijkl}^{(4b)}(\mathbf{k},\omega) \right)~,
    \label{eq:ExpressionAutocorrelationSpectrumC1}
\end{equation}
and
\begin{align}
    & \alpha_3 = \dfrac{1}{4 \mathcal{I}^2} \mathfrak{I}\left( \vphantom{\dfrac{1}{2}} \rho_0 F_i^{(1)} F_j^{(1) \star} \phi_{ij}^{(2)}(2\mathbf{k}, 2\omega) \right. \nonumber \\
	& + \rho_0 F_{ij}^{(2)} F_{kl}^{(2) \star} \phi_{ijkl}^{(4b)}(2\mathbf{k}, 2\omega) + \rho_0 F_{ijm}^{(3a)} F_{kln}^{(3a) \star} \phi_{ijklmn}^{(4c)}(2\mathbf{k}, 2\omega) \nonumber \\
	& \left. \vphantom{\dfrac{1}{2}} + 2\rho_0 ~ \mathrm{Re}\left[F_i^{(1)} F_{jkl}^{(3b) \star} \phi_{ijkl}^{(4a)}(2\mathbf{k}, 2\omega) + F_{ijm}^{(3a)} F_{kl}^{(2) \star} \phi_{ijklm}^{(4d)}(2\mathbf{k}, 2\omega) \right] \right)~,
    \label{eq:ExpressionAutocorrelationSpectrumC3}
\end{align}
where we recall that the operator $\mathfrak{I}$ is defined by \cref{eq:OperatorIntegral}, $\mathbf{k}$ is the (space-dependent) wave vector of the mode, and the functions $F^{(1)}$, $F^{(2)}$, $F^{(3a)}$ and $F^{(3b)}$ are given by\footnote{The subscript $_j$, used as a coordinate index, must not be confused with the in-line notation $\ii$, which refers to the imaginary unit.}
\begin{multline}
	F_i^{(1)} = 4 \ii k_j \xiosci{i} \xiosci{j} + \ii k_i \xiosci{j} \xiosci{j} + \dfrac{G_{ij,0}}{\omega} k_k \xiosci{j} \xiosci{k} \\
	+ \dfrac{\partial G_{ij}}{\partial \widetilde{u_k''u_l''}} \dfrac{\widetilde{u_k''u_l''}_0}{\omega} k_m \xiosci{m} \xiosci{j} + \dfrac{\partial G_{ij}}{\partial (\partial_k\widetilde{u_l})} j k_k \xiosci{j} \xiosci{l} \\
	+ \dfrac{\partial G_{ij}}{\partial \epsilon} \dfrac{\omega_t \widetilde{u_i''u_i''}_0}{2\omega} k_m \xiosci{j} \xiosci{m}~,
	\label{eq:DefinitionF1}
\end{multline}
\begin{equation}
    F_{ij}^{(2)} = \left(\dfrac{\partial G_{ki}}{\partial \widetilde{u_j''u_l''}} + \dfrac{\partial G_{ki}}{\partial \widetilde{u_l''u_j''}} \right) \xiosci{l} \xiosci{k} + \dfrac{\partial G_{ki}}{\partial \epsilon} \omega_t \xiosci{j} \xiosci{k}~,
    \label{eq:DefinitionF2}
\end{equation}
\begin{equation}
    F_{ijk}^{(3a)} = \dfrac{\partial G_{li}}{\partial (\partial_k\widetilde{u_j})} \dfrac{1}{\omega} \ii k_m \xiosci{l} \xiosci{m}
    \label{eq:DefinitionF3a}
\end{equation}
and
\begin{equation}
	F_{ijk}^{(3b)} = -\dfrac{\partial G_{li}}{\partial \widetilde{u_j''u_k''}} \dfrac{1}{\omega} k_m \xiosci{l} \xiosci{m} - \dfrac{1}{2} \dfrac{\partial G_{li}}{\partial \epsilon} \dfrac{\omega_t}{\omega} k_m \xiosci{l} \xiosci{m} \delta_{jk}~. \label{eq:DefinitionF3b}
\end{equation}

These four functions can be broken down into several contributions, thus allowing us to separate the effect of each physical process to the turbulence--oscillation coupling. In the present case, one can identify the turbulent pressure contribution, represented by the advection term in the stochastic velocity equation (see Eq. 26 in Paper I), and which is responsible for the entirety of $\alpha_1$, as well as the first two terms on the right-hand side of \cref{eq:DefinitionF1}. On the other hand, all the other contributions to $\alpha_3$ stem from the $G_{ij}$ term in the stochastic velocity equation, which we recall encompasses the collective effect of the buoyancy force, the fluctuating gas pressure force, and the turbulent dissipation (see Paper I, and more specifically Appendix A therein, for more details). We also note that, \cref{eq:ExpressionAutocorrelationSpectrumC1,eq:ExpressionAutocorrelationSpectrumC3} being written in the form of spatial integrals spanning across the entire volume of the star, this formalism allows us to determine which regions of the star are most responsible for the coupling between the turbulent motions and the oscillatory motions.

\subsection{Mode driving\label{subsec:Driving}}

Multiplying \cref{eq:SingleModeAmplitudeFinal} by $2A$ and taking the ensemble average, we obtain the following evolution equation for the mean\footnote{It must be understood that in this context, the word `mean' refers to ensemble average, and not time average. As such, $E_m$ is susceptible to depend on time.} energy of the mode $E_m(t) \equiv \langle A(t)^2 \rangle$:
\begin{equation}
    \dfrac{\d E_m}{\d t} = 2 E_m \left(\vphantom{f_1^\tau} \kappa + \mathrm{Re}(\alpha_3)\right) + \mathrm{Re}(\alpha_1)~.
    \label{eq:MeanEnergy}
\end{equation}
We note that, because $W_A(t)$ is a Wiener processes, the diffusion part of \cref{eq:SingleModeAmplitudeFinal} (i.e. the second, stochastic term on its right-hand side) is necessarily of zero mean, even when the diffusion coefficient depends explicitly on the stochastic variables themselves \citep[e.g.][]{gardinerBook}. The right-hand side of \cref{eq:MeanEnergy} contains a linear contribution that corresponds to the linear damping of the mode and an additive term, $\mathcal{P} \equiv \mathrm{Re}\left( \alpha_1 \right)$, which corresponds to the excitation rate of the mode (i.e the amount of energy injected into the mode per unit time). We focus on the latter for the moment. It can be seen that $\mathcal{P}$ only contains contributions from the turbulence-induced stochastic perturbation of the wave equation (\cref{eq:GeneralWaveEquation}), in the sense that if $c_1(t)$ was identically zero, $\mathcal{P}$ would also be zero. This is in accordance with the widely acknowledged picture of solar-like oscillations being stochastically excited by highly turbulent motions of the plasma at the top of the convection zone.

Using \cref{eq:AutocorrelationSpectrumC3}, the excitation rate $\mathcal{P}$ of the mode becomes
\begin{equation}
    \mathcal{P} = \dfrac{2}{\mathcal{I}} \displaystyle\int \d^3\mathbf{X} ~ \rho_0^2 k_j k_l \mathrm{Re}\left(\xiosci{i} \xiosci{k}^\star \phi_{ijkl}^{(4b)}(\mathbf{k},\omega) \right)~,
    \label{eq:ExcitationRateGLM}
\end{equation}
where $\mathbf{k}$ and $\omega$ denote the wave vector and angular frequency of the mode. This expression is fully similar to the formulation obtained by previous models for the excitation of solar-like oscillations \citep[see, for instance,][]{samadi01a,chaplin05}, which gives further support to the consistency and validity of the method presented here. First, it is inversely proportional to the inertia of the mode: the larger the mass flow entailed by an oscillating mode, the harder it is to get the flow to actually move. Secondly, it appears that the efficiency of mode driving by turbulence is directly related to the spectrum of the fourth-order moment of turbulent velocity. This is in accordance with widely accepted results of past studies. Finally, the integrand is weighted by a quantity that takes the form of the product of two different components of the wave vector and two different components of the modal fluid displacement. Qualitatively, it can be seen that this is closely related to the square of the mode compressibility $|\bm{\nabla}\cdot\xiosc|$. In other words, the turbulence drives the mode much more efficiently in regions where the compressibility of the mode is high. This is also not surprising, as the turbulent pressure must actually be able to transfer mechanical work to the mode in order to give it energy, and mechanical work is only transferable if the flow undergoes successive compression and dilatation phases.

\subsection{Mode damping and the modal surface effect\label{subsec:DampingSurfaceEffect}}

We now turn to the first term on the right-hand side of \cref{eq:MeanEnergy}. Solar-like oscillations being intrinsically stable, it gives the damping rate of the mode:
\begin{equation}
    \eta_\text{tot} \equiv -\left(\vphantom{f_1^\tau}\kappa + \mathrm{Re}(\alpha_3)\right)~.
    \label{eq:GeneralDampingRate}
\end{equation}
The total damping rate, $\eta_\text{tot}$, is directly related to the observed linewidth at half maximum $\Gamma_\text{obs}$ of the resonant peak corresponding to the mode in the Fourier power spectrum, through
\begin{equation}
    \Gamma_\text{obs} = \eta_\text{tot} / \pi~.
\end{equation}
In the Sun, $\Gamma_\text{obs}$ typically has the order of a few $\mu$Hz, while the eigenfrequency of the modes is much larger ($\sim 3000 ~ \mu$Hz at the maximum spectral height). \cref{eq:GeneralDampingRate} shows that the total damping rate of the mode -- and therefore its observed linewidth -- consists of two different contributions. The first one corresponds to the non-turbulent, deterministic contribution $\kappa$. It represents the effect of the non-adiabatic energy exchanges between the $p$-modes and the background in which they develop. Explicitly modelling $\kappa$ requires the inclusion of non-adiabatic effects in the model, through an energy equation, which is outside the scope of this study (see the discussion in \cref{sec:Conclusion} however). The second one represents the contribution from turbulence, and encompasses a whole array of individual physical contributions from turbulent pressure, turbulent dissipation or convective enthalpy flux for instance. In the following, we only investigate this second contribution, and which we denote as $\eta_\text{turb}$.

As a side note, we remark that taking $\d E_m / \d t = 0$ in \cref{eq:MeanEnergy} (i.e. assuming that the transient has died away and that the mode has reached a stationary state) yields the following expression for the stationary average energy of the mode:
\begin{equation}
    E_\text{stat} \equiv \dfrac{\mathcal{P}}{2\eta_\text{tot}}~.
    \label{eq:GeneralStationaryEnergy}
\end{equation}
This is in perfect accordance with Eq. 5 of \citet{goldreich77a} for instance, and reflects the well-established picture of the equilibrium energy of the mode resulting from a balance between driving and damping processes.

In parallel, taking the ensemble average of \cref{eq:SingleModePhaseFinal}, we obtain the following simple evolution equation for the mean phase $\Phi_m(t) \equiv \langle \Phi(t) \rangle$:
\begin{equation}
    \dfrac{\d\Phi_m}{\d t} = \mathrm{Im}\left( \alpha_3 \right)~,
    \label{eq:MeanPhase}
\end{equation}
which is easily integrated to yield
\begin{equation}
    \Phi_m(t) = \mathrm{Im}(\alpha_3) ~ t + \Phi_{m,0}~,
\end{equation}
where $\Phi_{m,0}$ is an arbitrary initial average phase. In turn, this yields the following expression for the average global phase of the mode $\phi(t) \equiv \omega t + \Phi_m(t)$:
\begin{equation}
    \phi(t) = \left(\vphantom{f_1^\tau}\omega + \mathrm{Im}(\alpha_3)\right) t + \Phi_{m,0}~.
\end{equation}
This amounts to a systematic shift, $\delta\omega$, in the angular frequency of the mode compared to the angular frequency in the absence of turbulence. This shift actually represents what is commonly referred to as the modal -- or `intrinsic' -- part of the surface effect \citep{balmforth92b}, that is, the contribution of turbulent convection not to the equilibrium structure but to the propagation of the waves themselves.

All in all, the turbulence-induced damping rate and the modal surface effect turn out to correspond respectively to the real and imaginary parts of the same complex autocorrelation spectrum $\alpha_3$, according to
\begin{align}
    & \eta_\text{turb} = -\mathrm{Re}\left( \alpha_3 \right)~, \\
    & \delta\omega = \mathrm{Im}\left( \alpha_3 \right)~.
\end{align}
This is not quite surprising, as the damping rate and oscillatory frequency of a mode are themselves simply two sides of the complex eigenvalue associated with the mode -- more specifically, its real and imaginary parts. This is further illustrated by the fact that, in the first term on the right-hand side of \cref{eq:SingleModeComplexFinal}, the linear contribution (i.e. the one proportional to $a$) involves directly the complex quantity $\alpha_3$. This is a very interesting output of this model as mode excitation on the one hand and mode damping and the surface effect on the other hand are usually treated separately. Furthermore, while mixing-length formalisms allow mode damping rates and the surface effect to be investigated simultaneously, they must nevertheless be the subject of separate fitting procedures, owing to the large number of free parameters involved \citep[see for example][]{houdek17}. By contrast, the model presented here allows us to treat all three aspects -- mode excitation, in addition to mode damping and the surface effect -- in the same consistent framework, and without having to resort to separate procedures.

Using the expression for $\alpha_3$ given by \cref{eq:ExpressionAutocorrelationSpectrumC3}, we find
\begin{align}
    & \eta_\text{turb} = -\dfrac{1}{4 \mathcal{I}^2} \displaystyle\int \d^3\mathbf{X} ~ \rho_0^2 ~ \mathrm{Re}(\mathcal{F})~, \label{eq:DampingRateGLM} \\
    & \delta\omega = \dfrac{1}{4 \mathcal{I}^2} \displaystyle\int \d^3\mathbf{X} ~ \rho_0^2 ~ \mathrm{Im}(\mathcal{F})~,
    \label{eq:SurfaceEffectGLM}
\end{align}
where we have defined
\begin{multline}
    \mathcal{F} \equiv \underbrace{F_i^{(1)} F_j^{(1) \star} \phi_{ij}^{(2)}(2\mathbf{k}, 2\omega)}_{\mathrm{second-order}} \\
	+ \underbrace{F_{ij}^{(2)} F_{kl}^{(2) \star} \phi_{ijkl}^{(4b)}(2\mathbf{k}, 2\omega)}_{\mathrm{fourth-order}} + \underbrace{F_{ijm}^{(3a)} F_{kln}^{(3a) \star} \phi_{ijkl}^{(4c)}(2\mathbf{k}, 2\omega)}_{\mathrm{fourth-order}} \\
	+ 2 ~ \mathrm{Re}\left( \underbrace{F_i^{(1)} F_{jkl}^{(3b) \star} \phi_{ijkl}^{(4a)}(2\mathbf{k}, 2\omega)}_{\mathrm{fourth-order}} + \underbrace{F_{ijm}^{(3a)} F_{kl}^{(2) \star} \phi_{ijkl}^{(4d)}(2\mathbf{k}, 2\omega)}_{\mathrm{fourth-order}} \right)~,
	\label{eq:BrokenDownOrders}
\end{multline}
and we recall that the wave vector, $\mathbf{k}$, and the angular frequency, $\omega$, are those of the mode under consideration.

Some general results can be drawn from \cref{eq:DampingRateGLM,eq:SurfaceEffectGLM}. First it can be seen that both quantities go as $\mathcal{I}^{-2}$: similarly to the excitation rate $\mathcal{P}$, the higher the mass flow pertaining to the oscillation, the harder it is to take energy from it. This is due to the fact that the wave variable $\xiosc$ appears to the fourth power in \cref{eq:DampingRateGLM,eq:SurfaceEffectGLM}. Since the inertia of the mode depends on the square of $\xiosc$, the normalisation condition on the eigenfunction naturally involves the square of the inertia (see \cref{sec:appB-Normalisation} for more details). This is similar to the work integral formulation of the damping rate \citep[e.g.][]{belkacem15review}, where the wave variables appear to the second power, and therefore the damping rate goes as $\mathcal{I}^{-1}$ instead. We note that the reason why the integrals on the right-hand side of \cref{eq:DampingRateGLM,eq:SurfaceEffectGLM} involve the fourth power of the wave variables is that we compute the mean effect of the damping on the mode, which corresponds to the autocorrelation of the instantaneous damping: since the latter depends on the square of the wave variables, the autocorrelation depends on their fourth power.

Secondly, while the driving source is a fourth-order quantity in terms of the turbulent velocity (see \cref{eq:ExcitationRateGLM} for the excitation rate $\mathcal{P}$), the damping rate and the modal surface effect, by contrast, have both a second-order and a fourth-order dependence in terms of the turbulent velocity. These are explicitly broken down in \cref{eq:BrokenDownOrders}.

Finally, it is possible to separate the different physical contributions to the turbulent damping rate $\eta_\text{turb}$ and the modal surface effect $\delta\omega$. To that end, we decomposed the function $F_i^1$ (defined by \cref{eq:DefinitionF1}) into three different terms, in the following way:
\begin{equation}
    F_i^{(1)} = F_{i,\text{pt}}^{(1)} + F_{i,\text{G}}^{(1)} + F_{i,\delta\text{G}}^{(1)}~,
    \label{eq:DecomposedF1}
\end{equation}
where
\begin{align}
    & F_{i,\text{pt}}^{(1)} \equiv 4 \ii k_j \xiosci{i} \xiosci{j} + \ii k_i \xiosci{j} \xiosci{j}~, \\
    & F_{i,\text{G}}^{(1)} \equiv \dfrac{G_{ij,0}}{\omega} k_k \xiosci{j} \xiosci{k}~, \\
	& F_{i,\delta\text{G}}^{(1)} \equiv \dfrac{\partial G_{ij}}{\partial \widetilde{u_k''u_l''}} \dfrac{\widetilde{u_k''u_l''}_0}{\omega} k_m \xiosci{m} \xiosci{j} + \dfrac{\partial G_{ij}}{\partial (\partial_k\widetilde{u_l})} \ii k_k \xiosci{j} \xiosci{l} \nonumber \\
	& \hspace{1cm} + \dfrac{\partial G_{ij}}{\partial \epsilon} \dfrac{\omega_t \widetilde{u_i''u_i''}_0}{2\omega} k_m \xiosci{j} \xiosci{m}~.
\end{align}
The first term represents the effect of turbulent pressure; the second term represents the joint effect of the pressure-rate-of-strain correlation and the dissipation of turbulent kinetic energy into heat, both of which are collectively modelled by means of the drift tensor, $G_{ij}$; and the third term stems from the variations in the drift tensor, $G_{ij}$, under the influence of both the turbulence and the oscillations. By contrast, the functions $F_{ij}^{(2)}$, $F_{ijk}^{(3a)}$, and $F_{ijk}^{(3b)}$, defined by \cref{eq:DefinitionF2,eq:DefinitionF3a,eq:DefinitionF3b}, respectively, entirely originate from the modulation of the drift tensor. Then \cref{eq:DampingRateGLM,eq:SurfaceEffectGLM} can be rewritten in the following way:
\begin{align}
    & \eta_\text{turb} = \eta_\text{pt} + \eta_\text{G} + \eta_{\delta\text{G}} + \eta_\text{cross}~, \label{eq:DecomposedDampingGLM} \\
    & \delta\omega = \delta\omega_\text{pt} + \delta\omega_\text{G} + \delta\omega_{\delta\text{G}} + \delta\omega_\text{cross}~,
    \label{eq:DecomposedSurfaceEffectGLM}
\end{align}
where the various physical contributions to the total damping rate are given by
\begin{align}
    & \eta_\text{pt} \equiv -\dfrac{1}{4\mathcal{I}^2} \displaystyle\int \d^3\mathbf{X} ~ \rho_0^2 ~ \mathrm{Re}\left( \mathcal{F}_\text{pt} \right)~, \label{eq:DampingPt} \\
    & \eta_\text{G} \equiv -\dfrac{1}{4\mathcal{I}^2} \displaystyle\int \d^3\mathbf{X} ~ \rho_0^2 ~ \mathrm{Re}\left( \mathcal{F}_\text{G} \right)~, \label{eq:DampingG} \\
    & \eta_{\delta\text{G}} \equiv -\dfrac{1}{4\mathcal{I}^2} \displaystyle\int \d^3\mathbf{X} ~ \rho_0^2 ~ \mathrm{Re}\left( \mathcal{F}_{\delta\text{G}} \right)~, \label{eq:DampingDeltaG} \\
    & \eta_\text{cross} \equiv -\dfrac{1}{4\mathcal{I}^2} \displaystyle\int \d^3\mathbf{X} ~ \rho_0^2 ~ \mathrm{Re}\left( \mathcal{F}_\text{cross} \right)~, \label{eq:DampingCross}
\end{align}
those of the modal surface effect are given by
\begin{align}
    & \delta\omega_\text{pt} \equiv \dfrac{1}{4\mathcal{I}^2} \displaystyle\int \d^3\mathbf{X} ~ \rho_0^2 ~ \mathrm{Im}\left( \mathcal{F}_\text{pt} \right)~, \label{eq:SurfaceEffectPt} \\
    & \delta\omega_\text{G} \equiv \dfrac{1}{4\mathcal{I}^2} \displaystyle\int \d^3\mathbf{X} ~ \rho_0^2 ~ \mathrm{Im}\left( \mathcal{F}_\text{G} \right)~, \label{eq:SurfaceEffectG} \\
    & \delta\omega_{\delta\text{G}} \equiv \dfrac{1}{4\mathcal{I}^2} \displaystyle\int \d^3\mathbf{X} ~ \rho_0^2 ~ \mathrm{Im}\left( \mathcal{F}_{\delta\text{G}} \right)~, \label{eq:SurfaceEffectDeltaG} \\
    & \delta\omega_\text{cross} \equiv \dfrac{1}{4\mathcal{I}^2} \displaystyle\int \d^3\mathbf{X} ~ \rho_0^2 ~ \mathrm{Im}\left( \mathcal{F}_\text{cross} \right)~, \label{eq:SurfaceEffectCross}
\end{align}
and we have defined
\begin{align}
    & \mathcal{F}_\text{pt} \equiv F_{i,\text{pt}}^{(1)} F_{j,\text{pt}}^{(1) \star} ~ \phi_{ij}^{(2)}~, \\
    & \mathcal{F}_\text{G} \equiv F_{i,\text{G}}^{(1)} F_{j,\text{G}}^{(1) \star} ~ \phi_{ij}^{(2)}~, \\
    & \mathcal{F}_{\delta\text{G}} \equiv F_{i,\delta\text{G}}^{(1)} F_{j,\delta\text{G}}^{(1) \star} ~ \phi_{ij}^{(2)} + F_{ij}^{(2)} F_{kl}^{(2) \star} ~ \phi_{ijkl}^{(4b)} + F_{ijm}^{(3a)} F_{kln}^{(3a) \star} ~ \phi_{ijkl}^{(4c)} \nonumber \\
    & \hspace{0.5cm} + 2 ~ \mathrm{Re}\left( F_{i,\delta\text{G}}^{(1)} F_{jkl}^{(3b) \star} ~ \phi_{ijkl}^{(4a)} + F_{ijm}^{(3a)} F_{kl}^{(2) \star} ~ \phi_{ijkl}^{(4d)} \right)~, \\
	& \mathcal{F}_\text{cross} \equiv \left( \mathlarger{\sum}_{\mu,\nu = \text{`pt', `G', `} \delta\text{G'}}^{\mu \neq \nu} F_{i,\mu}^{(1)} F_{j,\nu}^{(1) \star} \right) ~ \phi_{ij}^{(2)} \nonumber \\
	& \hspace{0.5cm} + 2 ~ \mathrm{Re}\left( F_{i,\text{pt}}^{(1)} F_{jkl}^{(3b) \star} ~ \phi_{ijkl}^{(4a)} + F_{i,\text{G}}^{(1)} F_{jkl}^{(3b) \star} ~ \phi_{ijkl}^{(4a)} \right)~.
\end{align}
All the functions $F^{(1)}$, $F^{(2)}$ and $F^{(3)}$ are implicitly evaluated at $\mathbf{X}$, and all the turbulent spectra $\phi^{(2)}$ and $\phi^{(4)}$ are implicitly evaluated at $(2\mathbf{k},2\omega)$, even though we dropped the notation for the sake of clarity. The contributions labelled `pt', `$\text{G}$', and `$\delta\text{G}$' have the same meaning as in \cref{eq:DecomposedF1}, and those labelled `cross' are cross terms between these different contributions. This decomposition shows that the first two contributions only depend on the second-order turbulent velocity spectrum, while the others depend on both the second- and fourth-order spectra.

\subsection{Contributions of turbulent dissipation and redistribution versus turbulent pressure\label{sec:SimpleApplication}}

In \cref{subsec:DampingSurfaceEffect} we derived general expressions for the various physical contributions to mode damping (see \cref{eq:DampingPt,eq:DampingG,eq:DampingDeltaG,eq:DampingCross}) and the modal surface effect (see \cref{eq:SurfaceEffectPt,eq:SurfaceEffectG,eq:SurfaceEffectDeltaG,eq:SurfaceEffectCross}). The resulting expressions are valid for any specification of the drift tensor $G_{ij}$, and regardless of whether the mode is radial or not. While we postpone more realistic applications of these expressions to a later paper in this series, it is still worthwhile to examine their output in the following simple case. First, we assume that the drift tensor is constant, in the sense that it does not depend on the Reynolds-stress tensor, $\widetilde{u_k''u_l''}$, the mean shear tensor, $\partial_k\widetilde{u_l}$, or the turbulent dissipation rate, $\epsilon$
\begin{equation}
    \dfrac{\partial G_{ij}}{\partial \widetilde{u_k'' u_l''}} = \dfrac{\partial G_{ij}}{\partial (\partial_k \widetilde{u_l})} = \dfrac{\partial G_{ij}}{\partial \epsilon} = 0~.
\end{equation}
Physically, this means that the rate at which the different components of the turbulent motions exchange energy, or are dissipated into heat, is fixed. We note that this assumption is not very realistic, but that it allows for a simple illustration of the usefulness of this formalism. However, we do not specify how the behaviour of these different components compare to each other, meaning that we do not make any assumption regarding the relative value of the different components of $G_{ij}$. Secondly, we consider a radial mode, such that both the displacement eigenfunction and the wave vector of the mode are radial:
\begin{align}
    & \xiosc = \xi_r ~ \mathbf{e_r}~, \\
    & \mathbf{k} = k_r ~ \mathbf{e_r}~.
\end{align}

Then \cref{eq:DampingPt,eq:DampingG,eq:DampingDeltaG} reduce to
\begin{align}
    & \eta_\text{pt} \equiv -\dfrac{1}{4\mathcal{I}^2} \displaystyle\int \d^3\mathbf{X} ~ 25 \rho_0^2 k_r^2 \left| \xi_r \right|^4 \mathrm{Re}\left( \phi_{rr}^{(2)} \right)~, \\
    & \eta_\text{G} \equiv -\dfrac{1}{4\mathcal{I}^2} \displaystyle\int \d^3\mathbf{X} ~ \rho_0^2 k_r^2 \left| \xi_r \right|^4 \dfrac{G_{ir} G_{jr}}{\omega^2} \mathrm{Re}\left( \phi_{ij}^{(2)} \right)~, \\
    & \eta_{\delta G} = 0~,
\end{align}
and \cref{eq:SurfaceEffectPt,eq:SurfaceEffectG,eq:SurfaceEffectDeltaG} become
\begin{align}
    & \delta\omega_\text{pt} \equiv \dfrac{1}{4\mathcal{I}^2} \displaystyle\int \d^3\mathbf{X} ~ 25 \rho_0^2 k_r^2 \left| \xi_r \right|^4 \mathrm{Im}\left( \phi_{rr}^{(2)} \right)~, \\
    & \delta\omega_\text{G} \equiv \dfrac{1}{4\mathcal{I}^2} \displaystyle\int \d^3\mathbf{X} ~ \rho_0^2 k_r^2 \left| \xi_r \right|^4 \dfrac{G_{ir} G_{jr}}{\omega^2} \mathrm{Im}\left( \phi_{ij}^{(2)} \right)~, \\
    & \delta\omega_{\delta G} = 0~.
\end{align}
We note that, to keep the discussion simple, we omit the crossed contributions $\eta_\text{cross}$ and $\delta\omega_\text{cross}$ in the following. The only remaining contributions to both the damping rate of the modes and their modal surface effect are the contribution of turbulent pressure, and the contribution of the drift tensor $G_{ij}$, which we recall here jointly models the pressure-rate-of-strain correlations and the turbulent dissipation.

These simplified expressions allow us to roughly estimate which process contributes predominantly to both the damping rate and the modal surface effect, depending on the frequency regime. Under the assumption that the region of the star where the damping of the modes and the surface effects occur is a thin subsurface layer, then we can write
\begin{align}
    & \dfrac{\eta_\text{pt}}{\eta_\text{G}} \equiv R_\eta \sim \dfrac{25 \omega^2 \mathrm{Re}\left( \phi_{rr}^{(2)} \right)}{G_{ir}G_{jr} \mathrm{Re}\left( \phi_{ij}^{(2)} \right)}~, \\
    & \dfrac{\delta\omega_\text{pt}}{\delta\omega_\text{G}} \equiv R_{\delta\omega} \sim \dfrac{25 \omega^2 \mathrm{Im}\left( \phi_{rr}^{(2)} \right)}{G_{ir}G_{jr} \mathrm{Im}\left( \phi_{ij}^{(2)} \right)}~.
\end{align}
Explicitly expanding the index summation in the denominator, we obtain
\begin{equation}
    G_{ir}G_{jr} \phi_{ij}^{(2)} = 2G_{hr}^2 \phi_{hh}^{(2)} + G_{rr}^2 \phi_{rr}^{(2)}~,
\end{equation}
where the index `h' refers to an arbitrary horizontal direction. Then
\begin{align}
    & R_\eta = \dfrac{25 \omega^2}{G_{rr}^2}\left( \vphantom{\dfrac{1}{2}} 1 + 2 \Phi_G^2 \Phi_{u,\eta} \right)^{-1}~, \\
    & R_{\delta\omega} = \dfrac{25 \omega^2}{G_{rr}^2}\left( \vphantom{\dfrac{1}{2}} 1 + 2 \Phi_G^2 \Phi_{u,\delta\omega} \right)^{-1}~,
\end{align}
where the anisotropy factors are defined by
\begin{align}
    & \Phi_{u,\eta} \equiv \left| \dfrac{\mathrm{Re}\left( \phi_{hh}^{(2)} \right)}{\mathrm{Re}\left( \phi_{rr}^{(2)} \right)} \right|~, \\
    & \Phi_{u,\delta\omega} \equiv \left| \dfrac{\mathrm{Im}\left( \phi_{hh}^{(2)} \right)}{\mathrm{Im}\left( \phi_{rr}^{(2)} \right)} \right|~, \\
    & \Phi_G \equiv \left| \dfrac{G_{rh}}{G_{rr}} \right|~.
\end{align}
The anisotropy factors $\Phi_{u,\eta}$ and $\Phi_{u,\delta\omega}$ concern the turbulent velocity field. By contrast, the anisotropy factor $\Phi_G$ concerns the drift tensor $G_{ij}$, and physically represents the ratio between the rate at which vertical and horizontal turbulent motions exchange energy, and the rate at which the vertical motions are dissipated into heat.

Regardless of the relative weight of the different components of $G_{ij}$, their order of magnitude is always $\sim \epsilon / k = \omega_t$ \citep[e.g.][]{pope94}, where we recall that $\epsilon$ is the turbulent dissipation rate, $k$ is the turbulent kinetic energy, and the ratio between the two defines the turbulent frequency $\omega_t$ already introduced above. We note that in the present model, $\omega_t$ is assumed constant, but not necessarily uniform. Physically, it represents the lifetime of the energy-bearing eddies in the turbulent cascade, and corresponds to the autocorrelation timescale of granulation. Typically, close to the surface of the Sun, this timescale is $\tau_\text{conv} \sim 300 - 500$ s \citep[e.g.][]{belkacem15review}, in which case $\omega_t \sim 0.01 - 0.02$ rad.s$^{-1}$. In particular, in the Simplified Langevin Model, every diagonal component of the drift tensor is equal to \citep[e.g.][]{popeBook}
\begin{equation}
    G_{ii} = - \left( \dfrac{1}{2} + \dfrac{3}{4} C_0 \right) \omega_t~,
\end{equation}
where $C_0 = 2.1$ is the Kolmogorov constant. Then, we finally obtain
\begin{equation}
    R_{\eta / \delta\omega} = \dfrac{25\omega^2}{\omega_t^2} \left( \dfrac{1}{2} + \dfrac{3}{4} C_0 \right)^{-2} \left( \vphantom{\dfrac{1}{2}} 1 + 2 \Phi_G^2 \Phi_{u,\eta / \delta\omega} \right)^{-1}~.
    \label{eq:ExpressionRatio}
\end{equation}

In the following, we consider indifferently either one of the two ratios $R_\eta$ or $R_{\delta\omega}$, which we simply denote as $R$. The turbulent velocity anisotropy factor will indifferently refer to $\Phi_{u,\eta}$ or $\Phi_{u,\delta\omega}$, which we simply denote as $\Phi_u$. It can be thought of as representing the ratio between the squared horizontal and vertical turbulent velocities, and can therefore be related to the perhaps more familiar anisotropy factor $\Phi$ introduced by \citet{gough77a} through
\begin{equation}
    \Phi \sim 1 + 2 \Phi_u~.
\end{equation}

\begin{figure*}
    \centering
    \includegraphics[width=\linewidth,trim={4cm 0 4cm 0}]{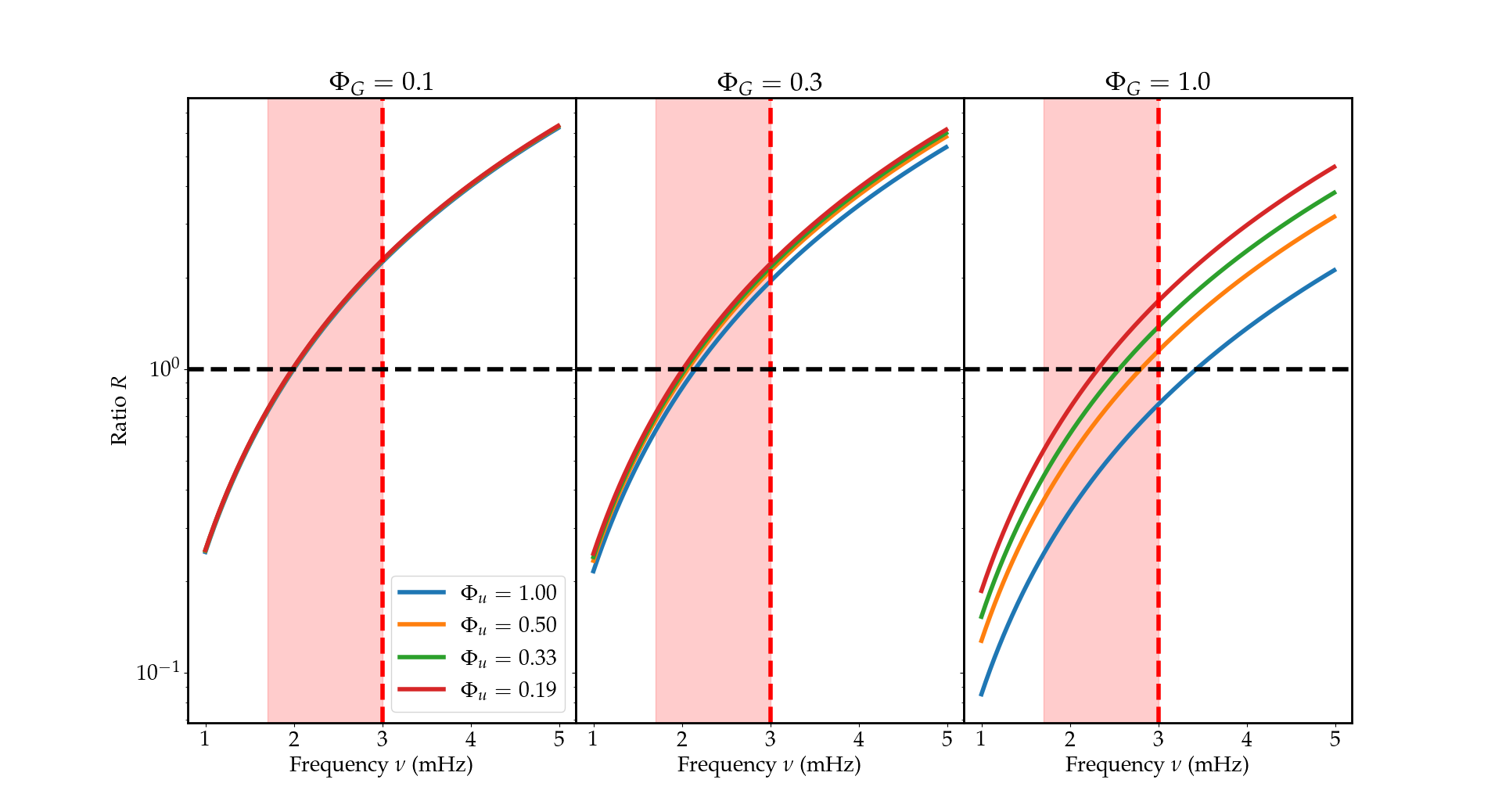} \\
    \caption{Ratio between the two physical contributions to either mode damping or the modal surface effect (turbulent pressure vs. turbulent dissipation and redistribution of kinetic energy), as a function of mode frequency. Each panel corresponds to a fixed value of the drift tensor anisotropy, $\Phi_G$, and the colour code refers to the turbulent velocity anisotropy factor, $\Phi_u$. In particular, $\Phi_u = 1$ corresponds to the isotropic case, and $\Phi_u = 0.5$ corresponds to the anisotropy measured in 3D large-eddy simulations \citep[e.g.][]{samadi03}. The turbulent frequency is set to $\omega_t = 0.03$ rad.s$^{-1}$. The region filled in red corresponds to the solar damping plateau, and the vertical dashed red line marks the frequency of maximum spectral height in the Sun, $\nu_\text{max}$. The horizontal line marks the threshold $R = 1$: above this line, the turbulent pressure contribution dominates, and below this line the joint effect of turbulent dissipation and pressure-rate-of-strain correlation prevails. The vertical axis is logarithmic.}
    \label{fig:RvsNu}
\end{figure*}

In \cref{fig:RvsNu}, we plot the ratio $R$ between the two contributions to either mode damping or the modal surface effect, as a function of mode frequency. While the turbulent frequency $\omega_t$ is fixed (we set it to $\omega_t = 0.03$ rad.s$^{-1}$), we vary the two anisotropy factors $\Phi_u$ and $\Phi_G$ (see caption). As expected from \cref{eq:ExpressionRatio}, the turbulent pressure contribution tends to dominate for high-frequency modes, while the joint effect of the pressure-rate-of-strain correlation and of the turbulent dissipation tends to prevail for low-frequency modes. However, the threshold between these two regimes drastically depends on $\Phi_u$ and $\Phi_G$. More specifically, for low values of $\Phi_G$, this threshold is on the low-frequency edge of the solar damping plateau, around $2$ mHz, and does not depend on $\Phi_u$. For higher values of $\Phi_G$ -- and in particular for $\Phi_G = 1$, which corresponds to a vertical-horizontal turbulent energy redistribution rate similar to the vertical turbulent dissipation rate --, this threshold is shifted to higher frequencies, and has a much stronger dependence on $\Phi_u$.

\begin{figure*}
    \centering
    \begin{subfigure}{0.49\textwidth}
        \centering
        \includegraphics[width=\linewidth,trim={4cm 0 2cm 0}]{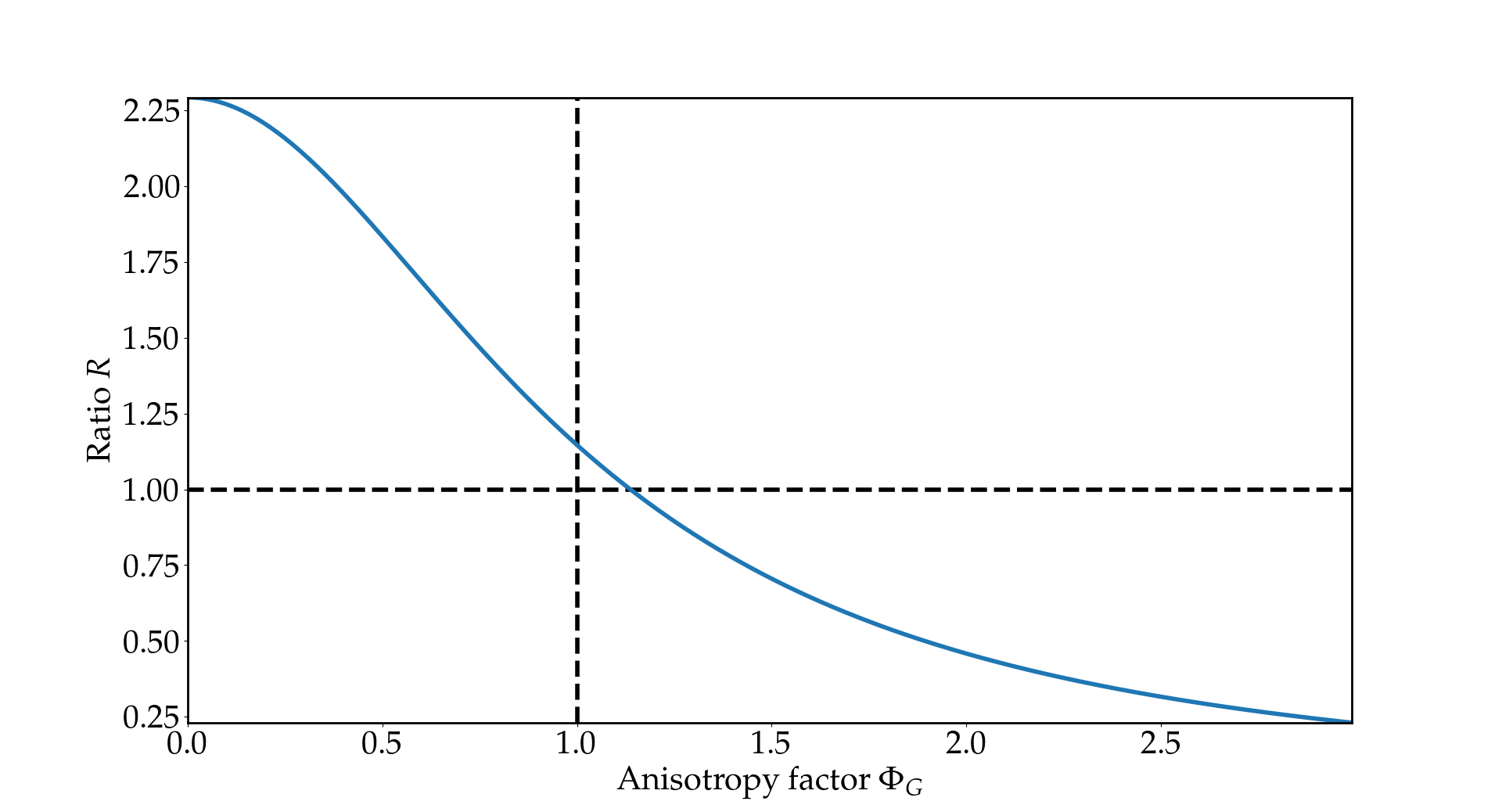}
        \caption{$R$ as a function of $\Phi_G$}
        \label{fig:RvsPhiG}
    \end{subfigure}
    \hfill
    \begin{subfigure}{0.49\textwidth}
        \centering
        \includegraphics[width=\linewidth,trim={4cm 0 2cm 0}]{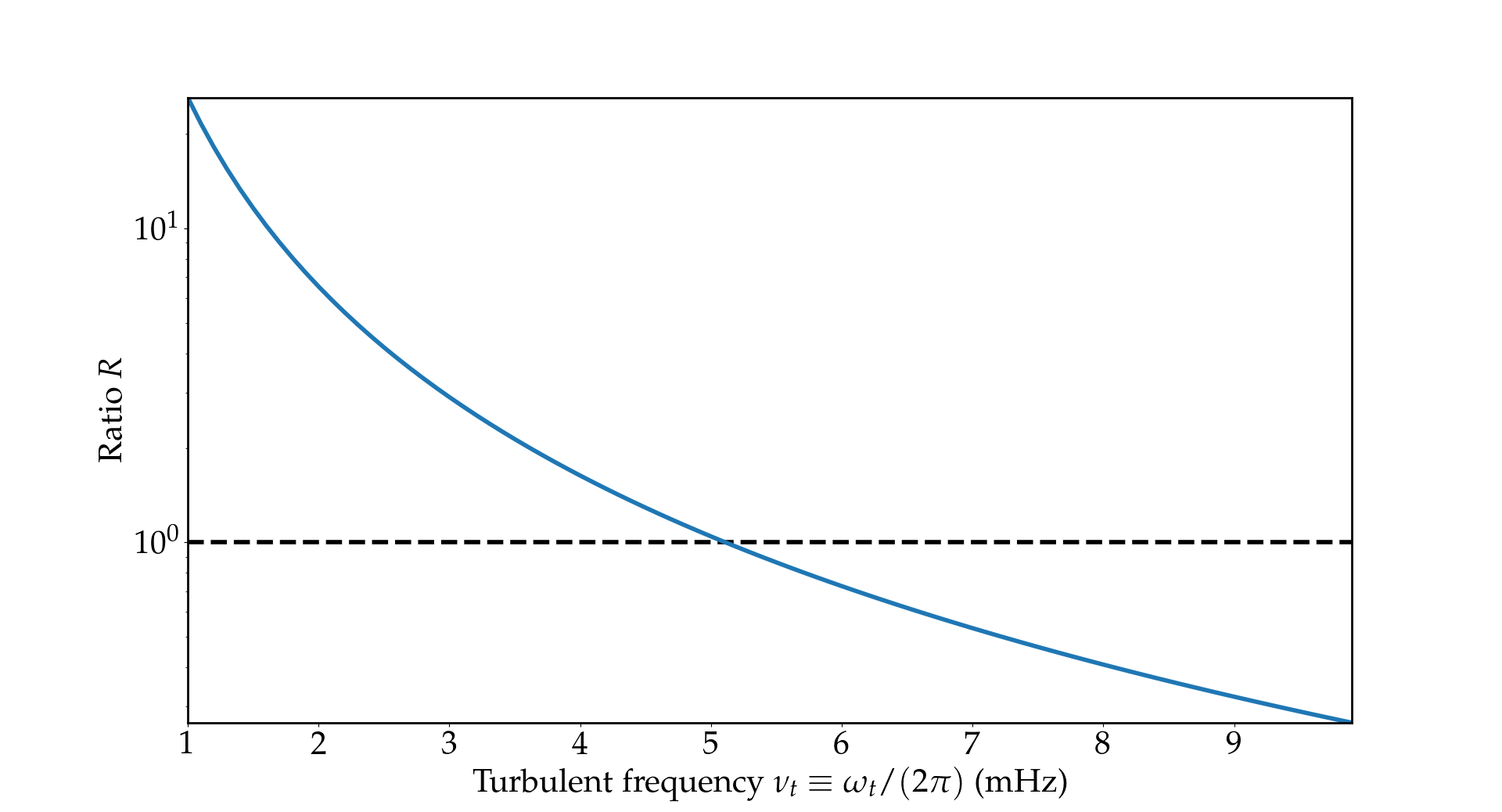}
        \caption{$R$ as a function of $\omega_t$}
        \label{fig:RvsNut}
    \end{subfigure}
    \caption{\textbf{Left:} Ratio between the two physical contributions to either mode damping or the surface effect (turbulent pressure vs. turbulent dissipation and redistribution of kinetic energy), as a function of the drift tensor anisotropy factor, $\Phi_G$, for a mode of frequency $\nu = \nu_\text{max} = 3$ mHz. The turbulent velocity anisotropy is set to $\Phi_u = 0.5$ and the turbulent frequency to $\omega_t = 0.03$ rad.s$^{-1}$. The vertical dashed line marks the limit $\Phi_G = 1$: to the left, the dissipation of the vertical turbulent motions prevails over the redistribution of turbulent kinetic energy between horizontal and vertical motions, and vice versa to the right. The horizontal dashed line is the same as in \cref{fig:RvsNu}. \textbf{Right:} Same ratio, as a function of the turbulent frequency, $\omega_t$, for a mode of frequency $\nu = \nu_\text{max} = 3$ mHz. The two anisotropy factors are set to $\Phi_u = 0.5$ and $\Phi_G = 1.0$. The horizontal dashed line is the same as in \cref{fig:RvsNu}.}
    \label{fig:RvsPhiGNut}
\end{figure*}

To better illustrate the effect of varying the model parameters, we plot in \cref{fig:RvsPhiGNut} the same ratio $R$, but this time as a function of the drift tensor anisotropy factor $\Phi_G$ (left panel) and turbulent frequency $\omega_t$ (right panel), for a fixed mode frequency $\nu = \nu_\text{max} = 3$ mHz. These figures show that, if the turbulent velocity anisotropy factor is fixed to the value observed in large-eddy simulations (i.e. $\Phi_u = 0.5$; \citep[e.g.][]{samadi03}), then the turbulent pressure contribution dominates for $\Phi_G \lesssim 1$ and $\nu_t \equiv \omega_t / (2\pi) \lesssim 5$ mHz, respectively. On the other hand, for a higher anisotropy factor $\Phi_G$ or a higher turbulent frequency $\nu_t$, the turbulent dissipation plays a much more important role in both mode damping and the modal surface effect. While we only considered a simplified case for illustration purposes, this still tends to show how important it is to account for turbulent dissipation in a realistic way when modelling the lifetime of the modes, or when trying to correct the surface effect.

Naturally, this picture is a simplistic one, as it stems primarily from hypothesis (H7) in Paper I: the reduction of the turbulent cascade to a single timescale, $\omega_t^{-1}$. In a more realistic picture, one can assume that within the continuous range of turbulent timescales, some eddies act on the modes through turbulent pressure, and others through turbulent dissipation and pressure-rate-of-strain correlation, depending on the criteria given by \cref{eq:ExpressionRatio} with $\omega_t^{-1}$ being replaced with their lifetime. Nevertheless, the results showcased in \cref{fig:RvsNu,fig:RvsPhiGNut} illustrate how the formalism presented in this paper allows us to break down the relative importance of the various physical contributions to mode damping and the modal surface effect.

\section{Conclusions\label{sec:Conclusion}}

This series of papers aims to investigate Lagrangian stochastic models of turbulence for the study of the coupling between solar-like oscillations and turbulent convection, as an alternative to more traditional methods based on mixing-length theory for example. In Paper I \citep{paper1} we established a linear stochastic wave equation designed to govern the oscillations while encompassing the full effect of the turbulence thereon. In this second paper we exploit this wave equation to obtain stochastic equations that govern the coupled evolution of the complex amplitudes of all the modes of the system simultaneously. As a first application of these equations, we consider the single-mode case, which yields explicit and simultaneous expressions for the excitation rate (\cref{eq:ExcitationRateGLM}), the damping rate (\cref{eq:DampingRateGLM}), and the modal -- or `intrinsic' -- surface effect (\cref{eq:SurfaceEffectGLM}) associated with any radial or non-radial solar-like $p$-mode, directly as a function of the statistical properties of the underlying turbulent velocity field.

The expressions for the excitation rate, damping rate, and modal surface effect provide a useful insight into the physical origin of the energetic aspects of the solar-like oscillations (at least those stemming from mechanical work), as well as the discrepancy between observed and theoretical $p$-mode frequencies. The main conclusions are as follows.
\begin{enumerate}
    \item The seismic observables -- amplitudes, linewidths, and the modal surface effect -- all simultaneously arise in a coherent manner from the same formalism. Furthermore, they are explicitly related to the statistical properties of the underlying turbulent convection.
    \item The excitation rate of the mode (\cref{eq:ExcitationRateGLM}) reduces to the same theoretical formulation obtained by previous studies on the subject \citep[e.g.][]{samadi01a,chaplin05}, which further supports the validity of this formalism.
    \item We also recover the fact that the damping rate and modal surface effect (\cref{eq:DampingRateGLM,eq:SurfaceEffectGLM}) are simply the real and imaginary part of the same complex quantity. This means that it is necessary to treat them not as separate phenomena, but as two aspects of the same physical process, which should be reflected in the way they are theoretically predicted or empirically treated.
    \item Our formalism allows for the separation of the different physical contributions to both the damping rate (\cref{eq:DecomposedDampingGLM}) and the modal surface effect (\cref{eq:DecomposedSurfaceEffectGLM}). In the present quasi-adiabatic formalism (i.e. under the assumption that fluid particles conserve their entropy during their motion), we can separate the effect of the turbulent pressure from the joint effect of the pressure-rate-of-strain correlation and that of the turbulent dissipation.
    \item We apply this formalism to a simple case where the relative importance of these two contributions (turbulent pressure vs. pressure-rate-of-strain correlation and turbulent dissipation) can be quantified (\cref{eq:ExpressionRatio}) in terms of 1) the turbulent frequency, 2) the degree of anisotropy of the Reynolds-stress tensor, and 3) the degree of anisotropy of the drift tensor, $G_{ij}$ (i.e. the ratio between the turbulent energy vertical--horizontal redistribution rate and its dissipation rate). We show that the turbulent pressure tends to dominate for high-frequency modes, and that the turbulent dissipation and pressure-rate-of-strain correlation tend to prevail for low-frequency modes.
\end{enumerate}

We stress that the Lagrangian stochastic model of turbulence from which we started in Paper I, and which we have continued to adopt in the present study, was deliberately oversimplified so as to provide a proof-of-concept to illustrate the validity of this formalism and its relevance for the study of turbulence--oscillation coupling in the context of solar-like oscillators. In particular, among the various simplifying assumptions that we adopted in Paper I, two of them stand out. The first is hypothesis H6, according to which the individual fluid particles conserve their entropy over time. This hypothesis is substantially drastic in the context of solar-like oscillations, as the local thermal timescale actually has the same order of magnitude as the period of the modes, especially in the superadiabatic region. Therefore, this approximation needs to be dropped to obtain realistic expressions for the driving rate, damping rate, and modal surface effect associated with solar-like oscillations. This can be done from the start by adding a stochastic differential equation for the internal energy associated with the individual fluid particles in the Lagrangian stochastic model. The second is hypothesis H7, according to which the lifetime of the turbulent eddies is assumed to be constant in time -- albeit not necessarily uniform -- and is parameterised by the turbulent frequency, $\omega_t$. This is an equally drastic approximation, as the range of timescales associated with the turbulent eddies is actually very large due to the very high Reynolds number characteristic of stellar turbulent convection at the top of the convective envelope. Like the quasi-adiabatic approximation, obtaining a realistic analytical prescription for the excitation rate, damping rate, and modal surface effect will require this assumption to be discarded. This can be done by treating $\omega_t$ as a physical property of each individual fluid particle separately. In doing so, it would also be necessary to add a stochastic differential equation for the turbulent frequency associated with a given fluid particle in the Lagrangian stochastic model from the start. Adding turbulent frequency and internal energy to the set of wave variables used to describe the solar-like oscillations is therefore an essential task in the development of this formalism, and will be the subject of a future paper in this series. While it does not constitute a conceptual challenge, and while many studies in the fluid dynamics community have already been devoted to the inclusion of turbulent frequency in turbulence modelling \citep[e.g.][]{pope90,pope91}, the application to solar-like oscillations is likely to involve many more theoretical developments and yield much more complex relations: it may become necessary, at some point, to proceed towards a numerical resolution of these equations instead.

It will also be necessary, in the long run, to discard the assumption that the PDF of the turbulent flow is Gaussian. Indeed, the Lagrangian stochastic model from which we started in Paper I is originally constructed under the implicit assumption that the equivalent Fokker-Planck equation yields a multivariate Gaussian PDF. By contrast, as shown by 3D hydrodynamic simulations, stellar turbulent convection is characterised by a typical structure composed of upflows and downdrafts, each separately exhibiting Gaussian turbulence, such that the total flow has a bimodal distribution. Because of the asymmetry between the upflows and the downdrafts -- the latter being colder and more turbulent than the former -- the total distribution is not Gaussian. A possible solution that would allow the non-Gaussian nature of turbulence to be accounted for would be to use a two-flow model, where the distribution is determined by the mean and variance of two Gaussian distributions, or equivalently by the first four moments of the whole distribution. It is possible to implement two-flow prescriptions in Lagrangian stochastic models \citep[see for instance][]{rodean96review} by using different sets of stochastic differential equations to model the evolution of the fluid particles, depending on whether their vertical velocity is positive or negative and on whether their temperature is higher or lower than the local mean temperature. Concretely, this would influence the relation between the turbulent fluctuations -- especially the turbulent velocity -- and the perturbations $\mathcal{L}^s$ and $\ket{\Theta}$ to the wave equation. This would constitute an important step towards making the present approach more realistic, and its results quantitatively applicable to solar-like oscillators and thus suitable for comparison with observations.

As a final, important note, we recall that the results presented from \cref{sec:SingleMode} onwards, and in particular the interpretation of the various terms in the averaged simplified amplitude equations in terms of mode excitation, mode damping, and the surface effect, were obtained by neglecting mode coupling, that is, by reducing the coupled amplitude and phase equations (\cref{eq:GeneralSimplifiedRealAmplitudeEquation,eq:GeneralSimplifiedPhaseEquation}) to two equations on the amplitude and phase of a single mode (\cref{eq:SingleModeAmplitudeFinal,eq:SingleModePhaseFinal}). While such coupling is deemed negligible in solar-type, main-sequence stars, where it is usually assumed that modes can be studied in isolation from one another, this assumption is in fact not warranted by any \textit{a priori} physical argument. This is a most important caveat of the findings presented in this study; in the future, it will be necessary to go beyond the single-mode case, and instead consider the full, coupled system of equations, especially when it comes to applying the present formalism to the case of a more evolved star. A later paper in this series will be dedicated to this task.

\bibliographystyle{aa}
\bibliography{biblio}

\begin{appendix}

\section{Derivation of the Fokker-Planck equation\label{appA}}

In this appendix, we detail the derivation of the probability fluxes $\mathcal{G}_\mu$ and $\mathcal{H}_\mu$ and the components of the diffusion matrix $\mathcal{D}_{\mu\nu}$, $\mathcal{E}_{\mu\nu}$ and $\mathcal{F}_{\mu\nu}$ appearing in the joint-amplitude-phase Fokker-Planck equation (\cref{eq:FokkerPlanckAmplitudePhase}). In order to account for the finite memory time of $G_\mu(A_\nu,\Phi_\nu,t)$ and $H_\mu(A_\nu,\Phi_\nu,t)$ (defined as the right-hand sides of \cref{eq:RealAmplitudeEquation,eq:RealPhaseEquation}, respectively), \citet{stratonovich65} showed that one can define effective probability fluxes and diffusion coefficients in the following way:
\begin{align}
    & \mathcal{G}_\mu = \langle G_\mu \rangle + \musum{\nu} \acc{\dfrac{\partial G_\mu}{\partial A_\nu}}{G_\nu^\tau} + \musum{\nu} \acc{\dfrac{\partial G_\mu}{\partial \Phi_\nu}}{H_\nu^\tau}~, \label{eq:appA-DefinitionProbabilityFluxAmplitude} \\
    & \mathcal{H}_\mu = \langle H_\mu \rangle + \musum{\nu} \acc{\dfrac{\partial H_\mu}{\partial A_\nu}}{G_\nu^\tau} + \musum{\nu} \acc{\dfrac{\partial H_\mu}{\partial \Phi_\nu}}{H_\nu^\tau}~, \label{eq:appA-DefinitionProbabilityFluxPhase} \\
    & \mathcal{D}_{\mu\nu} = \acc{G_\mu}{G_\nu^\tau} + \acc{G_\nu}{G_\mu^\tau}~, \label{eq:appA-DefinitionDiffusionCoefficientAmplitude} \\
    & \mathcal{E}_{\mu\nu} = \acc{G_\mu}{H_\nu^\tau} + \acc{H_\nu}{G_\mu^\tau}~, \label{eq:appA-DefinitionDiffusionCoefficientCross}\\
    & \mathcal{F}_{\mu\nu} = \acc{H_\mu}{H_\nu^\tau} + \acc{H_\nu}{H_\mu^\tau}~, \label{eq:appA-DefinitionDiffusionCoefficientPhase}
\end{align}
where $\langle . \rangle$ denotes an ensemble average, and the bilinear operator $\{.~;~.\}$ is defined by
\begin{equation}
    \left\{f_1~;~f_2^\tau\right\} \equiv \inttau \left[ \left\langle \vphantom{f_1^\tau} f_1(t)f_2(t+\tau) \right\rangle - \left\langle \vphantom{f_1^\tau} f_1(t) \right\rangle \left\langle \vphantom{f_1^\tau} f_2(t+\tau) \right\rangle \right]~.
    \label{eq:appA-DefinitionAcc}
\end{equation}
We can rewrite each of the stochastic processes $c_{1\mu}(t)$, $c_{2\mu\nu}(t)$ and $c_{3\mu\nu}(t)$ in the following polar form:
\begin{align}
    & c_{1\mu} = C_{1\mu}\exp^{j\phi_{1\mu}}~, \\
    & c_{2\mu\nu} = C_{2\mu\nu}\exp^{j\phi_{2\mu\nu}}~, \\
    & c_{3\mu\nu} = C_{3\mu\nu}\exp^{j\phi_{3\mu\nu}}~,
\end{align}
where the functions $C_i(t)$ and $\phi_i(t)$ are both real. Then the right-hand sides of \cref{eq:RealAmplitudeEquation,eq:RealPhaseEquation} can be rewritten
\begin{multline}
    G_\mu(A_\lambda, \Phi_\lambda, t) = \kappa_\mu A_\mu + C_{1\mu}\lcos{\omega_\mu t + \Phi_\mu - \phi_{1\mu}} \\
    + \musum{\lambda} A_\lambda C_{2\mu\lambda}\lcos{(\omega_\mu - \omega_\lambda) t + \Phi_\mu - \Phi_\lambda - \phi_{2\mu\lambda}} \\
    + \musum{\lambda} A_\lambda C_{3\mu\lambda}\lcos{(\omega_\mu + \omega_\lambda) t + \Phi_\mu + \Phi_\lambda - \phi_{3\mu\lambda}}
\end{multline}
and
\begin{multline}
    H_\mu(A_\lambda, \Phi_\lambda, t) = -\dfrac{1}{A_\mu} C_{1\mu} \lsin{\omega_\mu t + \Phi_\mu - \phi_{1\mu}} \\
    - \dfrac{1}{A_\mu} \musum{\lambda} A_\lambda C_{2\mu\lambda}\lsin{(\omega_\mu - \omega_\lambda) t + \Phi_\mu - \Phi_\lambda - \phi_{2\mu\lambda}} \\
    - \dfrac{1}{A_\mu} \musum{\lambda} A_\lambda C_{3\mu\lambda}\lsin{(\omega_\mu + \omega_\lambda) t + \Phi_\mu + \Phi_\lambda - \phi_{3\mu\lambda}}~.
\end{multline}
Their respective derivatives with respect to $A_\nu$ and $\Phi_\nu$ are
\begin{multline}
    \dfrac{\partial G_\mu}{\partial A_\nu} = \delta_{\mu\nu} \kappa_\mu + C_{2\mu\nu}\lcos{(\omega_\mu - \omega_\nu) t + \Phi_\mu - \Phi_\nu - \phi_{2\mu\nu}} \\
    + C_{3\mu\nu}\lcos{(\omega_\mu + \omega_\nu) t + \Phi_\mu + \Phi_\nu - \phi_{2\mu\nu}}~,
\end{multline}
\begin{multline}
    \dfrac{\partial G_\mu}{\partial \Phi_\nu} = -\delta_{\mu\nu} C_{1\mu} \lsin{\omega_\mu t + \Phi_\mu - \phi_{1\mu}} \\
    - \delta_{\mu\nu} \musum{\lambda} A_\lambda C_{2\mu\lambda}\lsin{(\omega_\mu - \omega_\lambda) t + \Phi_\mu - \Phi_\lambda - \phi_{2\mu\lambda}} \\
    + A_\nu C_{2\mu\nu} \lsin{(\omega_\mu - \omega_\nu) t + \Phi_\mu - \Phi_\nu - \phi_{2\mu\nu}} \\
    - \delta_{\mu\nu} \musum{\lambda} A_\lambda C_{3\mu\lambda}\lsin{(\omega_\mu + \omega_\lambda) t + \Phi_\mu + \Phi_\lambda - \phi_{3\mu\lambda}} \\
    - A_\nu C_{3\mu\nu} \lsin{(\omega_\mu + \omega_\nu) t + \Phi_\mu + \Phi_\nu - \phi_{3\mu\nu}}~,
\end{multline}
\begin{multline}
    \dfrac{\partial H_\mu}{\partial A_\nu} = \dfrac{\delta_{\mu\nu}}{A_\mu^2} C_{1\mu} \lsin{\omega_\mu t + \Phi_\mu - \phi_{1\mu}} \\
    + \dfrac{\delta_{\mu\nu}}{A_\mu^2} \musum{\lambda} A_\lambda C_{2\mu\lambda}\lsin{(\omega_\mu - \omega_\lambda) t + \Phi_\mu - \Phi_\lambda - \phi_{2\mu\lambda}} \\
    - \dfrac{1}{A_\mu} C_{2\mu\nu}\lsin{(\omega_\mu - \omega_\nu) t + \Phi_\mu - \Phi_\nu - \phi_{2\mu\nu}} \\
    + \dfrac{\delta_{\mu\nu}}{A_\mu^2} \musum{\lambda} A_\lambda C_{3\mu\lambda}\lsin{(\omega_\mu + \omega_\lambda) t + \Phi_\mu + \Phi_\lambda - \phi_{3\mu\lambda}} \\
    - \dfrac{1}{A_\mu} C_{3\mu\nu}\lsin{(\omega_\mu + \omega_\nu) t + \Phi_\mu + \Phi_\nu - \phi_{3\mu\nu}}~,
\end{multline}
and
\begin{multline}
    \dfrac{\partial H_\mu}{\partial \Phi_\nu} = -\dfrac{\delta_{\mu\nu}}{A_\mu} C_{1\mu} \lcos{\omega_\mu t + \Phi_\mu - \phi_{1\mu}} \\
    - \dfrac{\delta_{\mu\nu}}{A_\mu} \musum{\lambda} A_\lambda C_{2\mu\lambda}\lcos{(\omega_\mu - \omega_\lambda) t + \Phi_\mu - \Phi_\lambda - \phi_{2\mu\lambda}} \\
    + \dfrac{A_\nu}{A_\mu} C_{2\mu\nu}\lcos{(\omega_\mu - \omega_\nu) t + \Phi_\mu - \Phi_\nu - \phi_{2\mu\nu}} \\
    - \dfrac{\delta_{\mu\nu}}{A_\mu} \musum{\lambda} A_\lambda C_{3\mu\lambda}\lcos{(\omega_\mu + \omega_\lambda) t + \Phi_\mu + \Phi_\lambda - \phi_{3\mu\lambda}} \\
    - \dfrac{A_\nu}{A_\mu} C_{3\mu\nu}\lcos{(\omega_\mu + \omega_\nu) t + \Phi_\mu + \Phi_\nu - \phi_{3\mu\nu}}~.
\end{multline}

Plugging these into \cref{eq:appA-DefinitionProbabilityFluxAmplitude,eq:appA-DefinitionProbabilityFluxPhase,eq:appA-DefinitionDiffusionCoefficientAmplitude,eq:appA-DefinitionDiffusionCoefficientCross,eq:appA-DefinitionDiffusionCoefficientPhase}, it may seem at first glance that the expansion of the bilinear operators $\{~...~;~...~\}$ involves a large number of terms, all of which can be put in the following general form
\begin{equation}
    \acc{C_\alpha \lcos{\Omega_\alpha t + \phi_\alpha}}{C_\beta^\tau \lcos{\Omega_\beta (t+\tau) + \phi_\beta^\tau}}~,
\end{equation}
with various amplitudes $C_{\alpha/\beta}$, angular frequencies $\Omega_{\alpha/\beta}$ and phases $\phi_{\alpha/\beta}$ (we note that sines can always be written as cosines by redefining their phase). Plugging this general form into \cref{eq:appA-DefinitionAcc}, we find
\begin{align}
    & \acc{C_\alpha \lcos{\Omega_\alpha t + \phi_\alpha}}{C_\beta^\tau \lcos{\Omega_\beta (t+\tau) + \phi_\beta^\tau}} \nonumber \\
    & = \inttau \left\langle C_\alpha C_\beta^\tau \lcos{\Omega_\alpha t + \phi_\alpha} \lcos{\Omega_\beta (t+\tau) + \phi_\beta^\tau} \right\rangle \nonumber \\
    & = \dfrac{1}{2} \inttau \left\langle C_\alpha C_\beta^\tau \lcos{(\Omega_\alpha + \Omega_\beta) t + \Omega_\beta \tau + \phi_\alpha + \phi_\beta^\tau} \right\rangle \nonumber \\
    & \hspace{0.5cm} + \dfrac{1}{2} \inttau \left\langle C_\alpha C_\beta^\tau \lcos{(-\Omega_\alpha + \Omega_\beta) t + \Omega_\beta \tau - \phi_\alpha + \phi_\beta^\tau} \right\rangle~.
    \label{eq:appA-Example}
\end{align}
The first term oscillates in time $t$ with an angular frequency $\Omega_\beta + \Omega_\alpha$, while the second one oscillates with an angular frequency $\Omega_\beta - \Omega_\alpha$. However, since we are interested in the long-term variations in the complex amplitude of the modes, which occur on timescales much larger than $\omega_\lambda^{-1}$, all terms oscillating at frequencies comparable to any $\omega_\lambda$ can be averaged out. This means that \cref{eq:appA-Example} only yields non-vanishing contributions if either $\Omega_\alpha = \Omega_\beta$ or $\Omega_\alpha = -\Omega_\beta$. By redefining the phase inside the cosine, we can always restrict ourselves to the first case. Otherwise stated, this means that the only terms that can `interact' through the operator $\{~...~;~...~\}$ to yield a non-oscillating contribution are those that share the same frequency in the first place. This can happen either because the two frequencies are rigorously identical or because there is an accidental resonance of the type $\sum_i n_i \omega_i = 0$, where the sum concerns two or more modes, and $n_i$ are integers. In the following we refer to the first kind of contribution as `cross terms' and to the second kind as `resonances'. We treat each kind separately in the following two sections.

\subsection{Cross terms}

The contributions from the cross terms in the curly bracket operators appearing on the right-hand sides of \cref{eq:appA-DefinitionProbabilityFluxAmplitude,eq:appA-DefinitionProbabilityFluxPhase,eq:appA-DefinitionDiffusionCoefficientAmplitude,eq:appA-DefinitionDiffusionCoefficientCross,eq:appA-DefinitionDiffusionCoefficientPhase} are given by \cref{eq:appA-ExpansionG1,eq:appA-ExpansionG2,eq:appA-ExpansionH1,eq:appA-ExpansionH2,eq:appA-ExpansionD,eq:appA-ExpansionF} below\footnote{In addition, we also have $\mathcal{E}_{\mu\nu} = 0$. Indeed, all the integrands appearing in $\{G_\mu;H_\nu^\tau\}$ are odd functions of $\tau$. Therefore, since $\{H_\nu;G_\mu^\tau\} = \{G_\mu;H_\nu^{-\tau}\}$, we simply obtain
\begin{equation}
    \mathcal{E}_{\mu\nu} = \{G_\mu;H_\nu^\tau\} + \{G_\mu;H_\nu^{-\tau}\} = 0~.
\end{equation}
}. It can be seen that in each of the individual curly brackets therein, the frequencies in both cosines/sines are indeed rigorously identical.

\begin{figure*}
\hrule
\begin{multline}
    \acc{\dfrac{\partial G_\mu}{\partial A_\nu}}{G_\nu^\tau} = \acc{C_{2\mu\nu}\lcos{(\omega_\mu - \omega_\nu) t + \Phi_\mu - \Phi_\nu - \phi_{2\mu\nu}}}{A_\mu C_{2\nu\mu}^\tau \lcos{(\omega_\mu - \omega_\nu)(t+\tau) + \Phi_\mu - \Phi_\nu + \phi_{2\nu\mu}^\tau}} \\
    + \acc{C_{3\mu\nu}\lcos{(\omega_\mu + \omega_\nu) t + \Phi_\mu + \Phi_\nu - \phi_{3\mu\nu}}}{A_\mu C_{3\nu\mu}^\tau \lcos{(\omega_\mu + \omega_\nu)(t+\tau) + \Phi_\mu + \Phi_\nu - \phi_{3\nu\mu}^\tau}}~,
    \label{eq:appA-ExpansionG1}
\end{multline}
\begin{multline}
    \acc{\dfrac{\partial G_\mu}{\partial \Phi_\nu}}{H_\nu^\tau} = \acc{-\delta_{\mu\nu} C_{1\mu} \lsin{\omega_\mu t + \Phi_\mu - \phi_{1\mu}}}{-\dfrac{\delta_{\mu\nu}}{A_\mu} C_{1\mu}^\tau \lsin{\omega_\mu (t+\tau) + \Phi_\mu - \phi_{1\mu}^\tau}} \\
    + \musum{\lambda} \acc{-\delta_{\mu\nu} A_\lambda C_{2\mu\lambda}\lsin{(\omega_\mu - \omega_\lambda) t + \Phi_\mu - \Phi_\lambda - \phi_{2\mu\lambda}}}{-\dfrac{A_\lambda}{A_\mu} C_{2\mu\lambda}^\tau \lsin{(\omega_\mu - \omega_\lambda)(t+\tau) + \Phi_\mu - \Phi_\lambda - \phi_{2\mu\lambda}^\tau}} \\
    + \acc{A_\nu C_{2\mu\nu}\lsin{(\omega_\mu - \omega_\nu) t + \Phi_\mu - \Phi_\nu - \phi_{2\mu\nu}}}{\dfrac{A_\mu}{A_\nu} C_{2\nu\mu}^\tau \lsin{(\omega_\mu - \omega_\nu)(t+\tau) + \Phi_\mu - \Phi_\nu + \phi_{2\nu\mu}^\tau}} \\
    + \musum{\lambda} \acc{-\delta_{\mu\nu} A_\lambda C_{3\mu\lambda}\lsin{(\omega_\mu + \omega_\lambda) t + \Phi_\mu + \Phi_\lambda - \phi_{3\mu\lambda}}}{-\dfrac{A_\lambda}{A_\mu} C_{3\mu\lambda}^\tau \lsin{(\omega_\mu + \omega_\lambda)(t+\tau) + \Phi_\mu + \Phi_\lambda - \phi_{3\mu\lambda}^\tau}} \\
    + \acc{-A_\nu C_{3\mu\nu}\lsin{(\omega_\mu + \omega_\nu) t + \Phi_\mu + \Phi_\nu - \phi_{3\mu\nu}}}{-\dfrac{A_\mu}{A_\nu} C_{3\nu\mu}^\tau \lsin{(\omega_\mu + \omega_\nu)(t+\tau) + \Phi_\mu + \Phi_\nu - \phi_{3\nu\mu}^\tau}}~,
    \label{eq:appA-ExpansionG2}
\end{multline}
\begin{multline}
    \acc{\dfrac{\partial H_\mu}{\partial A_\nu}}{G_\nu^\tau} = \acc{\dfrac{\delta_{\mu\nu}}{A_\mu^2} C_{1\mu} \lsin{\omega_\mu t + \Phi_\mu - \phi_{1\mu}}}{\delta_{\mu\nu} C_{1\mu}^\tau \lcos{\omega_\mu (t+\tau) + \Phi_\mu - \phi_{1\mu}^\tau}} \\
    + \musum{\lambda} \acc{\dfrac{\delta_{\mu\nu}}{A_\mu^2} A_\lambda C_{2\mu\lambda}\lsin{(\omega_\mu - \omega_\lambda) t + \Phi_\mu - \Phi_\lambda - \phi_{2\mu\lambda}}}{A_\lambda C_{2\mu\lambda}^\tau \lcos{(\omega_\mu - \omega_\lambda)(t+\tau) + \Phi_\mu - \Phi_\lambda - \phi_{2\mu\lambda}^\tau}} \\
    + \acc{-\dfrac{1}{A_\mu} C_{2\mu\nu}\lsin{(\omega_\mu - \omega_\nu) t + \Phi_\mu - \Phi_\nu - \phi_{2\mu\nu}}}{A_\mu C_{2\nu\mu}^\tau \lcos{(\omega_\mu - \omega_\nu)(t+\tau) + \Phi_\mu - \Phi_\nu + \phi_{2\nu\mu}^\tau}} \\
    + \musum{\lambda} \acc{\dfrac{\delta_{\mu\nu}}{A_\mu^2} A_\lambda C_{3\mu\lambda}\lsin{(\omega_\mu + \omega_\lambda) t + \Phi_\mu + \Phi_\lambda - \phi_{3\mu\lambda}}}{A_\lambda C_{3\mu\lambda}^\tau \lcos{(\omega_\mu + \omega_\lambda)(t+\tau) + \Phi_\mu + \Phi_\lambda - \phi_{3\mu\lambda}^\tau}} \\
    + \acc{-\dfrac{1}{A_\mu} C_{3\mu\nu}\lsin{(\omega_\mu + \omega_\nu) t + \Phi_\mu + \Phi_\nu - \phi_{3\mu\nu}}}{A_\mu C_{3\nu\mu}^\tau \lcos{(\omega_\mu + \omega_\nu)(t+\tau) + \Phi_\mu + \Phi_\nu - \phi_{3\nu\mu}^\tau}}~,
    \label{eq:appA-ExpansionH1}
\end{multline}
\begin{multline}
    \acc{\dfrac{\partial H_\mu}{\partial \Phi_\nu}}{H_\nu^\tau} = \acc{-\dfrac{\delta_{\mu\nu}}{A_\mu} C_{1\mu} \lcos{\omega_\mu t + \Phi_\mu - \phi_{1\mu}}}{-\dfrac{\delta_{\mu\nu}}{A_\mu} C_{1\mu}^\tau \lsin{\omega_\mu (t+\tau) + \Phi_\mu - \phi_{1\mu}^\tau}} \\
    + \musum{\lambda} \acc{-\dfrac{\delta_{\mu\nu}}{A_\mu} A_\lambda C_{2\mu\lambda}\lcos{(\omega_\mu - \omega_\lambda) t + \Phi_\mu - \Phi_\lambda - \phi_{2\mu\lambda}}}{-\dfrac{A_\lambda}{A_\mu} C_{2\mu\lambda}^\tau \lsin{(\omega_\mu - \omega_\lambda)(t+\tau) + \Phi_\mu - \Phi_\lambda - \phi_{2\mu\lambda}^\tau}} \\
    + \acc{\dfrac{A_\nu}{A_\mu} C_{2\mu\nu}\lcos{(\omega_\mu - \omega_\nu) t + \Phi_\mu - \Phi_\nu - \phi_{2\mu\nu}}}{\dfrac{A_\mu}{A_\nu} C_{2\nu\mu}^\tau \lsin{(\omega_\mu - \omega_\nu)(t+\tau) + \Phi_\mu - \Phi_\nu + \phi_{2\nu\mu}^\tau}} \\
    + \musum{\lambda} \acc{-\dfrac{\delta_{\mu\nu}}{A_\mu} A_\lambda C_{3\mu\lambda}\lcos{(\omega_\mu + \omega_\lambda) t + \Phi_\mu + \Phi_\lambda - \phi_{3\mu\lambda}}}{-\dfrac{A_\lambda}{A_\mu} C_{3\mu\lambda}^\tau \lsin{(\omega_\mu + \omega_\lambda)(t+\tau) + \Phi_\mu + \Phi_\lambda - \phi_{3\mu\lambda}^\tau}} \\
    + \acc{-\dfrac{A_\nu}{A_\mu} C_{3\mu\nu}\lcos{(\omega_\mu + \omega_\nu) t + \Phi_\mu + \Phi_\nu - \phi_{3\mu\nu}}}{-\dfrac{A_\mu}{A_\nu} C_{3\nu\mu}^\tau \lsin{(\omega_\mu + \omega_\nu)(t+\tau) + \Phi_\mu + \Phi_\nu - \phi_{3\nu\mu}^\tau}}~,
    \label{eq:appA-ExpansionH2}
\end{multline}
\begin{multline}
    \acc{G_\mu}{G_\nu^\tau} = \acc{C_{1\mu} \lcos{\omega_\mu t + \Phi_\mu - \phi_{1\mu}}}{\delta_{\mu\nu} C_{1\mu}^\tau \lcos{\omega_\mu (t+\tau) + \Phi_\mu - \phi_{1\mu}^\tau}} \\
    + \delta_{\mu\nu} \musum{\lambda} \acc{A_\lambda C_{2\mu\lambda}\lcos{(\omega_\mu - \omega_\lambda) t + \Phi_\mu - \Phi_\lambda - \phi_{2\mu\lambda}}}{A_\lambda C_{2\mu\lambda}^\tau \lcos{(\omega_\mu - \omega_\lambda)(t+\tau) + \Phi_\mu - \Phi_\lambda - \phi_{2\mu\lambda}^\tau}} \\
    + (1-\delta_{\mu\nu}) \acc{A_\nu C_{2\mu\nu}\lcos{(\omega_\mu - \omega_\nu) t + \Phi_\mu - \Phi_\nu - \phi_{2\mu\nu}}}{A_\mu C_{2\nu\mu}^\tau \lcos{(\omega_\mu - \omega_\nu)(t+\tau) + \Phi_\mu - \Phi_\nu + \phi_{2\nu\mu}^\tau}} \\
    + \delta_{\mu\nu} \musum{\lambda} \acc{A_\lambda C_{3\mu\lambda}\lcos{(\omega_\mu + \omega_\lambda) t + \Phi_\mu + \Phi_\lambda - \phi_{3\mu\lambda}}}{A_\lambda C_{3\mu\lambda}^\tau \lcos{(\omega_\mu + \omega_\lambda)(t+\tau) + \Phi_\mu + \Phi_\lambda - \phi_{3\mu\lambda}^\tau}} \\
    + (1-\delta_{\mu\nu}) \acc{A_\nu C_{3\mu\nu}\lcos{(\omega_\mu + \omega_\nu) t + \Phi_\mu + \Phi_\nu - \phi_{3\mu\nu}}}{A_\mu C_{3\nu\mu}^\tau \lcos{(\omega_\mu + \omega_\nu)(t+\tau) + \Phi_\mu + \Phi_\nu - \phi_{3\nu\mu}^\tau}}~,
    \label{eq:appA-ExpansionD}
\end{multline}
\hrule
\end{figure*}

\begin{figure*}
\hrule
\begin{multline}
    \acc{H_\mu}{H_\nu^\tau} = \acc{-\dfrac{1}{A_\mu} C_{1\mu} \lsin{\omega_\mu t + \Phi_\mu - \phi_{1\mu}}}{-\dfrac{\delta_{\mu\nu}}{A_\mu} C_{1\mu}^\tau \lsin{\omega_\mu (t+\tau) + \Phi_\mu - \phi_{1\mu}^\tau}} \\
    + \delta_{\mu\nu} \musum{\lambda} \acc{-\dfrac{A_\lambda}{A_\mu} C_{2\mu\lambda}\lsin{(\omega_\mu - \omega_\lambda) t + \Phi_\mu - \Phi_\lambda - \phi_{2\mu\lambda}}}{-\dfrac{A_\lambda}{A_\mu} C_{2\mu\lambda}^\tau \lsin{(\omega_\mu - \omega_\lambda)(t+\tau) + \Phi_\mu - \Phi_\lambda - \phi_{2\mu\lambda}^\tau}} \\
    + (1-\delta_{\mu\nu}) \acc{-\dfrac{A_\nu}{A_\mu} C_{2\mu\nu}\lsin{(\omega_\mu - \omega_\nu) t + \Phi_\mu - \Phi_\nu - \phi_{2\mu\nu}}}{\dfrac{A_\mu}{A_\nu} C_{2\nu\mu}^\tau \lsin{(\omega_\mu - \omega_\nu)(t+\tau) + \Phi_\mu - \Phi_\nu + \phi_{2\nu\mu}^\tau}} \\
    + \delta_{\mu\nu} \musum{\lambda} \acc{-\dfrac{A_\lambda}{A_\mu} C_{3\mu\lambda}\lsin{(\omega_\mu + \omega_\lambda) t + \Phi_\mu + \Phi_\lambda - \phi_{3\mu\lambda}}}{-\dfrac{A_\lambda}{A_\mu} C_{3\mu\lambda}^\tau \lsin{(\omega_\mu + \omega_\lambda)(t+\tau) + \Phi_\mu + \Phi_\lambda - \phi_{3\mu\lambda}^\tau}} \\
    + (1-\delta_{\mu\nu}) \acc{-\dfrac{A_\nu}{A_\mu} C_{3\mu\nu}\lsin{(\omega_\mu + \omega_\nu) t + \Phi_\mu + \Phi_\nu - \phi_{3\mu\nu}}}{-\dfrac{A_\mu}{A_\nu} C_{3\nu\mu}^\tau \lsin{(\omega_\mu + \omega_\nu)(t+\tau) + \Phi_\mu + \Phi_\nu - \phi_{3\nu\mu}^\tau}}~.
    \label{eq:appA-ExpansionF}
\end{multline}
\hrule
\end{figure*}

Applying the same procedure underlined in \cref{eq:appA-Example}, \cref{eq:appA-ExpansionG1,eq:appA-ExpansionG2,eq:appA-ExpansionH1,eq:appA-ExpansionH2,eq:appA-ExpansionD,eq:appA-ExpansionF} become
\begin{multline}
    \acc{\dfrac{\partial G_\mu}{\partial A_\nu}}{G_\nu^\tau} \\
    = \dfrac{1}{2}\left(1 - \delta_{\mu\nu} \right) A_\mu \inttau \mathrm{Re}\left( \left\langle c_{2\mu\nu} c_{2\nu\mu}^\tau \right\rangle \exp^{\ii (\omega_\mu - \omega_\nu) \tau} \right) \\
    + \delta_{\mu\nu} A_\mu \inttau \left\langle \mathrm{Re}\left( c_{2\mu\mu} \right) \mathrm{Re}\left( c_{2\mu\mu}^\tau \right) \right\rangle \\
    + \dfrac{1}{2} A_\mu \inttau \mathrm{Re}\left( \left\langle c_{3\mu\nu} c_{3\nu\mu}^{\tau\star} \right\rangle \exp^{\ii (\omega_\mu + \omega_\nu) \tau} \right)~,
\end{multline}
\begin{multline}
    \acc{\dfrac{\partial G_\mu}{\partial \Phi_\nu}}{H_\nu^\tau} = \dfrac{1}{2}\dfrac{\delta_{\mu\nu}}{A_\mu} \inttau \mathrm{Re} \left( \left\langle c_{1\mu} c_{1\mu}^{\tau\star} \right\rangle \exp^{\ii \omega_\mu \tau} \right) \\
    + \dfrac{1}{2}\dfrac{\delta_{\mu\nu}}{A_\mu} \musum{\lambda\neq\mu} A_\lambda^2 \inttau \mathrm{Re}\left( \left\langle c_{2\mu\lambda} c_{2\mu\lambda}^{\tau\star} \right\rangle \exp^{\ii(\omega_\mu - \omega_\lambda)\tau} \right) \\
    + \delta_{\mu\nu} A_\mu \inttau \left\langle \mathrm{Im}\left( c_{2\mu\mu} \right) \mathrm{Im}\left( c_{2\mu\mu}^\tau \right) \right\rangle \\
    + \dfrac{1}{2}\left(1 - \delta_{\mu\nu} \right) A_\mu \inttau \mathrm{Re}\left( \left\langle c_{2\mu\nu} c_{2\nu\mu}^\tau \right\rangle \exp^{\ii (\omega_\mu - \omega_\nu) \tau} \right) \\
    - \delta_{\mu\nu} A_\mu \inttau \left\langle \mathrm{Im}\left( c_{2\mu\mu} \right) \mathrm{Im}\left( c_{2\mu\mu}^\tau \right) \right\rangle \\
    + \dfrac{1}{2} \dfrac{\delta_{\mu\nu}}{A_\mu} \musum{\lambda} A_\lambda^2 \inttau \mathrm{Re}\left( \left\langle c_{3\mu\lambda} c_{3\mu\lambda}^{\tau\star} \right\rangle \exp^{\ii (\omega_\mu + \omega_\lambda) \tau} \right) \\
    + \dfrac{1}{2} A_\mu \inttau \mathrm{Re}\left( \left\langle c_{3\mu\nu} c_{3\nu\mu}^{\tau\star} \right\rangle \exp^{\ii (\omega_\mu + \omega_\nu) \tau} \right)~,
\end{multline}
\begin{multline}
    \acc{\dfrac{\partial H_\mu}{\partial A_\nu}}{G_\nu^\tau} \\
    = -\dfrac{1}{2}\dfrac{\delta_{\mu\nu}}{A_\mu^2} \inttau \mathrm{Im} \left( \left\langle c_{1\mu} c_{1\mu}^{\tau\star} \right\rangle \exp^{\ii \omega_\mu \tau} \right) \\
    - \dfrac{1}{2}\dfrac{\delta_{\mu\nu}}{A_\mu^2} \musum{\lambda\neq\mu} A_\lambda^2 \inttau \mathrm{Im}\left( \left\langle c_{2\mu\lambda} c_{2\mu\lambda}^{\tau\star} \right\rangle \exp^{\ii(\omega_\mu - \omega_\lambda)\tau} \right) \\
    - \delta_{\mu\nu} \inttau \left\langle \mathrm{Im}\left( c_{2\mu\mu} \right) \mathrm{Re}\left( c_{2\mu\mu}^\tau \right) \right\rangle \\
    + \dfrac{1}{2}\left(1 - \delta_{\mu\nu} \right) \inttau \mathrm{Im}\left( \left\langle c_{2\mu\nu} c_{2\nu\mu}^\tau \right\rangle \exp^{\ii (\omega_\mu - \omega_\nu) \tau} \right) \\
    + \delta_{\mu\nu} \inttau \left\langle \mathrm{Im}\left( c_{2\mu\mu} \right) \mathrm{Re}\left( c_{2\mu\mu}^\tau \right) \right\rangle \\
    - \dfrac{1}{2} \dfrac{\delta_{\mu\nu}}{A_\mu^2} \musum{\lambda} A_\lambda^2 \inttau \mathrm{Im}\left( \left\langle c_{3\mu\lambda} c_{3\mu\lambda}^{\tau\star} \right\rangle \exp^{\ii (\omega_\mu + \omega_\lambda) \tau} \right) \\
    + \dfrac{1}{2} \inttau \mathrm{Im}\left( \left\langle c_{3\mu\nu} c_{3\nu\mu}^{\tau\star} \right\rangle \exp^{\ii (\omega_\mu + \omega_\nu) \tau} \right)~,
\end{multline}
\begin{multline}
    \acc{\dfrac{\partial H_\mu}{\partial \Phi_\nu}}{H_\nu^\tau} \\
    = \dfrac{1}{2}\dfrac{\delta_{\mu\nu}}{A_\mu^2} \inttau \mathrm{Im} \left( \left\langle c_{1\mu} c_{1\mu}^{\tau\star} \right\rangle \exp^{\ii \omega_\mu \tau} \right) \\
    + \dfrac{1}{2}\dfrac{\delta_{\mu\nu}}{A_\mu^2} \musum{\lambda\neq\mu} A_\lambda^2 \inttau \mathrm{Im}\left( \left\langle c_{2\mu\lambda} c_{2\mu\lambda}^{\tau\star} \right\rangle \exp^{\ii(\omega_\mu - \omega_\lambda)\tau} \right) \\
    - \delta_{\mu\nu} \inttau \left\langle \mathrm{Re}\left( c_{2\mu\mu} \right) \mathrm{Im}\left( c_{2\mu\mu}^\tau \right) \right\rangle \\
    + \dfrac{1}{2}\left(1 - \delta_{\mu\nu} \right) \inttau \mathrm{Im}\left( \left\langle c_{2\mu\nu} c_{2\nu\mu}^\tau \right\rangle \exp^{\ii (\omega_\mu - \omega_\nu) \tau} \right) \\
    + \delta_{\mu\nu} \inttau \left\langle \mathrm{Re}\left( c_{2\mu\mu} \right) \mathrm{Im}\left( c_{2\mu\mu}^\tau \right) \right\rangle \\
    + \dfrac{1}{2} \dfrac{\delta_{\mu\nu}}{A_\mu^2} \musum{\lambda} A_\lambda^2 \inttau \mathrm{Im}\left( \left\langle c_{3\mu\lambda} c_{3\mu\lambda}^{\tau\star} \right\rangle \exp^{\ii (\omega_\mu + \omega_\lambda) \tau} \right) \\
    + \dfrac{1}{2} \inttau \mathrm{Im}\left( \left\langle c_{3\mu\nu} c_{3\nu\mu}^{\tau\star} \right\rangle \exp^{\ii (\omega_\mu + \omega_\nu) \tau} \right)~,
\end{multline}
\begin{multline}
    \acc{G_\mu}{G_\nu^\tau} \\
    = \dfrac{1}{2} \delta_{\mu\nu} \inttau \mathrm{Re} \left( \left\langle c_{1\mu} c_{1\mu}^{\tau\star} \right\rangle \exp^{\ii \omega_\mu \tau} \right) \\
    + \dfrac{1}{2} \delta_{\mu\nu} \musum{\lambda\neq\mu} A_\lambda^2 \inttau \mathrm{Re}\left( \left\langle c_{2\mu\lambda} c_{2\mu\lambda}^{\tau\star} \right\rangle \exp^{\ii(\omega_\mu - \omega_\lambda)\tau} \right) \\
    + \delta_{\mu\nu} A_\mu^2 \inttau \left\langle \mathrm{Re}\left( c_{2\mu\mu} \right) \mathrm{Re}\left( c_{2\mu\mu}^\tau \right) \right\rangle \\
    + \dfrac{1}{2}\left(1 - \delta_{\mu\nu} \right) A_\mu A_\nu \inttau \mathrm{Re}\left( \left\langle c_{2\mu\nu} c_{2\nu\mu}^\tau \right\rangle \exp^{\ii (\omega_\mu - \omega_\nu) \tau} \right) \\
    + \dfrac{1}{2} \delta_{\mu\nu} \musum{\lambda} A_\lambda^2 \inttau \mathrm{Re}\left( \left\langle c_{3\mu\lambda} c_{3\mu\lambda}^{\tau\star} \right\rangle \exp^{\ii (\omega_\mu + \omega_\lambda) \tau} \right) \\
    + \dfrac{1}{2} \left(1 - \delta_{\mu\nu} \right) A_\mu A_\nu \inttau \mathrm{Re}\left( \left\langle c_{3\mu\nu} c_{3\nu\mu}^{\tau\star} \right\rangle \exp^{\ii (\omega_\mu + \omega_\nu) \tau} \right)~,
\end{multline}
and
\begin{multline}
    \acc{H_\mu}{H_\nu^\tau} \\
    = \dfrac{1}{2} \dfrac{\delta_{\mu\nu}}{A_\mu^2} \inttau \mathrm{Re} \left( \left\langle c_{1\mu} c_{1\mu}^{\tau\star} \right\rangle \exp^{\ii \omega_\mu \tau} \right) \\
    + \dfrac{1}{2} \dfrac{\delta_{\mu\nu}}{A_\mu^2} \musum{\lambda\neq\mu} A_\lambda^2 \inttau \mathrm{Re}\left( \left\langle c_{2\mu\lambda} c_{2\mu\lambda}^{\tau\star} \right\rangle \exp^{\ii(\omega_\mu - \omega_\lambda)\tau} \right) \\
    + \delta_{\mu\nu} \inttau \left\langle \mathrm{Im}\left( c_{2\mu\mu} \right) \mathrm{Im}\left( c_{2\mu\mu}^\tau \right) \right\rangle \\
    - \dfrac{1}{2}\left(1 - \delta_{\mu\nu} \right) \inttau \mathrm{Re}\left( \left\langle c_{2\mu\nu} c_{2\nu\mu}^\tau \right\rangle \exp^{\ii (\omega_\mu - \omega_\nu) \tau} \right) \\
    + \dfrac{1}{2} \dfrac{\delta_{\mu\nu}}{A_\mu^2} \musum{\lambda} A_\lambda^2 \inttau \mathrm{Re}\left( \left\langle c_{3\mu\lambda} c_{3\mu\lambda}^{\tau\star} \right\rangle \exp^{\ii (\omega_\mu + \omega_\lambda) \tau} \right) \\
    + \dfrac{1}{2} \left(1 - \delta_{\mu\nu} \right) \inttau \mathrm{Re}\left( \left\langle c_{3\mu\nu} c_{3\nu\mu}^{\tau\star} \right\rangle \exp^{\ii (\omega_\mu + \omega_\nu) \tau} \right)~.
\end{multline}

In turn, plugging these expressions into \cref{eq:appA-DefinitionProbabilityFluxAmplitude,eq:appA-DefinitionProbabilityFluxPhase,eq:appA-DefinitionDiffusionCoefficientAmplitude,eq:appA-DefinitionDiffusionCoefficientCross,eq:appA-DefinitionDiffusionCoefficientPhase} yields the final expressions for the probability flux and diffusion matrix elements, as given in the main body of this paper (\cref{eq:ProbabilityFluxAmplitude,eq:ProbabilityFluxPhase,eq:DiffusionCoefficientAmplitude,eq:DiffusionCoefficientOffDiagonal,eq:DiffusionCoefficientPhase}).

\subsection{Resonances}

To illustrate the emergence of these resonance terms in the coefficients of the Fokker-Planck equation, we consider the curly bracket operator acting on the second term on the right-hand side of $\partial G_\mu / \partial A_\nu$ and the third term on the right-hand side of $G_\nu$. These two terms oscillate at frequencies $\omega_\mu - \omega_\nu$ and $\omega_\nu - \omega_\lambda$, respectively. As we saw before, the resulting curly bracket only yields non-zero contributions to the long-term evolution of the mode amplitudes and phases if these two frequencies are either equal or opposite. Obviously, for $\lambda = \mu$ the two frequencies are the exact opposite of each other: from this stems one of the cross terms accounted for in the previous section (more specifically, the first term on the right-hand side of \cref{eq:appA-ExpansionG1}). But there may also be another mode $\lambda$ for which the two above frequencies coincide (i.e. $\omega_\mu + \omega_\lambda - 2\omega_\nu = 0$). To be perfectly accurate, this resonance condition only needs to be verified to within the sum of the inverse lifetimes of all the modes involved (i.e. about $10~\mu$Hz close to the frequency of maximum spectral height $\nu_\mathrm{max}$ for the Sun). Following the lead of \citet{kumar89} for instance, we argue that given the high density of modes per unit frequency in the solar $p$-mode spectrum, the sums over frequencies can be treated as integrals, and the resonance condition enforced with a Dirac distribution. The resulting contribution then reads
\begin{align}
    & \displaystyle\int \d\omega_\lambda ~ \delta\left( \omega_\mu + \omega_\lambda - 2\omega_\nu \right) A_\lambda \nonumber \\
    & \hspace{1cm} \times \left\{ \vphantom{\dfrac{a}{a}} C_{2\mu\nu} \lcos{(\omega_\mu - \omega_\nu)t + \Phi_\mu - \Phi_\nu - \phi_{2\mu\nu}} ~ ; \right. \nonumber \\
    & \hspace{1cm} \left. \vphantom{\dfrac{a}{a}} C_{2\nu\lambda}^\tau \lcos{(\omega_\mu - \omega_\nu)(t + \tau) + \Phi_\nu - \Phi_\lambda - \phi_{2\nu\lambda}} \right\} \nonumber \\
    & = \dfrac{1}{2} \displaystyle\int \d\omega_\lambda ~ \delta\left( \omega_\mu + \omega_\lambda - 2\omega_\nu \right) A_\lambda \nonumber \\
    & \hspace{1cm} \inttau \mathrm{Re}\left( \left\langle c_{2\mu\nu} c_{2\nu\lambda}^{\tau\star} \right\rangle \exp^{\ii(2\Phi_\nu - \Phi_\lambda - \Phi_\mu)} ~ \exp^{\ii(\omega_\mu - \omega_\nu)\tau} \right)~.
\end{align}
It can be seen that the major difference between this contribution and those computed in the previous section is the additional factor $\exp^{\ii( 2\Phi_\nu - \Phi_\lambda - \Phi_\mu)}$ inside the integral (this factor being rigorously equal to unity in each of the cross terms above). Under the hypothesis that the phase differences between the modes are fairly independent from one pair of modes to another, it is reasonable to assume that all these resonance terms cancel each other out, thus yielding a total net contribution that can be neglected compared to the cross terms computed in the previous section.

\section{Derivation of the coefficients $\alpha_i$\label{appB}}

The evolution equations on the mean mode energy and phase, given by \cref{eq:MeanEnergy,eq:MeanPhase}, only depend on the constant, complex values of the coefficients $\alpha_1$ and $\alpha_3$. We recall here, for clarity, that they are defined as the complex autocorrelation spectra of the stochastic processes $c_1(t)$ and $c_3(t)$ evaluated at angular frequencies $\omega$ and $2\omega$, respectively, where $\omega$ is the angular eigenfrequency of the mode, such that (see \cref{eq:AutocorrelationSpectrumC1,eq:AutocorrelationSpectrumC3})
\begin{align}
    & \alpha_1 = \displaystyle\int_{-\infty}^0 \left\langle c_1(t)c_1^\star(t+\tau) \right\rangle \exp^{\ii\omega\tau}~\d\tau~, \label{eq:appB-AutocorrelationSpectrumC1} \\
    & \alpha_3 = \displaystyle\int_{-\infty}^0 \left\langle c_3(t)c_3^\star(t+\tau) \right\rangle \exp^{2\ii\omega\tau}~\d\tau~.
    \label{eq:appB-AutocorrelationSpectrumC3}
\end{align}
In turn, these stochastic processes are given by (see \cref{eq:ExpressionC1,eq:ExpressionC3})
\begin{align}
    & c_1 = 2 ~ \mathfrak{I}\left( \vphantom{f_1^\tau} \omega\Psi_{\xi,i} \xiti{j} \partial_j \uti{i} + \Psi_{u,i} L_{0,i} \right) \label{eq:appB-ExpressionC1} \\
    & c_3 = \mathfrak{I}\left( \vphantom{f_1^\tau} \Psi_{\xi,i} \Psi_{\xi,j} \partial_j \uti{i} + \omega \Psi_{\xi,i} \xiti{j} \partial_j \Psi_{u,i} + \Psi_{u,i} L_{1,i}^s \right)~,
    \label{eq:appB-ExpressionC3}
\end{align}
where we recall that the operator $\mathfrak{I}$ is given by \cref{eq:OperatorIntegral}, $\bm{\Psi}_\xi$ and $\bm{\Psi}_u$ are the normalised displacement and velocity eigenfunctions respectively, and we have (see \cref{eq:StochasticLinearOperator,eq:StochasticDriving})
\begin{multline}
    L_{1,i}^s = -\Psi_{u,j} \partial_j \uti{i} - \uti{j} \partial_j \Psi_{u,i} - \dfrac{G_{ij,0}}{\omega \rho_0} \mathfrak{I}\left( \Psi_{\xi,k} \partial_k \left(\uti{j} K^{\mathbf{x}} \right) \right) \\
    + \left(\dfrac{\partial G_{ij}}{\partial \widetilde{u_k'' u_l''}} \widetilde{u_k'' u_l''}_{(1)} + \dfrac{\partial G_{ij}}{\partial (\partial_k\widetilde{u_l})} \partial_k \widetilde{u_l}_{(1)} + \dfrac{1}{2} \dfrac{\partial G_{ij}}{\partial \epsilon} \omega_t \widetilde{u_i''u_i''}_{(1)} \right) \uti{j} \\
    + \dfrac{1}{4}\sqrt{\dfrac{2 C_0 \omega_t}{\widetilde{u_i''u_i''}_0}} \widetilde{u_i''u_i''}_{(1)} \eta_i
    \label{eq:appB-StochasticLinearOperator}
\end{multline}
and
\begin{equation}
    L_{0,i} = -\dfrac{1}{\rho_0} \partial_j \left( \vphantom{\dfrac{1}{2}} \rho_0 \uti{i}\uti{j} - \rho_0 \overline{\uti{i} \uti{j}} \right)~,
    \label{eq:appB-StochasticDriving}
\end{equation}
and the perturbations of the Reynolds-stress tensor and mean shear tensor are given by (see \cref{eq:PerturbationReynoldsStressTensor,eq:PerturbationMeanShearTensor})
\begin{multline}
    \rho_0 \widetilde{u_i'' u_j''}_{(1)} = - \widetilde{u_i''u_j''}_0 \mathfrak{I}\left( \dfrac{\Psi_{\xi,k}}{\omega} \partial_k K^{\mathbf{x}} \right) \\
    + \mathfrak{I}\left( \vphantom{\dfrac{1}{2}} \dfrac{\Psi_{\xi,k}}{\omega} \partial_k \left(\uti{i} \uti{j} K^{\mathbf{x}}\right) + \uti{i} \Psi_{u,j} K^{\mathbf{x}} + \uti{j} \Psi_{u,i} K^{\mathbf{x}} \right)
    \label{eq:appB-PerturbationReynoldsStressTensor}
\end{multline}
and
\begin{multline}
    \rho_0 (\partial_i \widetilde{u_j})_{(1)} = - \mathfrak{I}\left( \Psi_{u,j} \partial_i K^{\mathbf{x}} + \dfrac{\Psi_{\xi,k}}{\omega} \partial_k \left(\uti{j} \partial_i K^{\mathbf{x}}\right) \right) \\
    - \dfrac{1}{\rho_0} \partial_i \rho_0 \mathfrak{I}\left( \Psi_{u,j} K^{\mathbf{x}} + \dfrac{\Psi_{\xi,k}}{\omega} \partial_k \left(\uti{j} K^{\mathbf{x}}\right) \right)~.
    \label{eq:appB-PerturbationMeanShearTensor}
\end{multline}
The apparition of the factor $\omega^{-1}$ in conjunction with every occurrence of $\bm{\Psi}_\xi$ stems from the fact that the latter is defined in terms of $\omega\xiosc$ rather than simply $\xiosc$.

\subsection{Contribution of the turbulent displacement field\label{sec:appB-SmallnessTurbulentDisplacement}}

In Paper I we used a certain number of approximations to derive our formalism. One of these approximations -- which we labelled (H3) -- consisted in adopting the anelastic approximation for turbulence, in the sense that we considered $\rho_t \ll \rho_0$, where $\rho_t$ is the turbulent fluctuation of density, and $\rho_0$ is the equilibrium density. Using the continuity equation, this amounts to neglecting the quantity $\bm{\nabla} \cdot \left(\rho_0 \bm{\xi}_t\right)$. As we will now see, this allows us to discard the first term on the right-hand side of \cref{eq:appB-ExpressionC1} and the second term on the right-hand side of \cref{eq:appB-ExpressionC3}. Considering the former and performing an integration by part, we can put it in the following form:
\begin{multline}
    c_1 = [...] + 2\displaystyle\int_\mathcal{S} \d^2\mathbf{x} ~ \rho_0 \Psi_{\xi,i} \uti{i} ~ \xit\cdot\mathbf{n} \\
    - 2\displaystyle\int \d^3\mathbf{x} ~ \uti{i} \bm{\nabla} \cdot \left(\rho_0 \Psi_{\xi,i} \xit \right)~.
\end{multline}
The surface integral (i.e. the first term on the right-hand side) systematically vanishes, on account of the product $\rho_0 \bm{\Psi_{\xi}}$ vanishing on the surface of the star. Furthermore, the typical length scale over which $\xit$ varies is much smaller than the wavelength of the mode. This allows us to pull $\Psi_{\xi,i}$ out of the gradient in the last term, and we recognise $\bm{\nabla} \cdot \left(\rho_0 \xit \right)$, which we neglected on account of our hypothesis (H3). The same procedure can be applied to the second term on the right-hand side of \cref{eq:appB-ExpressionC3}. As a result, in \cref{eq:appB-ExpressionC1,eq:appB-ExpressionC3}, the only source of stochasticity comes from the turbulent velocity field $\ut$ and the Wiener process $\bm{\eta}$, with no contribution from the turbulent displacement field $\xit$.

\subsection{Derivation of $\alpha_1$\label{sec:appB-Alpha1}}

Plugging \cref{eq:appB-StochasticDriving} into \cref{eq:appB-ExpressionC1}, neglecting the first term on the right-hand side (see \cref{sec:appB-SmallnessTurbulentDisplacement} for a justification), and performing an integration by parts, the quantity $c_1(t)$ can be rewritten in the following way:
\begin{equation}
    c_1 = -2 \displaystyle\int \d^3\mathbf{x} ~ \partial_j \left( \rho_0 \Psi_{u,i} \right) \left[ \uti{i}\uti{j} \right]'~,
\end{equation}
where the surface contribution of the integration by part vanishes, because it involves $\rho_0 \bm{\Psi}_u$ at the outer boundary of the star, where the oscillation is evanescent. This allows for a simple physical interpretation of this integral: in broad strokes, the quantity $\partial_j \left( \rho_0 \Psi_{u,i} \right) / \rho_0$ corresponds to the local compressibility of the mode, while the quantity $\left[ \rho_0 \uti{i}\uti{j} \right]'$ corresponds to the instantaneous fluctuating turbulent pressure. The product between the two can therefore be interpreted as a local instantaneous `$p\d V$' mechanical work exerted by the turbulent pressure on the mode; integrating over the entire stellar volume yields the total instantaneous work $c_1$; and forming the autocorrelation function of $c_1$ yields the effective work exerted on the mode over a large number of cycles, which corresponds to the excitation rate of the mode.

We remark that only the fluctuation of the turbulent pressure around its equilibrium state appears in this integral. Expanding it into the difference between the total turbulent pressure and its equilibrium value leads to four different integrals. However, it is readily seen that only one of them -- the one not involving the equilibrium turbulent pressure at all -- will have a non-zero $\omega$-component in its Fourier transform in time, and therefore will contribute to $\alpha_1$. As such, $\left[ \uti{i}\uti{j} \right]'$ can be replaced by $\uti{i}\uti{j}$. Then, forming the autocorrelation function of $c_1$, we obtain
\begin{multline}
    \left\langle \vphantom{f_1^\tau} c_1(t)c_1^\star(t+\tau) \right\rangle = 4 \displaystyle\iint \d^3\mathbf{x}\d^3\mathbf{x}' ~ \left.\partial_j \left( \rho_0 \Psi_{u,i} \right)\right|_{\mathbf{x}} \left. \partial_l \left( \rho_0 \Psi_{u,k}^\star \right)\right|_{\mathbf{x}'} \\
    \times \left\langle u_{t,i} u_{t,j}(\mathbf{x},t) u_{t,k} u_{t,l}(\mathbf{x}',t+\tau) \right\rangle~.
\end{multline}

In the scope of the Jeffreys-Wentzel–Kramers–Brillouin (JWKB) approximation, the velocity eigenfunction can be locally approximated by the following expression
\begin{equation}
    \Psi_{u,i}(\mathbf{x}) = \Psi_{u,i,0}(\mathbf{x})\exp^{\ii \mathbf{k}(\mathbf{x})\cdot\mathbf{x}}~,
    \label{eq:appB-JWKB}
\end{equation}
where $\bm{\Psi}_{u,0}(\mathbf{x})$ is the slowly varying amplitude in space of the velocity eigenfunction, and $\mathbf{k}(\mathbf{x})$ is the space-dependent wave vector of the mode. Both $\bm{\Psi}_{u,0}$ and $\mathbf{k}$ are slowly varying functions of space (meaning that they vary on length scales much larger than $|\mathbf{k}|^{-1}$). As such, we have
\begin{equation}
    \partial_j \left( \rho_0 \Psi_{u,i} \right) \sim \ii k_j(\mathbf{x}) \rho_0(\mathbf{x}) \Psi_{u,i,0}(\mathbf{x}) \exp^{\ii \mathbf{k}(\mathbf{x})\cdot\mathbf{x}}~,
\end{equation}
so
\begin{multline}
    \left\langle \vphantom{f_1^\tau} c_1(t)c_1^\star(t+\tau) \right\rangle = 4 \displaystyle\iint \d^3\mathbf{x}\d^3\mathbf{x}' ~ \left[ \vphantom{\dfrac{1}{2}} \rho_0 k_j \Psi_{u,i,0}(\mathbf{x}) \rho_0 k_l \Psi_{u,k,0}^\star(\mathbf{x}') \right. \\
    \left. \vphantom{\dfrac{1}{2}} \times \exp^{\ii (\mathbf{k}(\mathbf{x})\cdot\mathbf{x} - \mathbf{k}(\mathbf{x}')\cdot\mathbf{x}')} \left\langle u_{t,i} u_{t,j}(\mathbf{x},t) u_{t,k} u_{t,l}(\mathbf{x}',t+\tau) \right\rangle \right]~.
    \label{eq:appB-ExpressionC1Temp}
\end{multline}
Then, we implement the following change of variables:
\begin{align}
    & \mathbf{X} \equiv \mathbf{x}~, \\
    & \delta \mathbf{x} \equiv \mathbf{x}' - \mathbf{x}~,
\end{align}
where $\mathbf{X}$ and $\delta\mathbf{x}$ represent the slow and fast space variables, respectively. The implementation of this change of variables in \cref{eq:appB-ExpressionC1Temp} allows us to completely decouple the slowly varying contributions of $\bm{\Psi}_{u,0}$ and $\mathbf{k}$ on the one hand, and the rapidly varying contributions of $\exp(\ii \mathbf{k} \cdot \mathbf{x})$ and the turbulent velocity two-point correlation products on the other. We obtain\footnote{The Jacobian of the change of variable $(\mathbf{x},\mathbf{x}') \mapsto (\mathbf{X},\delta\mathbf{x})$ is straightforwardly estimated, and happens to equal unity.}
\begin{multline}
    \left\langle \vphantom{f_1^\tau} c_1(t)c_1^\star(t+\tau) \right\rangle = 4 \displaystyle\iint \d^3\mathbf{X}\d^3\delta\mathbf{x} ~ \left[ \vphantom{\dfrac{1}{2}} \rho_0^2 k_j k_l \Psi_{u,i,0} \Psi_{u,k,0}^\star(\mathbf{X}) \right. \\
    \left. \vphantom{\dfrac{1}{2}} \times \exp^{-\ii \mathbf{k}(\mathbf{X}) \cdot \delta\mathbf{x}} \left\langle u_{t,i} u_{t,j}\left(\mathbf{X},t\right) u_{t,k} u_{t,l}\left(\mathbf{X}+\delta\mathbf{x},t+\tau\right) \right\rangle \right]~.
\end{multline}
Taking advantage of the scale decoupling, we can separate the integrals over $\mathbf{X}$ and $\delta\mathbf{x}$,
\begin{multline}
    \left\langle \vphantom{f_1^\tau} c_1(t)c_1^\star(t+\tau) \right\rangle = 4 \displaystyle\int \d^3\mathbf{X} ~ \rho_0^2 k_j k_l \Psi_{u,i,0} \Psi_{u,k,0}^\star(\mathbf{X}) \\
    \displaystyle\int \d^3\delta\mathbf{x} \exp^{-\ii \mathbf{k}(\mathbf{X}) \cdot \delta\mathbf{x}} \left\langle u_{t,i} u_{t,j}\left(\mathbf{X},t\right) u_{t,k} u_{t,l}\left(\mathbf{X}+\delta\mathbf{x},t+\tau\right) \right\rangle~.
\end{multline}
Finally, using \cref{eq:appB-AutocorrelationSpectrumC1}, and taking advantage of the definition of the operator $\mathfrak{I}$ (see \cref{eq:OperatorIntegral}), the autocorrelation spectrum $\alpha_1$ can thus be expressed as
\begin{equation}
    \alpha_1 = 4 ~ \mathfrak{I}\left( \rho_0 k_j k_l \Psi_{u,i,0} \Psi_{u,k,0}^\star \phi_{ijkl}^{(4b)}(\mathbf{k},\omega) \right)~,
    \label{eq:appB-Alpha1Temp}
\end{equation}
where $\phi_{ijkl}^{(4b)}$ is the fourth-order correlation spectrum of the turbulent velocity $\mathbf{u}_t$, defined as
\begin{equation}
    \phi_{ijkl}^{(4b)}(\mathbf{k},\omega) \equiv \mathcal{C}_{\omega,\mathbf{k}}(\uti{i}\uti{j} ~;~ \uti{k}\uti{l})~,
\end{equation}
and the operator $\mathcal{C}_{\omega, \mathbf{k}}$ is defined by
\begin{multline}
    \mathcal{C}_{\omega, \mathbf{k}}(f_1 ~;~ f_2) \equiv \displaystyle\int_{-\infty}^0 \d\tau \displaystyle\int \d^3\delta\mathbf{x} ~ \left\langle f_1 f_2^{\delta\mathbf{x},\tau} \right\rangle \exp^{\ii (\omega\tau - \mathbf{k}\cdot\delta\mathbf{x})}~,
    \label{eq:appB-Phi4b}
\end{multline}
and
\begin{equation}
    f^{\delta\mathbf{x},\tau}(\mathbf{X},T) \equiv f(\mathbf{X}+\delta\mathbf{x}, T+\tau)~.
\end{equation}
Naturally, $\phi_{ijkl}^{(4b)}$ also depends on the slow variable $\mathbf{X}$ and time $t$, even though they do not appear explicitly in \cref{eq:appB-Alpha1Temp}.

\subsection{Derivation of $\alpha_3$\label{sec:appB-Alpha3}}

Plugging \cref{eq:appB-StochasticLinearOperator} into \cref{eq:appB-ExpressionC3} (from which the second term on the right-hand side was discarded, as per the argument developed in \cref{sec:appB-SmallnessTurbulentDisplacement}), we obtain
\begin{multline}
    c_3 = \mathfrak{I}\left( \vphantom{\dfrac{1}{2}} \Psi_{\xi,i} \Psi_{\xi,j} \partial_j u_{t,i} - \Psi_{u,i} \Psi_{u,j} \partial_j u_{t,i} - \Psi_{u,i} u_{t,j} \partial_j \Psi_{u,i} \right. \\
    - \dfrac{1}{\rho_0} \Psi_{u,i} \dfrac{G_{ij,0}}{\omega} \displaystyle\int \d^3\mathbf{y} ~ \rho_0 \Psi_{\xi,k} \partial_k \left(u_{t,j} K^{\mathbf{x}}\right) + \Psi_{u,i} u_{t,j} \dfrac{\partial G_{ij}}{\partial \widetilde{u_k'' u_l''}} \widetilde{u_k'' u_l''}_{(1)} \\
    + \Psi_{u,i} u_{t,j} \dfrac{\partial G_{ij}}{\partial (\partial_k \widetilde{u_l})} (\partial_k \widetilde{u_l})_{(1)} + \dfrac{1}{2} \Psi_{u,i} u_{t,j} \dfrac{\partial G_{ij}}{\partial \epsilon} \omega_t \widetilde{u_i''u_i''}_{(1)} \\
    + \left. \Psi_{u,i} \dfrac{\widetilde{u_i''u_i''}_{(1)}}{2\omega}\sqrt{\dfrac{2 C_0 \omega_t}{\widetilde{u_i''u_i''}_0}} \eta_i \right)~,
    \label{eq:appB-ExpressionC3Temp}
\end{multline}
where the perturbation of the Reynolds-stress tensor $\widetilde{u_k''u_l''}_{(1)}$ and mean shear tensor $(\partial_k \widetilde{u_l})_{(1)}$ are given by \cref{eq:appB-PerturbationReynoldsStressTensor,eq:appB-PerturbationMeanShearTensor}, respectively.

We consider the first four terms on the right-hand side of \cref{eq:appB-ExpressionC3Temp}. They can be rearranged with integrations by part to yield
\begin{align}
    & c_{3a} = \displaystyle\int \d^3\mathbf{x} ~ u_{t,i} \left( \vphantom{\dfrac{1}{2}} -\partial_j \left(\rho_0 \Psi_{\xi,i} \Psi_{\xi,j} \right) \right. \nonumber \\
    & \hspace{2.5cm} \left. \vphantom{\dfrac{1}{2}} + \partial_j \left(\rho_0 \Psi_{u,i} \Psi_{u,j} \right) - \rho_0 \Psi_{u,i} \partial_j \Psi_{u,i} \right)~, \\
    & c_{3b} = \displaystyle\int \d^3\mathbf{x} ~ \Psi_{u,i} \dfrac{G_{ij,0}}{\omega} \mathfrak{I}\left( u_{t,j} K^{\mathbf{x}} \partial_k \left(\ln\rho_0 \Psi_{\xi,k} \right) \right)~. \\
\end{align}
The last term can be further simplified by permuting the integral over $\mathbf{x}$ and the integral defining the operator $\mathfrak{I}$ (see \cref{eq:OperatorIntegral}), which yields
\begin{equation}
    c_{3b} = \displaystyle\int \d^3\mathbf{y} ~ u_{t,j} \partial_k \left(\rho_0 \Psi_{\xi,k} \right) \displaystyle\int \d^3\mathbf{x} ~ \Psi_{u,i}(\mathbf{x}) \dfrac{G_{ij,0}}{\omega} K^{\mathbf{y}}(\mathbf{x})~,
\end{equation}
where we have used the isotropy of the kernel function $K$ to write $K^{\mathbf{x}}(\mathbf{y}) = K^{\mathbf{y}}(\mathbf{x})$. It can be seen that the integral over $\mathbf{x}$ corresponds to the kernel estimate at point $\mathbf{y}$ of the quantity $\Psi_{u,i} G_{ij,0} / (\omega \rho_0)$, which only involves quantities that are not stochastic. As such, this kernel estimation equals the actual value of this quantity at $\mathbf{y}$, and $c_{3d}(t)$ reduces to
\begin{equation}
    c_{3b} = \displaystyle\int \d^3\mathbf{y} \dfrac{G_{ij,0}}{\omega} u_{t,j} \Psi_{u,i} \partial_k \left(\rho_0 \Psi_{\xi,k} \right)~.
\end{equation}

As for the fifth, sixth, and seventh terms on the right-hand side of \cref{eq:appB-ExpressionC3Temp}, we can expand them in the following way. For the sake of clarity, we define the following symbol:
\begin{equation}
    \int_{\mathbf{xy}} \equiv \int \d^3\mathbf{x}\int \d^3\mathbf{y}~.
\end{equation}
Then the fifth term on the right-hand side of \cref{eq:appB-ExpressionC3Temp} can be split into
\begin{equation}
    c_{3c} = \displaystyle\int_{\mathbf{xy}} \uti{j}(\mathbf{x}) \Psi_{u,i}(\mathbf{x}) \left.\dfrac{\partial G_{ij}}{\partial \widetilde{u_k''u_l''}}\right|_{\mathbf{x}} \dfrac{\widetilde{u_k''u_l''}_0}{\omega} \left.\partial_m \left( \rho_0 \Psi_{\xi,m} \right)\right|_{\mathbf{y}} K^{\mathbf{x}}(\mathbf{y})~,
\end{equation}
\begin{equation}
    c_{3d} = \displaystyle\int_{\mathbf{xy}} \uti{j}(\mathbf{x}) \uti{k}(\mathbf{y}) \Psi_{u,i}(\mathbf{x}) \left.\dfrac{\partial G_{ij}}{\partial \widetilde{u_k''u_l''}}\right|_{\mathbf{x}} \rho_0(\mathbf{y}) \Psi_{u,l}(\mathbf{y}) K^{\mathbf{x}}(\mathbf{y})~,
\end{equation}
\begin{equation}
    c_{3e} = \displaystyle\int_{\mathbf{xy}} \uti{j}(\mathbf{x}) \uti{l}(\mathbf{y}) \Psi_{u,i}(\mathbf{x}) \left.\dfrac{\partial G_{ij}}{\partial \widetilde{u_k''u_l''}}\right|_{\mathbf{x}} \rho_0(\mathbf{y}) \Psi_{u,k}(\mathbf{y}) K^{\mathbf{x}}(\mathbf{y})
\end{equation}
and
\begin{multline}
    c_{3f} = - \displaystyle\int_{\mathbf{xy}} \uti{j}(\mathbf{x}) \uti{k}(\mathbf{y}) \uti{l}(\mathbf{y}) \Psi_{u,i}(\mathbf{x}) \dfrac{1}{\omega} \left.\dfrac{\partial G_{ij}}{\partial \widetilde{u_k''u_l''}}\right|_{\mathbf{x}} \\
    \left.\partial_m \left( \rho_0 \Psi_{\xi,m} \right)\right|_{\mathbf{y}} K^{\mathbf{x}}(\mathbf{y})~;
\end{multline}
the sixth term can be split into
\begin{equation}
    c_{3g} = \displaystyle\int_{\mathbf{xy}} \uti{j}(\mathbf{x}) \Psi_{u,i}(\mathbf{x}) \left.\dfrac{\partial G_{ij}}{\partial (\partial_k \widetilde{u_l})}\right|_{\mathbf{x}} \left.\partial_i \left(\rho_0 \Psi_{u,l} \right)\right|_{\mathbf{y}} K^{\mathbf{x}}(\mathbf{y})~,
\end{equation}
\begin{equation}
    c_{3h} = - \displaystyle\int_{\mathbf{xy}} \uti{j}(\mathbf{x}) \dfrac{\rho_0(\mathbf{y})}{\rho_0(\mathbf{x})} \Psi_{u,i}(\mathbf{x}) \left.\dfrac{\partial G_{ij}}{\partial (\partial_k \widetilde{u_l})}\right|_{\mathbf{x}} \left.\partial_k \rho_0\right|_{\mathbf{x}} \Psi_{u,l}(\mathbf{y}) K^{\mathbf{x}}(\mathbf{y})~,
\end{equation}
\begin{equation}
    c_{3i} = \displaystyle\int_{\mathbf{xy}} \uti{j}(\mathbf{x}) \uti{k}(\mathbf{y}) \Psi_{u,i}(\mathbf{x}) \dfrac{1}{\omega} \left.\dfrac{\partial G_{ij}}{\partial (\partial_l \widetilde{u_k})}\right|_{\mathbf{x}} \left.\partial_m \left(\rho_0 \Psi_{u,m} \right)\right|_{\mathbf{y}} \left.\partial_l K^{\mathbf{x}}\right|_{\mathbf{y}}
\end{equation}
and
\begin{multline}
    c_{3j} = \displaystyle\int_{\mathbf{xy}} \uti{j}(\mathbf{x}) \uti{k}(\mathbf{y}) \dfrac{1}{\omega \rho_0(\mathbf{x})} \Psi_{u,i}(\mathbf{x}) \left.\dfrac{\partial G_{ij}}{\partial (\partial_l \widetilde{u_k})}\right|_{\mathbf{x}} \\
    \left.\partial_l \left(\rho_0\right) \right|_{\mathbf{x}} \left.\partial_m \left( \rho_0 \Psi_{u,m} \right)\right|_{\mathbf{y}} K^{\mathbf{x}}(\mathbf{y})~;
\end{multline}
and the seventh term can be split into
\begin{equation}
    c_{3k} = \displaystyle\int_{\mathbf{xy}} \uti{j}(\mathbf{x}) \Psi_{u,i}(\mathbf{x}) \left.\dfrac{\partial G_{ij}}{\partial \epsilon}\right|_{\mathbf{x}} \dfrac{\omega_t \widetilde{u_n''u_n''}_0(\mathbf{x})}{2 \omega} \left.\partial_m \left( \rho_0 \Psi_{\xi,m} \right)\right|_{\mathbf{y}} K^{\mathbf{x}}(\mathbf{y})~,
\end{equation}
\begin{equation}
    c_{3l} = \displaystyle\int_{\mathbf{xy}} \uti{j}(\mathbf{x}) \uti{k}(\mathbf{y}) \omega_t \Psi_{u,i}(\mathbf{x}) \left.\dfrac{\partial G_{ij}}{\partial \epsilon}\right|_{\mathbf{x}} \rho_0(\mathbf{y}) \Psi_{u,k}(\mathbf{y}) K^{\mathbf{x}}(\mathbf{y})~,
\end{equation}
\begin{multline}
    c_{3m} = - \displaystyle\int_{\mathbf{xy}} \uti{j}(\mathbf{x}) \uti{k}(\mathbf{y}) \uti{k}(\mathbf{y}) \Psi_{u,i}(\mathbf{x}) \dfrac{1}{2} \dfrac{\omega_t}{\omega} \left.\dfrac{\partial G_{ij}}{\partial \epsilon}\right|_{\mathbf{x}} \\
    \left.\partial_m \left( \rho_0 \Psi_{\xi,m} \right)\right|_{\mathbf{y}} K^{\mathbf{x}}(\mathbf{y}) \delta_{kl}~.
\end{multline}

Finally, in the last term on the right-hand side of \cref{eq:appB-ExpressionC3Temp}, the quantity under the integral is proportional to the stochastic process $\eta_i(\mathbf{x},t)$, which, by construction, is $\delta$-correlated in both space and time. In particular, its correlation length scale is infinitesimally small compared to that of the turbulent velocity $\ut(\mathbf{x},t)$. But as we saw in \cref{sec:appB-Alpha1}, it is precisely the spatial coherence of the stochastic perturbations to the wave equation that explains its ability to impact the complex amplitude of the modes. As such, this part will not actually contribute to the final expression of $\alpha_3$, or to the stochastic amplitude equations in any way, and will be discarded in the following.

Formally, $c_3(t)$ can be written as a sum of contributions that are first-, second-, or third-order in terms of the turbulent velocity, $\ut$,
\begin{multline}
    c_3(t) = \displaystyle\int_{\mathbf{xy}} \left[ \vphantom{\dfrac{1}{2}} f_{1,i}(\mathbf{x},\mathbf{y}) u_{t,i}(\mathbf{x},t) + f_{2,ij}(\mathbf{x},\mathbf{y}) u_{t,i}(\mathbf{x},t) u_{t,j}(\mathbf{y},t) \right. \\
    \left. \vphantom{\dfrac{1}{2}} + f_{3,ijk}(\mathbf{x},\mathbf{y}) u_{t,i}(\mathbf{x},t) u_{t,j} u_{t,k}(\mathbf{y},t) \right]~,
    \label{eq:appB-ExpressionC3Formal}
\end{multline}
where
\begin{align}
    f_{1,i}(\mathbf{x},\mathbf{y}) \equiv
    & \left.\left[ \vphantom{\dfrac{G_{ij}}{\omega}} -\partial_j \left(\rho_0 \Psi_{\xi,i} \Psi_{\xi,j} \right) + \partial_j \left(\rho_0 \Psi_{u,i} \Psi_{u,j} \right) + \rho_0 \Psi_{u,j} \partial_i \Psi_{u,j} \right.\right. \nonumber \\
    & \left.\left. + \dfrac{G_{ji,0}}{\omega} \Psi_{u,j} \partial_k \left(\rho_0 \Psi_{\xi,k}\right) \right]\right|_{\mathbf{x}} \delta(\mathbf{x} - \mathbf{y}) \nonumber \\
    & + \left[\Psi_{u,j}(\mathbf{x}) \left.\dfrac{\partial G_{ji}}{\partial \widetilde{u_k''u_l''}}\right|_{\mathbf{x}} \dfrac{\widetilde{u_k''u_l''}_0}{\omega} \left.\partial_m \left( \rho_0 \Psi_{\xi,m} \right)\right|_{\mathbf{y}} \right. \nonumber \\
    & + \Psi_{u,j}(\mathbf{x}) \left.\dfrac{\partial G_{ji}}{\partial (\partial_k \widetilde{u_l})}\right|_{\mathbf{x}} \left.\partial_k \left( \rho_0 \Psi_{u,l} \right)\right|_{\mathbf{y}} \nonumber \\
    & - \dfrac{\rho_0(\mathbf{y})}{\rho_0(\mathbf{x})} \Psi_{u,j}(\mathbf{x}) \left.\dfrac{\partial G_{ji}}{\partial (\partial_k \widetilde{u_l})}\right|_{\mathbf{x}} \left.\partial_k \left( \rho_0 \right)\right|_{\mathbf{x}} \Psi_{u,l}(\mathbf{y}) \nonumber \\
    & \left. + \Psi_{u,j}(\mathbf{x}) \left.\dfrac{\partial G_{ji}}{\partial \epsilon}\right|_{\mathbf{x}} \dfrac{\omega_t \widetilde{u_n''u_n''}_0(\mathbf{x})}{2 \omega} \left.\partial_m \left( \rho_0 \Psi_{\xi,m} \right)\right|_{\mathbf{y}} \right] K^{\mathbf{x}}(\mathbf{y})~,
\end{align}
\begin{align}
    f_{2,ij}(\mathbf{x},\mathbf{y}) \equiv
    & \left[ \Psi_{u,k}(\mathbf{x}) \left.\dfrac{\partial G_{ki}}{\partial \widetilde{u_j''u_l''}}\right|_{\mathbf{x}} \rho_0(\mathbf{y}) \Psi_{u,l}(\mathbf{y}) \right. \nonumber \\
    & + \Psi_{u,k}(\mathbf{x}) \left.\dfrac{\partial G_{ki}}{\partial \widetilde{u_l''u_j''}}\right|_{\mathbf{x}} \rho_0(\mathbf{y}) \Psi_{u,l}(\mathbf{y}) \nonumber \\
    & + \dfrac{1}{\omega \rho_0(\mathbf{x})} \Psi_{u,k}(\mathbf{x}) \left.\dfrac{\partial G_{ki}}{\partial (\partial_l \widetilde{u_j})}\right|_{\mathbf{x}} \left.\partial_l \left( \rho_0 \right)\right|_{\mathbf{x}} \left.\partial_m \left( \rho_0 \Psi_{u,m} \right)\right|_{\mathbf{y}} \nonumber \\
    & \left. + \omega_t \Psi_{u,k}(\mathbf{x}) \left.\dfrac{\partial G_{ki}}{\partial \epsilon}\right|_{\mathbf{x}} \rho_0(\mathbf{y}) \Psi_{u,j}(\mathbf{y}) \right] K^{\mathbf{x}}(\mathbf{y}) \nonumber \\
    & + \Psi_{u,k}(\mathbf{x}) \dfrac{1}{\omega} \left.\dfrac{\partial G_{ki}}{\partial (\partial_l \widetilde{u_j})}\right|_{\mathbf{x}} \left.\partial_m \left( \rho_0 \Psi_{u,m} \right)\right|_{\mathbf{y}} \left.\partial_l \left( K^{\mathbf{x}} \right)\right|_{\mathbf{y}}
\end{align}
and
\begin{align}
    f_{3,ijk}(\mathbf{x},\mathbf{y}) \equiv
    & - \left[ \dfrac{1}{\omega} \Psi_{u,l}(\mathbf{x}) \left.\dfrac{\partial G_{li}}{\partial \widetilde{u_j''u_k''}}\right|_{\mathbf{x}} \left.\partial_m \left( \rho_0 \Psi_{\xi,m} \right)\right|_{\mathbf{y}} \right. \nonumber \\
    & \left. + \dfrac{1}{2} \dfrac{\omega_t}{\omega} \Psi_{u,l}(\mathbf{x}) \left.\dfrac{\partial G_{li}}{\partial \epsilon}\right|_{\mathbf{x}} \left.\partial_m \left( \rho_0 \Psi_{\xi,m} \right)\right|_{\mathbf{y}} \delta_{jk} \right] K^{\mathbf{x}}(\mathbf{y})~.
\end{align}

Forming the autocorrelation product of $c_3(t)$ from \cref{eq:appB-ExpressionC3Formal}, we see that the expansion involves correlation products of the turbulent velocity field $\ut$ of various orders, ranging from $2$ to $6$. In the following, we make the assumptions, often used in the context of Gaussian turbulence, that 1) these moments can be cut at fourth order, and 2) the contribution of the third-order moment can be neglected compared to that of the second- or fourth-order. With this approximation in mind, we then proceed to adopt the JWKB approximation in the same form as in \cref{sec:appB-Alpha1} (see \cref{eq:appB-JWKB}), which yields
\begin{align}
	& \langle c_3(t) c_3^\star(t+\tau) \rangle = \displaystyle\int \d^3\mathbf{X} \d^3\delta\mathbf{x} \d^3\delta\mathbf{y_1} \d^3\delta\mathbf{y_2} \nonumber \\
	& \exp\left( \vphantom{\dfrac{1}{1}} \ii ~ \mathbf{k} \cdot (-2\delta\mathbf{x} + \delta\mathbf{y_1} - \delta\mathbf{y_2})\right) \rho_0^2 \left[ \vphantom{\underbrace{1}_{1}} \left( \vphantom{\dfrac{1}{2}} F_i^{(1a)} F_j^{(1a) \star} ~  \delta(\delta\mathbf{y_1})\delta(\delta\mathbf{y_2}) \right. \right. \nonumber \\
	& + F_i^{(1a)} F_j^{(1b) \star} ~ \delta(\delta\mathbf{y_1}) K(\delta\mathbf{y_2}) + F_i^{(1b)} F_j^{(1a) \star} ~ K(\delta\mathbf{y_1}) \delta(\delta\mathbf{y_2}) \nonumber \\
	& \left. \vphantom{\dfrac{1}{2}} + F_i^{(1b)} F_j^{(1b) \star} ~ K(\delta\mathbf{y_1}) K(\delta\mathbf{y_2}) \right) \left\langle \uti{i}(\mathbf{X})\uti{j}^\tau(\mathbf{X}+\delta\mathbf{x}) \right\rangle \nonumber \\
	& + \left( \vphantom{\dfrac{1}{2}} F_i^{(1a)} F_{jkl}^{(3b) \star} ~ \delta(\delta\mathbf{y_1}) K(\delta\mathbf{y_2}) + F_i^{(1b)} F_{jkl}^{(3b) \star} ~ K(\delta\mathbf{y_1}) K(\delta\mathbf{y_2}) \right) \nonumber \\
	& \hspace{1cm} \times \left\langle \uti{i}(\mathbf{X})\uti{j}^\tau(\mathbf{X}+\delta\mathbf{x}) \uti{k}^\tau \uti{l}^\tau(\mathbf{X} + \delta\mathbf{x} - \delta\mathbf{y_2}) \right\rangle \nonumber \\
	& + \left( \vphantom{\dfrac{1}{2}} F_i^{(1a) \star} F_{jkl}^{(3b)} ~ K(\delta\mathbf{y_1}) \delta(\delta\mathbf{y_2}) + F_i^{(1b) \star}( F_{jkl}^{(3b)} ~ K(\delta\mathbf{y_1}) K(\delta\mathbf{y_2}) \right) \nonumber \\
	& \hspace{1cm} \times \left\langle \uti{i}^\tau(\mathbf{X}+\delta\mathbf{x})\uti{j}(\mathbf{X}) \uti{k} \uti{l}(\mathbf{X} - \delta\mathbf{y_1}) \right\rangle \nonumber \\
	& + \left( \vphantom{\dfrac{1}{2}} F_{ij}^{(2)} F_{kl}^{(2) \star} ~ K(\delta\mathbf{y_1}) K(\delta\mathbf{y_2}) + F_{ijm}^{(3a)} F_{kl}^{(2) \star} ~ \partial_m K(\delta\mathbf{y_1}) K(\delta\mathbf{y_2}) \right. \nonumber \\
	& \left. \vphantom{\dfrac{1}{2}} + F_{ij}^{(2)} F_{klm}^{(3a) \star} ~ K(\delta\mathbf{y_1}) \partial_m K(\delta\mathbf{y_2}) + F_{ijm}^{(3a)} F_{kln}^{(3a) \star} ~ \partial_m K(\delta\mathbf{y_1}) \partial_n K(\delta\mathbf{y_2}) \right) \nonumber \\
	& \hspace{1cm} \left. \vphantom{\underbrace{1}_{1}} \times \left\langle \uti{i}(\mathbf{X})\uti{j}(\mathbf{X}-\delta\mathbf{y_1}) \uti{k}^\tau(\mathbf{X} + \delta\mathbf{x}) \uti{l}^\tau(\mathbf{X} + \delta\mathbf{x} - \delta\mathbf{y_2}) \right\rangle \right]~,
	\label{eq:appB-AutocorrelationC3Temp}
\end{align}
where the subscript $\tau$ means that the turbulent velocity field is evaluated at time $t + \tau$, and we have introduced
\begin{align}
	& F_i^{(1a)} = 4 \ii k_j \Psi_{u,i,0} \Psi_{u,j,0} + \ii k_i \Psi_{u,j,0} \Psi_{u,j,0} + \dfrac{G_{ij,0}}{\omega} k_k \Psi_{u,j,0} \Psi_{u,k,0}~, \\
	& F_i^{(1b)} = \dfrac{\partial G_{ij}}{\partial \widetilde{u_k''u_l''}} \dfrac{\widetilde{u_k''u_l''}_0}{\omega} k_m \Psi_{u,m,0} \Psi_{u,j,0} + \dfrac{\partial G_{ij}}{\partial (\partial_k\widetilde{u_l})} \ii k_k \Psi_{u,j,0} \Psi_{u,l,0} \nonumber \\
	& \hspace{1cm} + \dfrac{\partial G_{ij}}{\partial \epsilon} \dfrac{\omega_t \widetilde{u_n''u_n''}_0}{2 \omega} k_m \Psi_{u,j,0} \Psi_{u,m,0}~, \\
	& F_{ij}^{(2)} = \left(\dfrac{\partial G_{ki}}{\partial \widetilde{u_j''u_l''}} + \dfrac{\partial G_{ki}}{\partial \widetilde{u_l''u_j''}} \right) \Psi_{u,l,0} \Psi_{u,k,0} + \dfrac{\partial G_{ki}}{\partial \epsilon} \omega_t \Psi_{u,j,0} \Psi_{u,k,0}~, \\
	& F_{ijk}^{(3a)} = \dfrac{\partial G_{li}}{\partial (\partial_k\widetilde{u_j})} \dfrac{1}{\omega} \ii k_m \Psi_{u,l,0} \Psi_{u,m,0}~, \\
	& F_{ijk}^{(3b)} = - \dfrac{\partial G_{li}}{\partial \widetilde{u_j''u_k''}} \dfrac{1}{\omega} k_m \Psi_{u,l,0} \Psi_{u,m,0} - \dfrac{1}{2} \dfrac{\partial G_{li}}{\partial \epsilon} \dfrac{\omega_t}{\omega} k_m \Psi_{u,l,0} \Psi_{u,m,0} \delta_{jk}~.
\end{align}

\Cref{eq:appB-AutocorrelationC3Temp} can be drastically simplified by remarking that any integral involving the product of a function $f$ with the kernel function $K$ correspond, by construction, to the ensemble average of said function $f$ (see Paper I for more details). Similarly, if the integral is weighted by the gradient of the kernel function, then it corresponds to the ensemble average of the gradient of $f$ (with a minus sign). But every quantity appearing in \cref{eq:appB-AutocorrelationC3Temp} is either already an ensemble average or an equilibrium quantity, or else the normalised velocity eigenfunction $\bm{\Psi}_u$. None of these are stochastic quantities, which means they are equal to their own ensemble average. This allows us to perform the kind of simplification illustrated by Eq. 50 of Paper I, and we eventually find
\begin{align}
	& \langle c_3(t) c_3(t+\tau) \rangle = \displaystyle\int \d^3\mathbf{X} \displaystyle\int \d^3 \delta\mathbf{x} ~ \rho_0^2 \exp^{-2\ii\mathbf{k} \cdot \delta\mathbf{x}} \left[ \vphantom{\underbrace{1}_{1}} \right. \nonumber \\
	& \hspace{0.4cm} F_i^{(1)} F_j^{(1) \star} \left\langle \uti{i}(\mathbf{X}) \uti{j}^\tau(\mathbf{X} + \delta\mathbf{x}) \right\rangle \nonumber \\
	& + F_i^{(1)} F_{jkl}^{(3b) \star} \left\langle \uti{i}(\mathbf{X}) \uti{j}^\tau \uti{k}^\tau \uti{l}^\tau(\mathbf{X} + \delta\mathbf{x}) \right\rangle \nonumber \\
	& + F_i^{(1) \star} F_{jkl}^{(3b)} \left\langle \uti{i}^\tau(\mathbf{X}) \uti{j} \uti{k} \uti{l}(\mathbf{X} + \delta\mathbf{x}) \right\rangle \nonumber \\
	& + F_{ij}^{(2)} F_{kl}^{(2) \star} \left\langle \uti{i} \uti{j}(\mathbf{X}) \uti{k}^\tau \uti{l}^\tau(\mathbf{X} + \delta\mathbf{x}) \right\rangle \nonumber \\
	& + F_{ijm}^{(3a)} F_{kln}^{(3a) \star} \left\langle \uti{i} \partial_m \uti{j}(\mathbf{X}+\delta\mathbf{x}) \uti{k}^\tau \partial_n \uti{l}^\tau(\mathbf{X} + \delta\mathbf{x}) \right\rangle \nonumber \\
	& + F_{ijm}^{(3a)} F_{kl}^{(2) \star} \left\langle \uti{i} \partial_m \uti{j}(\mathbf{X}) \uti{k}^\tau \uti{l}^\tau(\mathbf{X} + \delta\mathbf{x}) \right\rangle \nonumber \\
	& \left. \vphantom{\underbrace{1}_{1}} + F_{ij}^{(2)} F_{klm}^{(3a) \star} \left\langle \uti{i} \uti{j}(\mathbf{X}) \uti{k}^\tau \partial_m \uti{l}^\tau(\mathbf{X} + \delta\mathbf{x}) \right\rangle \right]~,
\end{align}
where we defined $F^{(1)} \equiv F^{(1a)} + F^{(1b)}$.

Finally, plugging this into \cref{eq:appB-AutocorrelationSpectrumC3}, we find the following expression:
\begin{multline}
	\alpha_3 = \displaystyle\int \d^3\mathbf{X} ~ \rho_0^2 \left(\vphantom{\dfrac{1}{2}} F_i^{(1)} F_j^{(1) \star} \phi_{ij}^{(2)}(2\mathbf{k}, 2\omega) \right. \\
	+ 2 \mathrm{Re}\left[F_i^{(1)} F_{jkl}^{(3b) \star} \phi_{ijkl}^{(4a)}(2\mathbf{k}, 2\omega) \right] + F_{ij}^{(2)} F_{kl}^{(2) \star} \phi_{ijkl}^{(4b)}(2\mathbf{k}, 2\omega) \\
	+ \left. \vphantom{\dfrac{1}{2}} F_{ijm}^{(3a)} F_{kln}^{(3a) \star} \phi_{ijkl}^{(4c)}(2\mathbf{k}, 2\omega) + 2 \mathrm{Re}\left[F_{ijm}^{(3a)} F_{kl}^{(2) \star} \phi_{ijkl}^{(4d)}(2\mathbf{k}, 2\omega) \right] \right)~,
	\label{eq:appB-Alpha3Temp}
\end{multline}
where we recall that the wave vector $\mathbf{k}$ and the angular frequency, $\omega$, are those of the mode under consideration, and we have defined the following spectra on the same template as \cref{eq:appB-Phi4b}:
\begin{align}
    & \phi_{ij}^{(2)}(\mathbf{k},\omega) \equiv \mathcal{C}_{\omega,\mathbf{k}}( \uti{i} ~;~ \uti{j} )~, \\
    & \phi_{ijkl}^{(4a)}(\mathbf{k},\omega) \equiv \mathcal{C}_{\omega,\mathbf{k}}( \uti{i} ~;~ \uti{j} \uti{k} \uti{l} )~, \\
    & \phi_{ijkl}^{(4b)}(\mathbf{k},\omega) \equiv \mathcal{C}_{\omega,\mathbf{k}}( \uti{i} \uti{j} ~;~ \uti{k} \uti{l} )~, \\
    & \phi_{ijklmn}^{(4c)}(\mathbf{k},\omega) \equiv \mathcal{C}_{\omega,\mathbf{k}}( \uti{i} \partial_m \uti{j} ~;~ \uti{k} \partial_n \uti{l} )~, \\
    & \phi_{ijklm}^{(4d)}(\mathbf{k},\omega) \equiv \mathcal{C}_{\omega,\mathbf{k}}( \uti{i} \partial_m \uti{j} ~;~ \uti{k} \uti{l} )~.
\end{align}

We note, as we did in \cref{sec:appB-Alpha1}, that $F^{(1)}$, $F^{(2)}$, $F^{(3a)}$, and $F^{(3b)}$ are also functions of $\mathbf{X}$, and that $\phi_{ij}^{(2)}$, $\phi_{ijkl}^{(4a)}$, $\phi_{ijkl}^{(4b)}$, $\phi_{ijkl}^{(4c)}$, and $\phi_{ijkl}^{(4d)}$ also depend on both $\mathbf{X}$ and $t$, even though these dependences do not appear explicitly in \cref{eq:appB-Alpha3Temp}.

\subsection{Mode normalisation and final form of $\alpha_i$\label{sec:appB-Normalisation}}

The last remaining modification to the explicit expressions of $\alpha_1$ and $\alpha_3$ concerns the normalisation condition on the mode $\ket{\Psi}$. Indeed, we need to relate $\bm{\Psi}_{u,0}(\mathbf{x})$ to the actual modal velocity fluctuation $\uosc(\mathbf{x})$, as it can be obtained through an oscillation code for instance. By construction of the ket $\ket{\Psi}$, we have this very simple proportionality relation:
\begin{equation}
    \bm{\Psi}_u = \dfrac{\uosc}{\sqrt{2\mathcal{I}\omega^2}}~,
    \label{eq:appB-ProportionalityRelation}
\end{equation}
where the proportionality factor $1 / \sqrt{2 \omega^2 \mathcal{I}}$ is given by the condition that $\ket{\Psi}$ must be normalised to unity, such that
\begin{equation}
    \braket{\Psi|\Psi} = 1~.
    \label{eq:appB-NormalisationCondition}
\end{equation}
Plugging \cref{eq:DefinitionKet,eq:ScalarProduct} into \cref{eq:appB-NormalisationCondition}, this becomes
\begin{equation}
    \displaystyle\int \d^3\mathbf{x} ~ \rho_0(\mathbf{x}) \left(\vphantom{f_1^\tau} \left|\bm{\Psi}_\xi\right|^2 + \left|\bm{\Psi}_u\right|^2\right) = 1~,
\end{equation}
and since $\bm{\Psi}_u = \ii \bm{\Psi}_\xi$,
\begin{equation}
    2 \displaystyle\int \d^3\mathbf{x} ~ \rho_0(\mathbf{x}) \left|\bm{\Psi}_u(\mathbf{x})\right|^2 = 1~.
\end{equation}
Finally, plugging \cref{eq:appB-ProportionalityRelation}, we find
\begin{equation}
    \mathcal{I} = \dfrac{1}{\omega^2} \displaystyle\int \d^3\mathbf{x} ~ \rho_0(\mathbf{x}) \left|\uosc(\mathbf{x})\right|^2 = \displaystyle\int \d^3\mathbf{x} ~ \rho_0(\mathbf{x}) \left|\xiosc(\mathbf{x})\right|^2~,
\end{equation}
which we recognise as the inertia of the mode. We then find the relation between $\bm{\Psi}_{u,0}$ and $\uosc$ by plugging \cref{eq:appB-JWKB} into \cref{eq:appB-ProportionalityRelation}. In turn, plugging \cref{eq:appB-ProportionalityRelation} into \cref{eq:appB-Alpha1Temp,eq:appB-Alpha3Temp}, we find the final expressions for $\alpha_1$ and $\alpha_3$ reproduced in the main body of the paper (\cref{eq:ExpressionAutocorrelationSpectrumC1,eq:ExpressionAutocorrelationSpectrumC3}).

\end{appendix}

\end{document}